\PassOptionsToPackage{table}{xcolor}
\documentclass{aa}  

\usepackage{graphicx}
\usepackage{txfonts}
\usepackage{tikz}
\usepackage{booktabs}
\usepackage{longtable}
\usepackage{threeparttable}
\usepackage{amsmath,siunitx}
\usepackage{lipsum}
\usepackage{amsmath}
\usepackage{adjustbox}
\usepackage{lscape}             
\usepackage{placeins}           
\usepackage{hyperref}

\newcommand{\asym}[3]{\ensuremath{#1^{+#2}_{-#3}}}

\begin{document}

   \title{ALMA visits the QSO MUSEUM: Connecting molecular gas and the cool circumgalactic medium around 37 $z\sim 3$ quasars}

   \author{Jelena Ritter\inst{1}\thanks{E-mail: jritter@mpa-garching.mpg.de}
        \and 
        Fabrizio Arrigoni Battaia\inst{1}
        \and 
        Bo Peng\inst{1}
        \and 
        Jay González Lobos\inst{2}
        \and
        Chian-Chou Chen\inst{3}
        \and 
        Aura Obreja\inst{4, 5}
        \and
        Nahir Mu\~noz-Elgueta\inst{1}
        \and
        Chiara Circosta\inst{6}
        }

   \institute{Max-Planck-Institut f\"ur Astrophysik, Karl-Schwarzschild-Str. 1, D-85748 Garching bei M\"unchen, Germany
   \and 
   Max-Planck-Institut f\"ur Astronomie, K\"onigstuhl 17, D-69117 Heidelberg, Germany
   \and
   Academia Sinica Institute of Astronomy and Astrophysics (ASIAA), 11F of Astronomy-Mathematics Building, AS/NTU, No. 1,
    Section 4, Roosevelt Road, Taipei 106319, Taiwan
    \and 
    Universit\"at Heidelberg, Interdisziplin\"ares Zentrum f\"ur Wissenschaftliches Rechnen (IWR), Im Neuenheimer Feld 205, D-69120 Heidelberg, Germany
   \and
   Universit\"at Heidelberg, Zentrum f\"ur Astronomie, Institut f\"ur Theoretische Astrophysik, Albert-Ueberle-Straße 2, D-69120 Heidelberg, Germany
   \and
    Institut de Radioastronomie Millimétrique (IRAM), 300 Rue de la Piscine, F-38400 Saint-Martin-d’Hères, France
   }
   
   \date{Received September xxxx, accepted xxxx}

  \abstract
   {Extended Ly$\alpha$ emission is ubiquitously observed around quasars. This emission traces the cool gas within the circumgalactic medium and provides insights into the complex interplay between gas dynamics and active galactic nucleus feedback. However, its connection to the cold molecular gas of the host galaxies remains largely unexplored.}
   {We aim to characterize the molecular gas reservoirs in quasars at cosmic noon and investigate how they are linked to extended Ly$\alpha$ emission. } 
   {We present ALMA observations of the CO(4-3) transition in 37 quasars at $z\sim3$ from the QSO MUSEUM survey, which have been previously mapped in Ly$\alpha$ with VLT/MUSE. We derive molecular gas masses and gas fractions, explore correlations with Ly$\alpha$ nebula and quasar properties, and search for CO-emitting companions in the fields.}
   {Of 37 quasars, 21 are detected in CO(4-3), with gas masses of $M_\mathrm{gas}\approx(3-40) \times10^9\,\mathrm{M_\odot}$. Quasars with the most massive molecular gas reservoirs are associated with the centrally dimmest Ly$\alpha$ nebulae, while those hosting the centrally brightest Ly$\alpha$ nebulae are generally not detected in CO. This suggests that gas and dust in the hosts regulate Ly$\alpha$ escape and consequently affect the emission from halo gas. We find evidence that quasars with lower Eddington ratios harbor more massive gas reservoirs, while strongly accreting quasars ($\lambda_\mathrm{Edd} \gtrapprox 0.9$) likely deplete their gas; for example, through powerful quasar-driven outflows. Despite their higher molecular gas masses within the sample, the low-Eddington quasars with CO detections exhibit low gas fractions with a median of $M_\mathrm{gas}/M_* \sim 0.10$, below what is typically found for inactive star-forming galaxies. Six quasars are marginally resolved in CO, with effective radii up to $\sim 8\,\mathrm{kpc}$. In addition, we detect 14 companion galaxies at high fidelity, indicating an overall overdense environment in the quasar fields with a derived quasar-galaxy cross-correlation length of $9.81^{+2.22}_{-2.05}\,h^{-1}\mathrm{cMpc}$. }
  {}


\titlerunning{Molecular gas and the cool CGM around $z\sim3$ quasars}
\maketitle
\nolinenumbers

\section{Introduction}

The growth and activity of supermassive black holes (SMBHs) are tightly coupled to the evolution of their host galaxies \citep{Fabian2012, KormendyHo2013}. Periods of rapid accretion onto SMBHs power active galactic nuclei (AGNs), whose radiative and mechanical feedback can regulate star formation and drive large-scale outflows \citep[e.g.,][]{DiMatteo2005,Fabian2012, Weinberger2017,Zinger2020}.

Quasars, the most luminous class of AGN, reside in massive dark matter (DM) halos of $\sim 10^{12.5}\,\mathrm{M}_\odot$ \citep{White2012, Timlin2018, Farina2019, DeBeer2023, Costa2024}, placing them in overdense regions of the Universe. Galaxy interactions and mergers, along with the enhanced gas infall in these environments, fuel both black hole accretion and star formation in the quasar hosts \citep{KauffmannHaehnelt2000}, resulting in host galaxies that are typically more massive and actively star-forming than their contemporaries \citep[e.g.,][]{Pitchford2016, Molina2023}. At redshifts of $>2$ and for halo masses characteristic of quasar hosts, a substantial fraction of the gas supply is expected to occur through accretion of cool gas (“cold-mode”) from the intergalactic medium (IGM) \citep{DekelBirnboim2006}. This process naturally gives rise to a multiphase circumgalactic medium (CGM), in which cool, dense gas streams coexist with hotter, more diffuse halo gas.

Studying the CGM of quasars is crucial to understanding the cycle of feedback and gas recycling. 
One particularly important emission line for the study of the CGM is Ly$\alpha$, tracing the cool  ($\sim 10^4\,\mathrm{K}$) halo gas.
With the advent of sensitive integral field unit (IFU) spectrographs such as the Multi-Unit Spectroscopic Explorer (MUSE; \citealt{Bacon2010}) and the Keck Cosmic Web Imager (KCWI; \citealt{Morrissey2012}), extended Ly$\alpha$ emission can be detected down to unprecedentedly low surface brightness levels. These observations have established that quasars and AGN at $z\,>\,2$ are ubiquitously surrounded by extended Ly$\alpha$ nebulae, prompting large dedicated surveys \citep[e.g.,][]{Borisova2016, Farina2019, Cai2019, ArrigoniBattaia2019, Osullivan2020, Fossati2021, Mackenzie2021, GonzalezLobos2025}. The nebulae typically extend to scales of $\sim100\,\mathrm{kpc}$, and in exceptional cases reach several hundred kiloparsecs, forming the so-called enormous Ly$\alpha$ nebulae (ELANe; e.g., \citealt{Cantalupo2014, Hennawi2015, Cai2017, ArrigoniBattaia2018}).

Active galactic nucleus activity is thought to play a key role in shaping the observed Ly$\alpha$ nebulae, primarily through recombination after quasar photoionization \citep[e.g.,][]{Heckman1991, Cantalupo2005, Costa2022, GonzalezLobos2025} and resonant scattering of photons from the broad-line region (BLR) \citep[e.g.,][]{Cantalupo2014, Costa2022, GonzalezLobos2025}. Collisional excitation through gravitational cooling  \citep[e.g.,][]{Haiman2000, Dijkstra2006, DijkstraLoeb2009, Faucher-Giguere2010} is likely subdominant in the presence of a bright AGN \citep{Costa2022, GonzalezLobos2023}.
Simulations further suggest that AGN feedback, particularly quasar-driven outflows, plays a key role in facilitating Ly$\alpha$ photon escape and producing the bright, large-scale nebulae observed around quasars \citep{Costa2022}. 

The Ly$\alpha$ emitting gas is likely composed of cool and compact gas clumps of densities similar to those in the interstellar medium (ISM; $n_\mathrm{H}>1\,\mathrm{cm}^{-3}$) as invoked from photoionization scenarios \citep{Cantalupo2014, Hennawi2015, ArrigoniBattaia2015, Borisova2016}. A density distribution function with a tail at these high densities is indeed needed in cosmological simulations to be able to reproduce the extended Ly$\alpha$ emission around quasars \citep{Costa2022,Obreja2024}.
Given these conditions, the cool halo gas could shield and harbor a cold, dense molecular phase (especially while it approaches the host galaxy), motivating efforts to establish a connection between the two gas phases and how they relate to AGN feedback processes \citep{Drake2022}. A pilot study by \citet{MunozElgueta2022}, targeting nine quasars in CO(7-6), CO(6-5), and [\ion{C}{i}](2-1), with the Atacama Pathfinder Experiment (APEX) telescope, revealed that the quasars with the most massive molecular gas reservoirs are embedded in the dimmest, least extended Ly$\alpha$ nebulae. These findings suggest that obscuration in the host galaxy could play a role in suppressing the reprocessed Ly$\alpha$ emission.

Beyond the host galaxies, extended molecular gas reservoirs of quasars out to CGM scales have also been detected, albeit only in rare cases \citep[e.g.,][]{Emonts2019, Cicone2021, Li2021, Jones2023, Scholtz2023, Li2023}. Molecular gas in the CGM could arise from close satellites, dense accreting gas, or gas expelled from the host galaxy by feedback processes, tidal interactions, or ram-pressure stripping. In this context, “cloud-crushing” simulations \citep[e.g.,][]{McCourt2018, GronkeOh2018, GronkeOh2020, Kanjilal2021} predict that cold gas can survive and even grow within a hot halo under favorable conditions. However, the latest observations suggest the presence of molecular gas only out to distances of about $10-15\,\mathrm{kpc}$ from the quasar \citep[e.g.,][]{Scholtz2023}. Such scales might still correspond to ISM extents, when considering predictions from cosmological simulations \citep[e.g.,][]{Obreja2024}.

Cold molecular gas is typically traced through the rotational ($J$) transitions of CO \citep[e.g.,][]{Bolatto2013}. With the development of advanced interferometric submillimeter facilities, such as ALMA (Atacama Large Millimeter/Submillimeter Array; \citealt{WoottenThompson2009}), quasar host galaxies can now be studied in detail at high redshift, revealing molecular gas masses in the range of $10^9$–$10^{11}\,\mathrm{M}_\odot$ \citep[e.g.,][]{2013CarilliWalter,Feruglio2018, Hill2019, Pensabene2020, Bischetti2021, Decarli2022, Li2023}.

At the same time, strong AGN activity, and the consequent radiation and outflows, can heat, dissociate, and expel molecular gas. Indeed, high-redshift ($z>2$) AGN host galaxies often show depleted gas reservoirs in the form of low gas fractions (gas mass to stellar mass ratio) compared to inactive (non-AGN) galaxies \citep[e.g.,][]{Brusa2018, Bischetti2021, Circosta2021, Bertola2024, Molyneux2025}. On the other hand, quasars in the local Universe are often found to be gas-rich and star-forming \citep[e.g.,][]{Rosario2018, Saintonge2012, Jarvis2020, Koss2021} with no clear evidence of AGN activity significantly impacting the ISM of their hosts. This contrast highlights the complex interplay between molecular gas and AGN feedback and hints at an evolution of the impact of feedback across cosmic time.

On larger scales, an important complementary perspective is provided by the environments that quasars inhabit. High-redshift quasars are found to be strongly clustered \citep[e.g.,][]{Shen2007,Shen2010}.
Overdensities surrounding quasars have been demonstrated in the literature across a range of tracers, including rest-frame UV-selected galaxies and dust-obscured submillimeter galaxies \citep[SMGs; e.g.,][]{Silva2015, Wylezalek2016, Fan2016, GarciaVergara2017, Decarli2017, Banerji2017,  Banerji2018, DiazSantos2018, Ota2018, GarciaVergara2019, Fogasy2020, Bischetti2021, GarciaVergara2022, Li2023, ArrigoniBattaia2023, Herwig2025}.
In particular, quasars at $z\sim3-4.5$ hosting extended Ly$\alpha$ nebulae were found to show significant evidence for overdensities of Ly$\alpha$ emitters (LAEs) in the vicinity of the quasars \citep{Fossati2021}. Together, these results indicate that quasars at $z\sim3$ inhabit environments where companion galaxies are common, motivating systematic searches for such systems in the submillimeter.

In this work, we present a sample of 37 quasars at $z\sim3$ observed with ALMA to study their molecular gas reservoirs via the CO($J$=4–3) rotational transition. The quasars are part of the QSO MUSEUM (Quasar Snapshot Observations with MUse: Search for Extended Ultraviolet eMission) survey \citep{ArrigoniBattaia2019, Herwig2024, GonzalezLobos2025}. This survey targets a population of 120 $z\sim3$ quasars with VLT/MUSE to study their extended Ly$\alpha$ emission and search for ELANe. It has so far revealed that (i) the extended Ly$\alpha$ emission is likely powered by a combination of photoionization, resonant scattering, and shocks, (ii) the average kinematics are consistent with gravitational motions within quasar-hosting DM halos (however, the velocity dispersion in the inner 40~kpc increases with higher quasar bolometric luminosities), (iii) similarly to the velocity dispersion, the Ly$\alpha$ surface brightness also varies with quasar bolometric luminosity, hinting at instantaneous AGN feedback effects on CGM scales, and (iv) the surface brightness level of Ly$\alpha$ emission evolves from $z\sim2$ to $z\sim3$.
Following the work of \citet{MunozElgueta2022}, we aim to investigate the connection between the cool halo gas traced by Ly$\alpha$ and the cold molecular gas, using a statistically more representative sample. In particular, we aim to constrain the molecular gas properties in a large sample of $z\sim3$ quasars and establish a connection between the molecular phase, the large-scale halo gas, and AGN feedback processes. Additionally, we investigate the small-scale clustering by searching for line-emitting companion galaxies.

This paper is structured as follows. In Section \ref{sec:Observations}, we provide an overview of the sample, observations, and data reduction. In Section \ref{sec:Results}, we present our analysis of the molecular gas and a comparison with quasar and Ly$\alpha$ properties. In Section \ref{sec:Discussion}, we discuss our findings and summarize the results in Section \ref{sec:Conclusions}. 
Throughout this paper, we adopt a flat $\Lambda$ cold dark matter cosmology with $\mathrm{H}_0 = 67.7\,\mathrm{km}\,\mathrm{s}^{-1}\,\mathrm{Mpc}^{-1}$, $\Omega_{\mathrm{m}} = 0.31$, and $\Omega_{\Lambda} = 0.69$ \citep{Planck2018}.

\section{Observations and data reduction}
\label{sec:Observations}
\subsection{Sample selection}
Our sample of quasars was selected from the QSO MUSEUM survey \citep{ArrigoniBattaia2019, GonzalezLobos2025}. The QSO MUSEUM quasars were selected from the Sloan Digital Sky Survey (SDSS, DR14 \citealt{Paris2018}) and the 13th edition of the \citet{VeronCettyVeron2010} catalog of quasars and AGNs. The sample is divided into a bright subsample \citep{ArrigoniBattaia2019, GonzalezLobos2025} with 61 quasars of absolute i-band magnitude normalized at $z = 2$ in a range $-29.67 \leq M_i(z = 2) \leq -27.03$ and a fainter subsample \citep{GonzalezLobos2025} with 59 quasars of magnitudes $-27 < M_i(z = 2) < -24$.

\begin{figure}[!ht]
    \centering
    \begin{adjustbox}{width=1.0\linewidth,center}
        \includegraphics{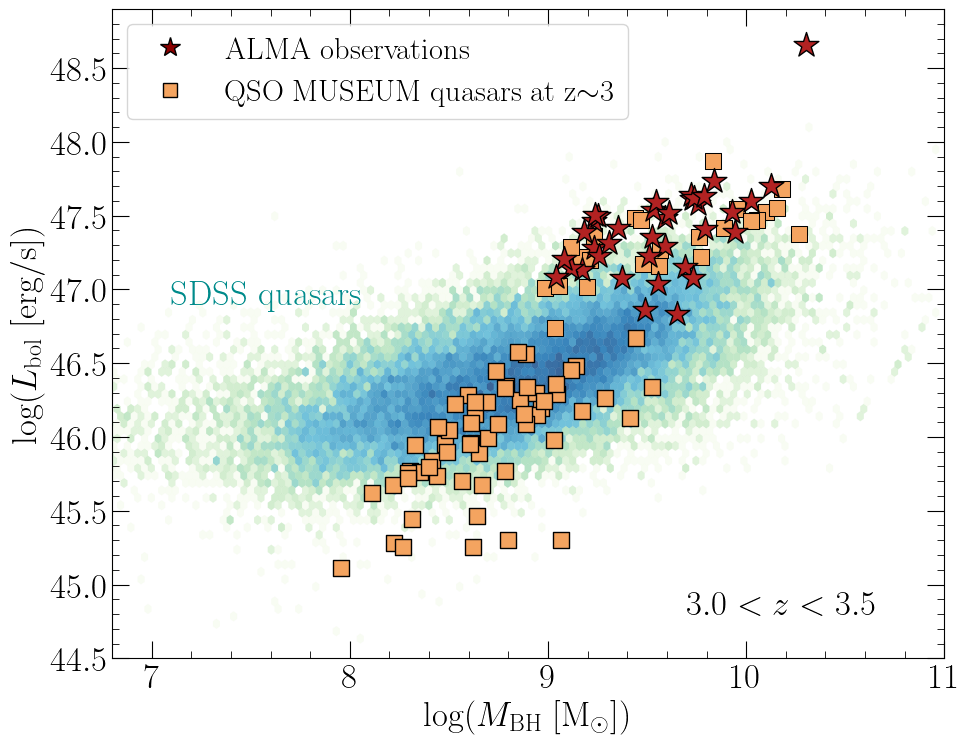}
    \end{adjustbox}
\caption[]{Bolometric luminosity as a function of black hole mass for the QSO MUSEUM quasars. The red stars indicate the ALMA follow-up observations. The SDSS quasars in the same redshift range of $3 < z < 3.5$ are plotted in blue as 2D number density bins \citep{Rakshit2020}. }
    \label{fig:LbolvsMBH}
\end{figure}

We selected 37 quasars from the bright subsample (with high bolometric luminosity ($L_{\mathrm{bol}}$) and nebular surface brightness) to be followed up with ALMA to explore a connection between their Ly$\alpha$ and CO emission (see Figure \ref{fig:LbolvsMBH} for the location of the selected quasars in the bolometric luminosity ($L_{\mathrm{bol}}$) versus black hole mass ($M_{\mathrm{BH}}$) plane). The targets were chosen such that they cover a wide range of Ly$\alpha$ nebula properties and surface brightnesses.

\subsection{ALMA observations} \label{sec:ALMA_observations}

The observations of the 37 quasars were carried out in ALMA Cycle 10 under the program ID 2023.1.00963.S (PI: Fabrizio Arrigoni Battaia) between October 3 2023 (UTC) and April 22 2024. The data were taken in Band 3 (84 to 116\,GHz) with the compact 12-m array configuration (maximum of 50 antennas with baselines ranging from 15\,m up to a maximum of 952\,m across our observations), as well as the 7-m Atacama Compact Array (ACA) configuration. It provides a primary beam of half power beam width (HPBW) of approximately $53\arcsec$ at the observed frequency of the sample, comparable to the $60\arcsec \times 60\arcsec$ MUSE field of view. The program targeted the CO(4-3) ($\nu_\mathrm{rest}$=461.041 GHz) line emission. The total on-source exposure times varied between about 30 and 90 minutes for the 12-m array and were chosen to be sensitive to a molecular mass as low as $ 5.5\times10^9\,\mathrm{M_\odot}$, on average. The target properties and observing conditions are summarized in Table \ref{tab:observing_summary}. The ACA observations were automatically carried out to capture the largest angular scales given in the proposal, but their depth is insufficient for our analysis and therefore not included here.
The spectral setup comprised four spectral windows with a bandwidth of 1.875~GHz ($\Delta v\sim 5100\,\mathrm{km\,s^{-1}}$ at the median redshift of the sample of $z\sim3.2$), one containing the CO(4-3) line and the other three the dust continuum.

The data were reduced using standard procedures with the Common Astronomy Software Applications (CASA) package version 6.5.4.9 \citep{CASA}. The calibration of the data was performed using the ALMA calibration pipeline 2023.1.0.124 \citep{Hunter2023}, together with the calibration scripts that were supplied with the archival data by the ALMA observatory. The pipeline calibration was used to create the measurement set (MS) for each observing run, which were then combined into a single MS for each source when multiple runs were available.

Using the task \texttt{uvcontsub}, we fit and subtracted the continuum from the MS in the uv domain using the three spectral windows unaffected by CO(4-3) emission. The datacube containing CO(4-3) was created from the task \texttt{tclean} with Briggs weighting and robustness parameter 2 and an automated cleaning mask with noisethreshold=3.5, lownoisethreshold=1.5, sidelobethreshold=2.0, and minbeamfrac=0.3. Robustness parameters of 2 (which maps to natural weighting) are used for high-redshift quasars to maximize the signal-to-noise ratio (S/N; see e.g., \citealt{Decarli2018}). We cleaned down to five times the sensitivity in the cube, but verified that reducing the threshold to three or two times does not noticeably change the final datacubes. The channel width was specified to be 50\,km/s for all cubes (with a native velocity resolution of $\sim2.7\,\mathrm{km/s}$). Across the full sample, the synthesized beam major axis has a median value of $1.6\arcsec$. The observing conditions are summarized in Table \ref{tab:observing_summary}.

We created continuum images by producing multifrequency synthesis (MFS) images from all line-free channels (i.e., the three spectral windows which do not contain the CO(4-3) emission line) via the CASA task \texttt{tclean} using Briggs weighting with the robustness parameter set to 2.
For each target, we additionally created MFS images of the CO emission over the line-emitting channels only using \texttt{tclean}, with the deconvolution performed to a threshold of three times the root-mean-square (rms) sensitivity. These images served as continuum-subtracted moment 0 maps, which we used in particular to examine the spatial extent of the emission region. Primary beam correction was applied to all datacubes and MFS images. 
The depth of each observation is summarized in Table \ref{tab:observing_summary}, with a median value of $\sim 370\,\mu\mathrm{Jy}$ in line rms per $\Delta v= 50\, \mathrm{km\,s^{-1}}$ and $\sim 12\,\mu\mathrm{Jy}\,{\mathrm{beam}}^{-1}$ in continuum rms.

\begin{figure}[!ht]
    \centering
    \begin{adjustbox}{width=1.0\linewidth,center}
        \includegraphics{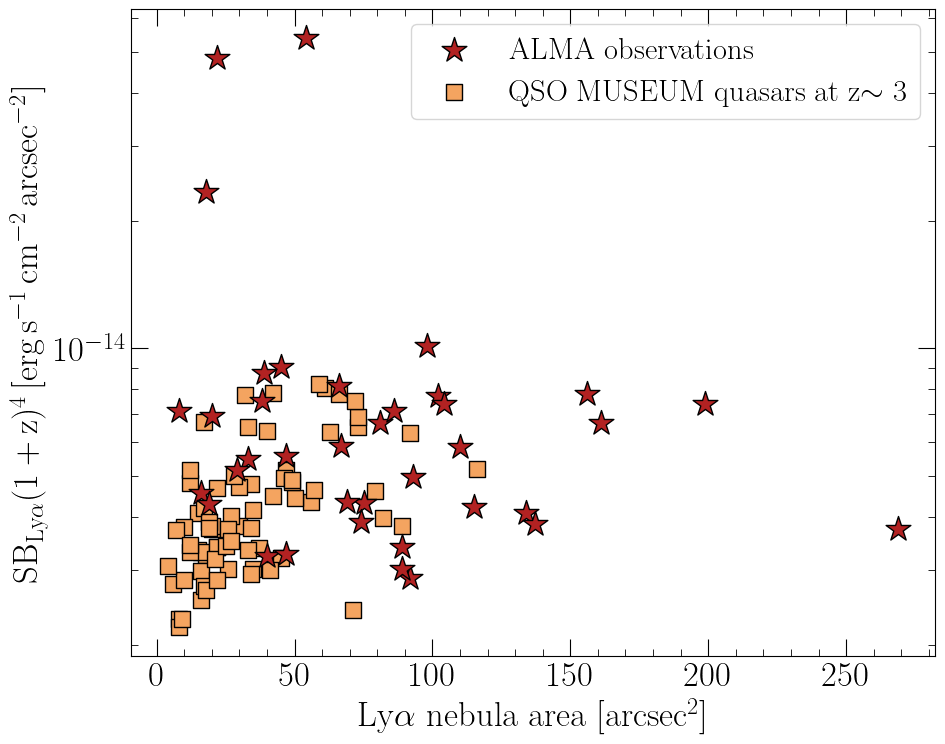}
    \end{adjustbox}
\caption[]{Average Ly$\alpha$ surface brightness corrected for cosmological dimming of the QSO MUSEUM quasars as a function of Ly$\alpha$ nebula area within the $2\sigma$ isophotes (from \citealt{GonzalezLobos2025}). The red stars indicate the ALMA follow-up observations in this work.}
    \label{fig:SBvsArea}
\end{figure}

\subsection{Quasar properties}
All targets have complementary ancillary data available. Black hole masses, bolometric luminosities, and Eddington ratios were taken from \cite{GonzalezLobos2025}. The black hole masses were derived from the \ion{C}{iv} emission line measurements using the \cite{VestergaardPeterson2006} single-epoch method. Bolometric luminosities were computed based on the monochromatic luminosity $L_{\lambda}(1350\,\AA)$, following the approach of \cite{Rakshit2020} and adopting the bolometric correction from \cite{Shen2011}. Appendix B in \citet{GonzalezLobos2025} compares the black hole properties derived in this way with the SDSS-based estimates from \citet{Rakshit2020}. The typical uncertainties are on the order of 0.5 dex for black hole masses and 0.3 dex for bolometric luminosities.

Additionally, all quasars have ancillary MUSE data available to study the extended Ly$\alpha$ emission \citep{ArrigoniBattaia2019, GonzalezLobos2025}. Ly$\alpha$ surface brightness measurements are obtained from \cite{GonzalezLobos2025}. Figure \ref{fig:LbolvsMBH} shows the QSO MUSEUM quasars as well as the quasars followed up with ALMA on the $L_\mathrm{bol}$-$M_\mathrm{BH}$ plot together with SDSS quasars \citep{Rakshit2020} in the same redshift range. Figure \ref{fig:SBvsArea} shows the nebula averaged Ly$\alpha$ surface brightness corrected for cosmological dimming as a function of the nebula area within the 2$\sigma$ isophotes.

\renewcommand{\arraystretch}{1.1}
\begin{table*}[htbp]
\centering
\small
\begin{threeparttable}
\caption{Summary of target properties and observing conditions.}
\label{tab:observing_summary}
\begin{tabular}{|c|c|c|c|c|c|c|c|}
\hline
ID \tnote{\textdagger} & Name & RA & DEC  & Exp. Time  & Beam size &  Line rms \tnote{*} & continuum rms \\
 &  & [J2000] &  [J2000] &  [min] & [arcsec$^2$] &  [$\mu$Jy] & [$\mu$Jy\,beam$^{-1}$]\\
\hline
1 & SDSS J2319-1040  & 23:19:34.771 & -10:40:36.94 & 39 & $1.3\times1.0$ & 405 & 12 \\
2 & UM 24  & 00:15:27.277 & +06:40:11.09 & 46 &  $1.4\times1.3$  & 354  & 10\\
3 & J 0525-233 & 05:25:06.506 & -23:38:10.81 & 31 & $1.4\times1.0$  & 385 & 113 \tnote{\textdaggerdbl} \\
4 & Q-0347-383 & 03:49:43.660 & -38:10:30.83 & 35 & $1.4\times1.1$ & 349 & 12\\
6 & SDSS J0947+1421 & 09:47:34.197 & +14:21:16.94 & 56 &$2.1\times1.4$ & 403 & 11 \\
7 & SDSS J1209+1138 & 12:09:17.939 & +11:38:30.37 & 34& $2.2\times1.9$& 408 & 14\\
8 & UM 683 & 03:36:26.993 & -20:19:40.07 & 31 &$1.4\times1.0$ & 415 & 13\\
10 & SDSS J1025+0452 & 10:25:09.635 & +04:52:46.82 & 36 &$1.9\times1.5$ & 315 & 13 \\
11 & Q-N1097.1 & 02:46:34.145 & -30:04:54.82 & 41 & $1.3\times1.0$ & 385 & 11\\
13 & PKS-1017+109 & 10:20:09.995 & +10:40:02.69 & 46 & $2.8\times1.6$& 309 & 13 \\
14 & SDSS J2100-0641 & 21:00:25.030 & -06:41:45.97 & 31 & $1.2\times1.1$& 306 & 12\\
15 & SDSS J1550+0537 & 15:50:36.802 & +05:37:50.08 & 33 & $2.3\times2.0$& 362 & 13 \\
17 & SDSS J0001-0956 & 00:01:44.887 & -09:56:30.81 & 40 &$1.5\times1.1$ & 329 & 14 \\
18 & SDSS J1557+1540 & 15:57:43.264 & +15:40:20.81 & 41 &$2.1\times1.9$ & 447 & 12 \\
20 & SDSS J1429-0145 & 14:29:03.038 & -01:45:19.37 & 43 & $2.3\times1.9$& 336 & 12 \\
21 & CT-669 & 20:34:26.336 & -35:37:27.01 & 34& $1.2\times1.0$& 351 & 11 \\
24 & SDSS J1342+1702 & 13:42:33.243 & +17:02:47.02 & 97& $2.3\times2.0$ & 275 & 8 \\
26 & Q-2204-408 &22:07:34.415 & -40:36:56.01 & 39& $1.2\times1.1$& 370 &   13\\
27 & Q-2348-4025 & 23:51:16.076 & -40:08:35.87 & 40& $1.4\times1.2$ & 301 & 12 \\
29 & Q-0115-30 & 01:17:34.013 & -29:46:28.78 & 34 & $1.3\times1.1$ & 356 & 13\\
30 & SDSS J1427-0029 & 14:27:55.852 & -00:29:51.14 & 41 &$2.3\times1.8$ & 333 & 13 \\
32 & Q-0058-292 & 01:01:04.650 & -28:58:01.68 & 41 &$1.4\times1.1$ & 597 &  11 \\
 34 & Q-0057-3948 & 00:59:53.219 & -39:31:57.44 & 36 &$1.5\times1.2$ &575 & 12 \\
36 & Q-0052-3901A & 00:54:45.303 & -38:44:14.94& 34 & $1.3\times1.1$& 374 & 12 \\
38 & SDSS J0125-1027 & 01:25:30.860 & -10:27:39.73 & 39&$1.6\times1.1$ & 271 & 10 \\
39 & SDSS J0100+2105 & 01:00:27.664 & +21:05:41.55 & 85 & $2.3\times2.1$ &  341 &  10 \\
43 & CTSH22.05 & 01:48:18.118 & -53:27:02.03 & 33 & $1.4\times1.2$&  580 & 11 \\
44 & SDSS J2321+1558 & 23:21:54.985 & +15:58:34.20 & 39& $2.3\times2.0$& 405 & 14 \\
45 & FBQS J2334-0908 & 23:34:46.405 & -09:08:12.25 & 40& $1.3\times1.1$ & 586 & 13 \\
46 & Q2355+0108 & 23:58:08.540 & +01:25:07.23 & 43 & $2.0\times1.8$& 295 & 11 \\
47 & 6dF J0032-0414 & 00:32:05.384 & -04:14:16.14 & 33 &$1.6\times1.1$ & 401 & 14 \\
49 & PKS0537-286 & 05:39:54.281 & -28:39:55.95 & 31 &$1.3\times1.0$ & 493 & 142 \tnote{\textdaggerdbl} \\
50 & SDSS J0819+0823 & 08:19:40.577 & +08:23:58.03 & 36& $1.9\times1.2$& 378 & 11 \\
51 & SDSS J0814+1950 & 08:14:53.453 & +19:50:18.68 & 36& $1.7\times1.3$ & 365 & 14 \\
52 & SDSS J0827+0300 & 08:27:21.966 & +03:00:54.71 & 32& $2.2\times1.3$& 441 & 15\\
56 & TEX1033+137 & 10:36:26.885 & +13:26:51.76 & 50 & $2.1\times1.8$& 432 & 12\\
58 & Q1205-30 & 12:08:12.751 & -30:31:07.59 &  76 &$2.6\times1.9$ & 362 & 10\\
\hline
\end{tabular}
\begin{tablenotes}
\footnotesize
\item[\textdagger] The IDs refer to the identification in \citet{ArrigoniBattaia2019}. 
\item[*] rms noise per $\Delta v= 50\, \mathrm{km\,s^{-1}}$ of the quasar spectra in an emission-free region.
\item[\textdaggerdbl] ID3 and ID49 exhibit elevated continuum noise, even after performing additional phase self-calibration, likely due to the brightness of the blazar continuum emission.
\end{tablenotes}
\end{threeparttable}
\end{table*}

\section{Results} \label{sec:Results}
\subsection{ALMA CO(4-3) line detections from quasars} \label{sec:Linedetections}
To robustly identify sources with CO(4–3) emission in the data cubes, we used the code \texttt{Lineseeker} \citep{Gonzalez-Lopez2017, Gonzalez-Lopez2019}, which searches for emission-line candidates in the ALMA cubes and provides S/N estimates for each detection. The code was run starting from the recommended \verb|MinSN| and \verb|MaxSigmas| settings. Sources were considered detections if their S/N exceeded that of any detection in the negative data. We additionally inspected the spectra at the quasar positions for potential line detections not identified by \texttt{Lineseeker}. Using this criterion, CO(4-3) is robustly detected (with fidelity $\sim100\%$) in 20 out of the 37 quasars. We additionally find very narrow line emission at the quasar position and close to the peak Ly$\alpha$ redshift of ID3. As continuum subtraction was more challenging in this source due to strong synchrotron emission, we verified that the line was also present in the unsubtracted cube, where it was clearly visible on top of the continuum. Although of low fidelity, its spatial and redshift agreement with the quasar led us to classify it as a tentative detection. Including this source as a detection does not affect our results, as its molecular gas mass is very low and comparable to the upper limits of the non-detections (see, e.g., Figure~\ref{fig:MH2vsMBH}).

For the detections, we extracted spectra at the location of the peak emission as given from the \texttt{Lineseeker} code with aperture sizes of two or three times the beam size. The choice of aperture size was based on a curve-of-growth analysis. We measured the integrated flux within increasing aperture sizes on the MFS image of the CO emission and determined the beam size at which the cumulative flux begins to plateau. This approach ensures that we capture the total flux associated with each detection while minimizing contamination from noise or nearby sources.

Throughout this work, velocities in the CO(4-3) spectra are defined relative to the centroid of the CO(4-3) emission (determined from a first-moment analysis), such that $0\,\mathrm{km}\mathrm{s}^{-1} $ corresponds to the systemic redshift of the host galaxy. 
To determine the velocity range for flux integration, we fit a Gaussian profile to each extracted spectrum. The number of Gaussian components (one, two, or three) was selected using the Akaike information criterion (AIC; \citealt{Liddle2007}). We then integrated the flux over the velocity interval defined by the line center $\pm$ one FWHM. 

\begin{figure}[!ht]
    \centering
    \begin{adjustbox}{width=1.0\linewidth,center}
        \includegraphics{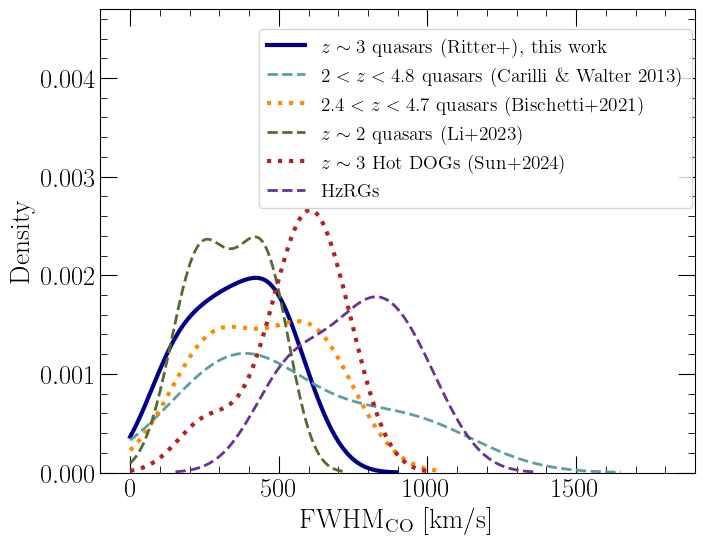}
    \end{adjustbox}
\caption[]{Kernel density estimation of CO(4–3) line widths (FWHMs) for the quasars in the sample analyzed in this work, compared with the line widths of quasars at similar redshifts detected in CO(4–3) from \citet{2013CarilliWalter}, \citet{Bischetti2021} and \citet{Li2023} and dust-obscured quasars from \citet{Sun2024}. We also show line widths of high-redshift radio galaxies ($z>1$) from \citet{DeBreuck2003, DeBreuck2005, Ivison2012, Emonts2023}, which include multiple CO transitions.}
    \label{fig:FWHMs}
\end{figure}

As several CO profiles require multiple Gaussian components, we measured CO line widths using a second-moment analysis, scaled to match the FWHM of a Gaussian, and derived CO redshifts from the first moment of the spectra. The line widths range between $\sim70$ to $640\,\mathrm{km\,s^{-1}}$ with a median of $360\,\mathrm{km\,s^{-1}}$. Figure \ref{fig:FWHMs} shows the distribution of line widths compared with other quasar studies and high-redshift radio galaxies (HzRGs). For direct comparison with our sample, the quasar samples are restricted to CO(4-3) line-width measurements, while the HzRG sample includes multiple CO transitions and is shown for broader context. In the literature, typical molecular line widths of high-redshift quasars are found to be $\sim 150 - 1000\,\mathrm{km\,s^{-1}}$ \citep[e.g.,][]{2013CarilliWalter, Venemans2017, Bischetti2021, Li2023, Molyneux2025}. We find that the detected quasars have line widths similar to previously observed unobscured objects, but lower values than hot dust-obscured systems 
(Hot DOGs; \citealt{Sun2024}) and HzRGs \citep{DeBreuck2003, DeBreuck2005, Ivison2012, Emonts2023}. The difference in CO line widths may arise from a combination of geometric effects and host-galaxy conditions. HzRGs are expected to be observed close to edge-on and are associated with dynamically disturbed environments, while dust-obscured quasars may exhibit broader CO profiles due to a more turbulent ISM within the hosts \citep[e.g.,][]{Martin2024}. 

Interestingly, two quasars with unusually narrow CO line widths in our sample, ID3 and ID49, are both classified as blazars (more specifically “flat-spectrum radio quasars, with an optical spectrum showing broad emission lines and dominant blazar characteristics”) in \citet{Massaro2009}. CO has been only sporadically detected in blazars. \citet{Fumagalli2012} searched for CO emission in three $z\sim0.1$ blazars, detecting it in only one case, and inferred that blazars may host relatively low molecular gas reservoirs (of $\lesssim  10^{9}\,\mathrm{M}_\odot$) compared with quasars. Additionally, the inferred CO(1-0) and CO(2-1) line widths are narrow at $\Delta v = 121$
and $110\,\mathrm{km\,s^{-1}}$, similar to the line widths we obtain. The narrow CO line widths in blazars resemble those of early-type galaxies, where the molecular gas, when present, is often settled in regular rotating disks \citep{young2011}, and the near face-on orientation of blazars may further suppress the observed velocity spread. 
We note, however, that the comparison for ID3 should be treated with caution, as it is a tentative detection and is affected by strong synchrotron emission.

We find offsets between the quasar systemic redshifts determined from the peak of the \ion{C}{iv} line (\citealt{Shen2016}), as reported in \citet{ArrigoniBattaia2019}, and the CO redshifts of up to $3000\,\mathrm{km\,s^{-1}}$. The intrinsic uncertainty on systemic redshifts inferred from \ion{C}{iv} is estimated to be $\sim 415\,\mathrm{km\,s^{-1}}$, and is therefore insufficient to fully account for the observed offsets. In contrast, the Ly$\alpha$ redshifts show better agreement with the CO redshifts (see also Section \ref{sec:COvsLyalpha}).

The CO properties of the quasars are shown in Table \ref{tab:CO_properties}. The spectra, CO flux maps, and Ly$\alpha$ surface brightness maps (after PSF- and continuum-subtraction) are shown in Figure \ref{fig:flux_maps_spectra}.

\subsection{CO line detections in the fields} \label{sec:Linedetections_comp}

In addition to the CO(4–3) detections associated with the target quasars, \textsc{LineSeeker} identifies 14 further line candidates at high fidelity ($\geq80\%$), where the fidelity is estimated from comparison with negative detections \citep[see][]{Gonzalez-Lopez2019}. We note, however, that we did not search for companion galaxies outside the spw containing the CO(4-3) emission line of the quasars and could therefore only detect potential companion galaxies that are within $\sim2000\,\mathrm{km}\,\mathrm{s^{-1}}$ of the quasars' systemic redshifts.

The companion galaxies are listed together with their CO properties in Appendix \ref{sec:companion_galaxies} in Table \ref{tab:companion_properties}. The CO flux maps and spectra are shown in Figure \ref{fig:flux_maps_spectra_companions}. We also list and show the companion galaxies discovered in \citet{ArrigoniBattaia2018} that are detected in CO(4-3). They are a second quasar (QSO2), a strong Ly$\alpha$ emitter (LAE1) and a type-2 AGN (AGN1), following their naming conventions. These companion galaxies have also been detected in CO(5-4) in \citet{ArrigoniBattaia2022}. 

We searched for corresponding optical counterparts for all the companion galaxies in optical imaging but did not find any detections. We additionally inspected the available MUSE data for continuum and emission-line counterparts but found none that would be indicative of lower redshift interlopers (see Section \ref{sec:companions} for details). The 14 CO emitters detected in the 37 fields are found at projected distances of 21 to 282\, kpc from the quasars. For comparison, in the sample of eight hyperluminous quasars studied by \citet{Bischetti2021}, CO(4–3) observations reveal three potential companion galaxies, with the closest at a projected distance of only $6\,\mathrm{kpc}$. The companions of ID2, ID11, ID34, ID45 and ID46 and the type-2 AGN in ID13's vicinity are more luminous in CO than the quasars themselves. However, we note that the companions of ID2 and ID46 lie close to the primary-beam edge, where the noise increases substantially, and their inferred CO luminosities are therefore more uncertain. In three cases (ID8, ID10 and ID50), the quasar itself is not detected. Two companions, ID11\_comp and ID13 LAE1 are also detected in continuum. Figure \ref{fig:companions_distr} shows the distribution of the CO emitters identified in our 37 fields with respect to the quasar on the sky as well as in line-of-sight velocity space.

\subsection{Continuum detections}
The continuum maps correspond to rest-frame $\sim 400 - 460\,\mathrm{GHz}\, (\sim 0.65-0.75\,\mathrm{mm})$. Of 37 quasars, 20 show ALMA dust continuum detections at $> 3\,\sigma$ significance. Of 21 quasars detected in CO, 13 are also detected in continuum. Figure~\ref{fig:flux_maps_spectra} shows the continuum emission contours overlaid on the CO flux maps, while figure~\ref{fig:cont_maps} shows the continuum flux maps for the quasars that are not detected in CO. The spatial locations of the CO emission are consistent with those of the continuum emission. To measure the continuum flux densities, we fit a 2D Gaussian to the emission region using the CASA task \textsc{imfit} and integrated the continuum flux density over the whole emission region. Owing to the limited angular resolution, however, we cannot place strong constraints on the morphology of the continuum emission. For the targets not detected in continuum, we derived $3\,\sigma$ upper limits. The continuum flux for each quasar is listed in Table \ref{tab:CO_properties}. Seven targets (ID3, ID26, ID36, ID45, ID47, ID49 and ID56) show significant continuum emission at the mJy level, which may indicate a contribution from synchrotron emission. All of these targets (except for ID26 which was not covered by any radio surveys) are also confirmed radio-loud quasars, as determined from the FIRST survey \citep{Becker1994} for northern sources and NRAO VLA Sky Survey (NVSS, \citealt{Condon1998}) for southern sources (see also Table \ref{tab:CO_properties}).

\subsection{Molecular gas masses}
To derive the molecular gas masses ($M_\mathrm{gas}$), we calculated the line luminosities, $L'_\mathrm{CO}$, in units of $\mathrm{K\,km\,s^{-1}\,pc^2}$ of the CO(4-3) emission from the integrated flux following the formalism of \citet{Solomon1992, 2013CarilliWalter}:
\begin{equation}
    L'_\mathrm{CO}=3.25\times 10^7\cdot S\Delta v\, \cdot \frac{D_
    L^2}{\nu_\mathrm{obs}^2 (1+z)^{3}},
\end{equation}
where $S\Delta v$ is the velocity-integrated line intensity in units of $\mathrm{Jy\,km\,s^{-1}}$, $D_L$ is the luminosity distance in Mpc, and $\nu_\mathrm{obs}$ is the observed frequency of the CO line in GHz.
We then used the equation $M_\mathrm{gas} = \alpha\,L'_\mathrm{CO(1-0)}$ assuming a CO-to-H$_2$ conversion factor of $\alpha_{\mathrm{CO}} = 0.8\, \mathrm{M_\odot}\,(\mathrm{K\,km\,s^{-1}\,pc^2})^{-1}$. This value is commonly adopted for high-redshift quasars ($z\,\sim\,2-7$, \citealt[e.g.,][]{Bischetti2021, Molyneux2025}) and is based on measurements in local ultra-luminous infrared galaxies (ULIRGs; \citealt{DownesSolomon1998}). Since our observations target the CO(4–3) transition, we adopted a
ratio of $r_{41}= L'_{\mathrm{CO}(4-3)} / L'_{\mathrm{CO}(1-0)} = 0.87$, a value typically assumed for quasar host galaxies \citep{2013CarilliWalter}, to convert the observed line luminosities to CO(1–0) equivalents before applying the CO-to-H$_2$ conversion factor. 

\begin{figure}[!ht]
    \centering
    \begin{adjustbox}{width=0.8\linewidth,center}
        \includegraphics{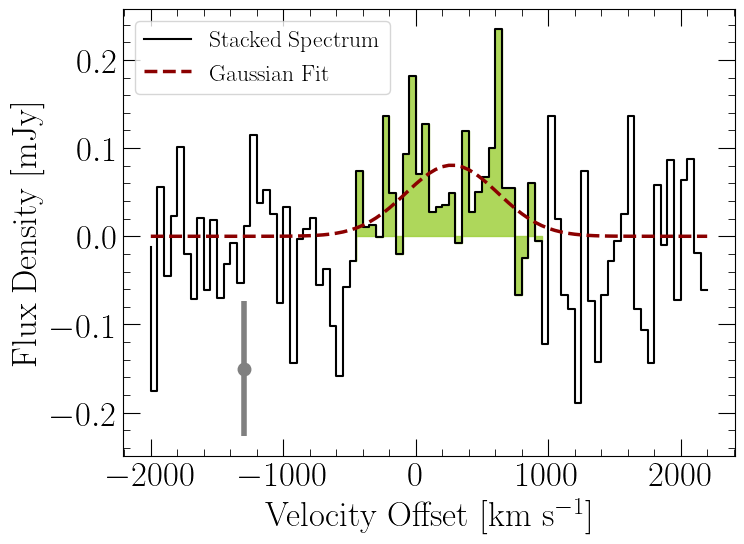}
    \end{adjustbox}
\caption[]{Stacked spectrum of the 16 non-detected quasars. The dashed red line indicates the Gaussian fit and the shaded green region highlights the line emission. The median per-channel uncertainty in the stacked flux density from Monte Carlo sampling is indicated by the error bar in the lower left corner.} 
    \label{fig:stacked_spectrum}
\end{figure}

For the quasars not detected in CO(4-3), we computed $3\,\sigma$ upper limits from the rms noise measured over the spectral region where the line is expected, assuming a velocity width of $300\,\mathrm{km}\,\mathrm{s}^{-1}$, consistent with the expected average line width \citep[e.g.,][]{Walter2011} and the peak of the line width distribution of our detected sources (see Figure \ref{fig:FWHMs}). To further constrain the average molecular gas content of the non-detections, we performed a variance-weighted mean stack of their spectra, aligning them using the peak Ly$\alpha$ nebula redshift. This redshift generally shows good agreement with the CO(4-3) redshifts of the detected sources in our sample (see also \ref{sec:COvsLyalpha}). To quantify the uncertainty introduced by this assumption, we additionally carried out 1000 Monte Carlo realizations of the stack, in which the redshift of each target was randomly varied within a window of $\Delta z=0.009$. This redshift interval corresponds to the standard deviation between the CO and Ly$\alpha$ redshifts measured for the detected sources. From the stacked spectrum, which yields a signal with integrated S/N $\sim4$, we infer an average molecular gas mass of log$(M_{\rm{gas}}/\rm{M}_{\odot})=9.37\pm0.11$. The uncertainty accounts for both the noise in the final stacked spectrum and the uncertainty introduced by the redshift variations in the Monte Carlo realizations. The stacked spectrum is presented in Figure \ref{fig:stacked_spectrum}.

We obtain molecular gas masses for the detected quasars in the range of $2.8\times10^9$ to $3.8\times10^{10}\,\mathrm{M}_{\odot}$, with a median value of $1.7\times10^{10}\,\mathrm{M}_{\odot}$, similar to other quasars in the same redshift range \citep{Bischetti2021, Li2023, Molyneux2025}. The inferred gas masses and upper limits are listed in Table \ref{tab:CO_properties}.

\renewcommand{\arraystretch}{1.1}
\begin{table*}[htbp]
\centering
\small
\begin{threeparttable}
\caption{Properties of the CO(4-3) and continuum emission.}
\label{tab:CO_properties}
\begin{tabular}{|c|c|c|c|c|c|c|}
\hline
ID & $z_\mathrm{CO}$\tnote{a} & FWHM \tnote{b}  & $S\Delta v$ \tnote{c} & $L'_\mathrm{CO(4-3)}$  \tnote{d}& log($\mathrm{M}_{\mathrm{gas}})$ &$S_\mathrm{cont}$  \tnote{e} \\
 &  &  [$\mathrm{km\,s^{-1}}$] &  [$\mathrm{Jy}\,\mathrm{kms^{-1}}$] & [$10^9\,\mathrm{K\,km\,s^{-1}\,pc^2}$] & [$\mathrm{M}_{\odot}$] &[mJy]  \\
\hline
1 &  -  & - & $<0.11$ & $<3.2$ & $<9.47$ & $<0.04$   \\
2 & $3.1626\pm0.0002$  & $297\pm42$ &  $0.77\pm0.06$ & $21.7\pm1.8$  & $10.30\pm0.04$ & $0.09\pm0.02$  \\
3 \tnote{\textdagger}&$3.1253\pm0.0003$ & $71\pm33$ &  $0.11\pm0.04$ & $3.1\pm1.0$ &$9.45\pm0.15$ &$246\pm1$ \tnote{*} \\
4 & $3.233\pm0.002$ & $635\pm248$ &  $0.38\pm0.07$ & $11.2\pm2.1$ & $10.01\pm0.08$ &$<0.05$ \\
6 & $3.0678\pm0.0008$ & $533\pm159$ & $0.73\pm0.09$ & $19.7\pm2.4$ & $10.26\pm0.05$ &$<0.06$ \\
7 & - & - & $<0.15$ & $<4.0$ & $<9.57$ &$<0.05$\\
8 & - & - & $<0.10$ & $<2.7$& $<9.40$ &$<0.06$ \\
10 & - & - & $<0.13$ & $<3.8$&$<9.55$ &$<0.06$ \\
11 & $3.1009\pm0.0003$ & $197\pm48$ & $0.29\pm0.04$ & $7.8\pm1.2$ &$9.85\pm0.07$ &$<0.06$ \\
13 & $3.1687\pm0.0005$ & $570\pm68$ & $0.59\pm0.07$ & $16.8\pm2.1$ & $10.19\pm0.05$ &$0.10\pm0.02$ \\
14 & $3.1357\pm0.0003$ & $404\pm54$ & $0.71\pm0.04$ &$19.7\pm1.2$ & $10.26\pm0.03$ &$0.06\pm0.03$     \\
15 & - & - & $<0.11$ & $<3.2$ & $<9.47$ &$<0.04$ \\
17 & $3.3512\pm0.0003$& $454\pm59$ &$0.57\pm0.04$ & $17.5\pm1.3$ & $10.21\pm0.03$ &$<0.08$ \\
18 & $3.2889\pm0.0005$ & $334\pm94$ & $0.35\pm0.06$ & $10.4\pm1.8$ &  $9.98\pm0.07$ &$<0.03$ \\
20 & $3.4266\pm0.0006$ & $480\pm118$ &$0.73\pm0.07$ & $23.5\pm2.3$ & $10.34\pm0.04$ &$0.14\pm0.03$ \\
21 & - & - & $<0.11$ & $<3.3$ & $<9.48$ &$0.10\pm0.03$ \\
24 & - & - & $<0.11$ & $<3.0$ &$<9.44$ &$<0.04$ \\
26 & - & - & $<0.11$ & $<3.0$ & $<9.45$ &$2.37\pm0.04$\\
27 & $3.3326\pm0.0004$ & $364\pm79$ & $0.78\pm0.06$ & $23.9\pm1.9$  &  $10.34\pm0.03$ &$0.12\pm0.03$ \\
29 & $3.2220\pm0.0005$ & $308\pm88$ & $0.45\pm0.06$ & $13.0\pm1.9$ & $10.08\pm0.06$ &$<0.06$  \\
30 & - & - & $<0.17$ & $<5.2$ & $<9.68$ &$0.11\pm0.03$\\
32 & $3.1031\pm0.0004$ & $487\pm67$ & $1.08\pm0.12$ & $29.5\pm3.2$ & $10.43\pm0.05$ &$0.06\pm0.02$ \\
34 & $3.256\pm0.001$ & $498\pm230$ & $0.71\pm0.13$ & $20.9\pm4.0$  & $10.28\pm0.08$ & $<0.05$ \\
36 & $3.2049\pm0.0003$ & $186\pm46$ &$0.25\pm0.04$ & $7.1\pm1.2$ & $9.81\pm0.08$ & $1.76\pm0.02$ \tnote{*} \\
38 & - & - & $<0.11$ & $<3.4$ & $<9.49$ & $<0.05$\\
39 & - & - & $<0.13$ & $<3.5$ & $<9.50$ &$<0.06$\\
43 & $3.1304\pm0.0005$ & $364\pm86$ & $0.85\pm0.12$ & $23.6\pm3.2$ & $10.34\pm0.06$ &$<0.05$  \\
44 & $3.2367\pm0.0002$ & $261\pm39$  & $0.63\pm0.06$ & $18.5\pm1.7$ & $10.23\pm0.04$ & $0.15\pm0.04$ \\
45 & $3.3577\pm0.0004$ & $469\pm64$ & $1.31\pm0.12$ & $40.9\pm3.8$ & $10.58\pm0.04$ & $3.43\pm0.03$ \tnote{*} \\
46 &$3.4024\pm0.0004$ & $141\pm41$ & $0.16\pm0.04$ & $5.2\pm1.2$ & $9.68\pm0.10$ & $0.18\pm0.02$ \\
47 & - & - & $<0.14$ & $<4.0$ &$<9.57$ & $1.24\pm0.05$ \tnote{*} \\
49 & $3.13836\pm0.00007$ & $113\pm13$ & $0.74\pm0.05$ & $20.6\pm1.5$ & $10.28\pm0.03$ & $373\pm0.4$ \tnote{*} \\
50 & - & - & $<0.21$ & $<6.0$ & $<9.74$ &$0.09\pm0.03$\\
51 & - & - & $<0.10$ & $<2.6$ & $<9.39$ &$<0.04$ \tnote{*}\\
52 & - & - & $<0.11$ & $<3.1$ & $<9.45$ &$0.37\pm0.02$ \tnote{*}\\
56 & - & - & $<0.13$ & $<3.6$ & $<9.52$ &$2.2\pm0.03$ \tnote{*} \\
58 & $3.0467\pm0.0002$ & $213\pm34$ & $0.31\pm0.03$ & $8.3\pm0.8$ &  $9.88\pm0.04$ & $0.08\pm0.02$ \\
\hline
\end{tabular}
\begin{tablenotes}
\footnotesize
\item[a] CO redshift determined from the line centroid, measured via a first-moment analysis.
\item[b] Line width determined from a second-moment analysis, scaled to match the full width at half maximum of a Gaussian profile.
\item[c] Line intensity.
\item[d] CO line luminosity (expressed via the source brightness temperature) for the CO(4-3) transition.
\item[e] Continuum.
\item[\tnote{\textdagger}] ID3 is considered a tentative detection due to its very narrow linewidth. 
\item[*] Confirmed radio-loud quasars as determined from the 1.4 GHz flux from the FIRST and NRAO VLA Sky Survey, as given in 
\cite{ArrigoniBattaia2019}.
\end{tablenotes}
\end{threeparttable}
\end{table*}

\subsection{CO versus QSO properties} \label{sec:COvsQSOproperties}
\begin{figure}[!ht]
    \centering
    \begin{adjustbox}{width=1.0\linewidth,center}
        \includegraphics{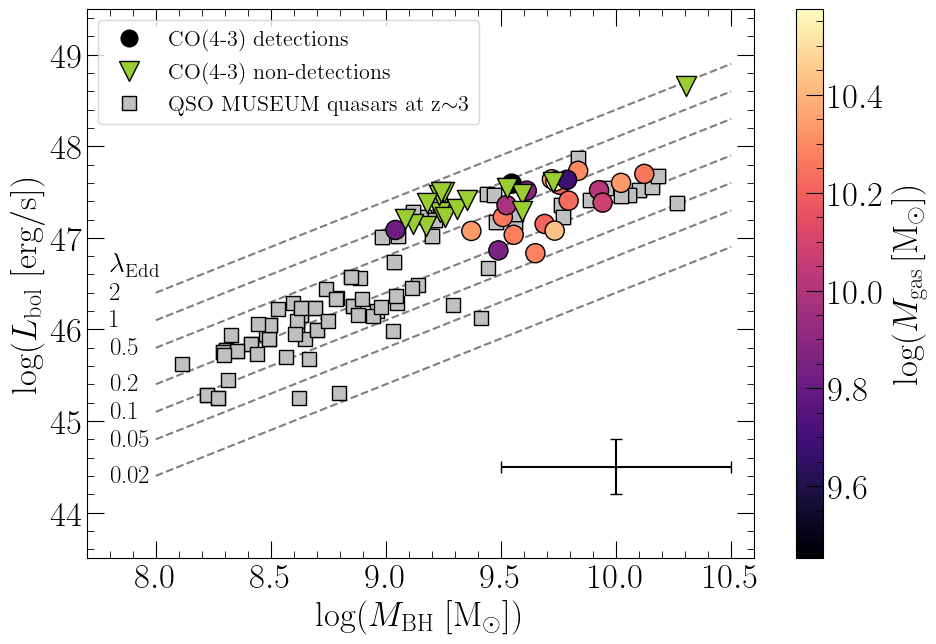}
    \end{adjustbox}
\caption[]{Bolometric luminosity as a function of black hole mass for the QSO MUSEUM quasars with available Ly$\alpha$ nebula properties (gray squares). The CO non-detections are overplotted in green triangles and the CO detections in circles, color-coded by their molecular mass. Different values of constant Eddington ratios ($\lambda_{\mathrm{Edd}}$) are shown with dashed lines. The intrinsic uncertainties on $M_{\mathrm{BH}}$ and $L_{\mathrm{bol}}$ are indicated by the black error bars in the lower right corner. }
    \label{fig:LbolvsMBH_MH2}
\end{figure}

\begin{figure}[!ht]
    \centering
    \begin{adjustbox}{width=1.0\linewidth,center}
        \includegraphics{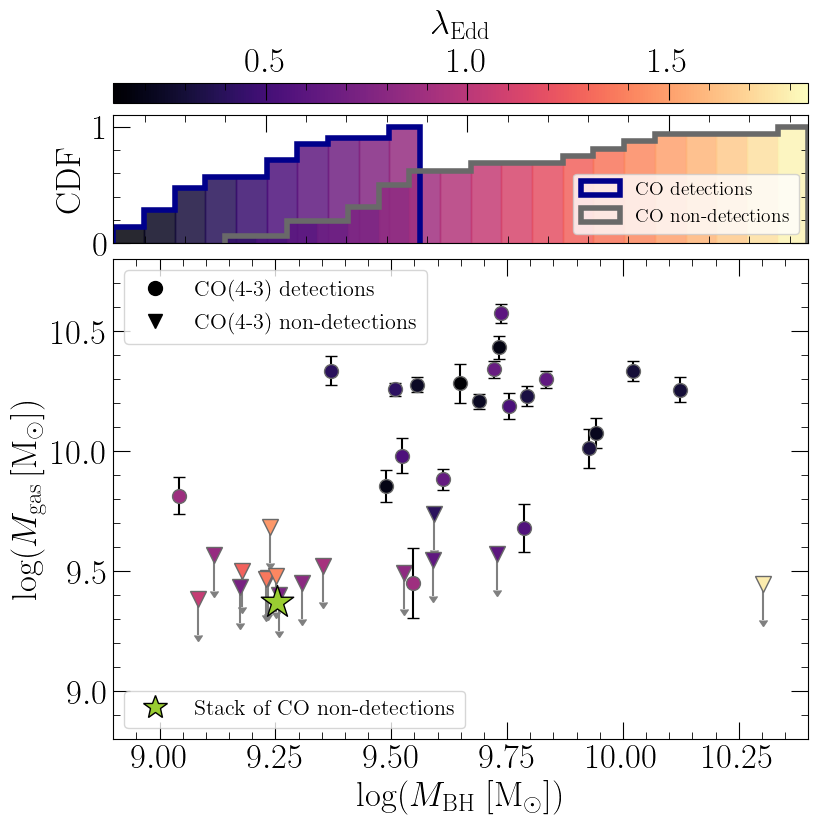}
    \end{adjustbox}
\caption[]{Molecular gas mass of the CO detections (in circles) and $3\,\sigma$ upper limits on the CO non-detections (in triangles) as a function of black hole mass. The points are color-coded by Eddington ratio, $\lambda_{\mathrm{Edd}}$. The green star shows the gas mass inferred from stacking the CO non-detections, placed at the median black hole mass of the non-detections. ID3 is included as a tentative detection and represents the lowest-mass system. We additionally show the cumulative distribution (CDF) of the Eddington ratios for the CO-detections and non-detections. }  
    \label{fig:MH2vsMBH}
\end{figure}

We examined potential correlations of quasar properties, such as the bolometric luminosity, black hole mass, and Eddington ratio, and the characteristics of their CO reservoirs. Figure \ref{fig:LbolvsMBH_MH2} shows the relation between $L_\mathrm{bol}$ and $M_\mathrm{BH}$ for quasars from the QSO MUSEUM survey, including both CO(4–3) detections and non-detections. Within the sample, we do not identify significant trends between molecular gas masses and either bolometric luminosities or black hole masses. However, we do find evidence that the CO-detected quasars, with more massive inferred molecular gas reservoirs, have lower Eddington ratios compared to the non-detected quasars. This is shown in Figure~\ref{fig:MH2vsMBH}, which presents $M_\mathrm{gas}$ as a function of $M_\mathrm{BH}$, with quasars color-coded by Eddington ratio, together with the cumulative Eddington-ratio distributions of the CO detections and non-detections. In particular, we do not find CO-detections for the quasars in our sample with Eddington ratios larger than $\sim0.9$. The median Eddington ratio of the CO detections is $0.40^{+0.23}_{-0.18}$, compared to $0.87^{+0.52}_{-0.21}$ for the non-detections. A Mann-Whitney U test comparing the Eddington ratios of detections and non-detections yields a
$p$ value of $\sim3\times10^{-5}$, indicating that there is a significant difference between the Eddington ratios of detections and non-detections.

We note that the CO-detected quasars generally show higher black hole masses. This could, however, be attributed to a selection bias, as the quasars with smaller Eddington ratios in the ALMA-observed sample generally tend to have higher black hole masses (see Figure \ref{fig:LbolvsMBH_MH2}). 
All these findings need to be confirmed 
given the substantial uncertainties in black hole masses ($\sim$ 0.5 dex), bolometric luminosities ($\sim$0.3 dex), and hence Eddington ratios. Indeed, within these error bars, all the quasars could be considered to have roughly comparable black hole masses. Furthermore, we note that differences in gas excitation could affect the detectability of the CO(4-3) line in our sample. In systems with highly excited molecular gas, a larger fraction of the CO emission may exist in higher $J$ transitions, suppressing the CO(4-3) flux expected from the standard assumed CO SLED even at fixed molecular gas mass.

We investigated potential correlations between CO line widths and quasar properties, but no significant trends are apparent. This suggests that the kinematics of the molecular gas traced by CO are not directly linked to the global properties of the central AGN.

\subsection{CO versus Ly$\alpha$ properties} \label{sec:COvsLyalpha}
\begin{figure}[!ht]
    \centering
        \includegraphics[width=\linewidth]{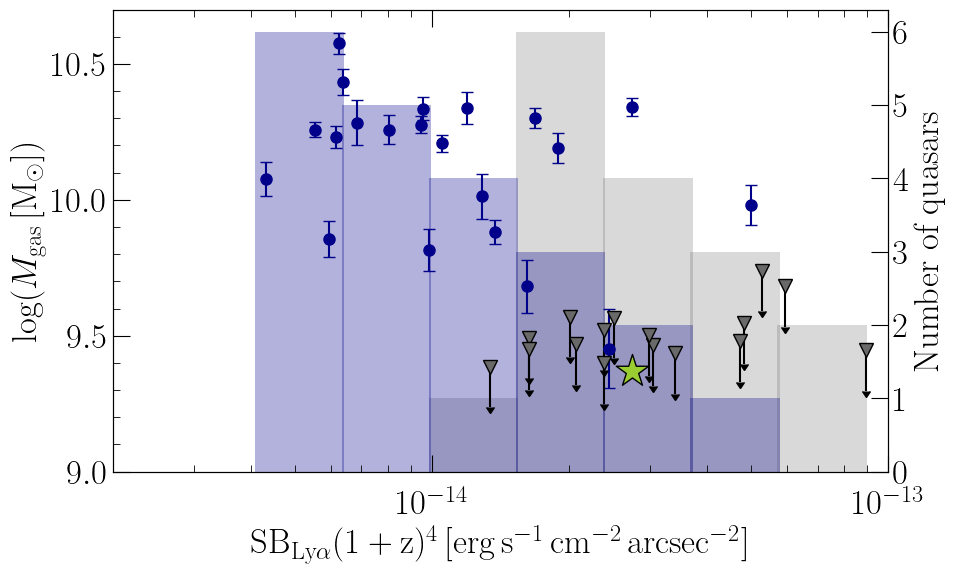}

        \vspace{-0.1cm}
        \includegraphics[width=\linewidth]{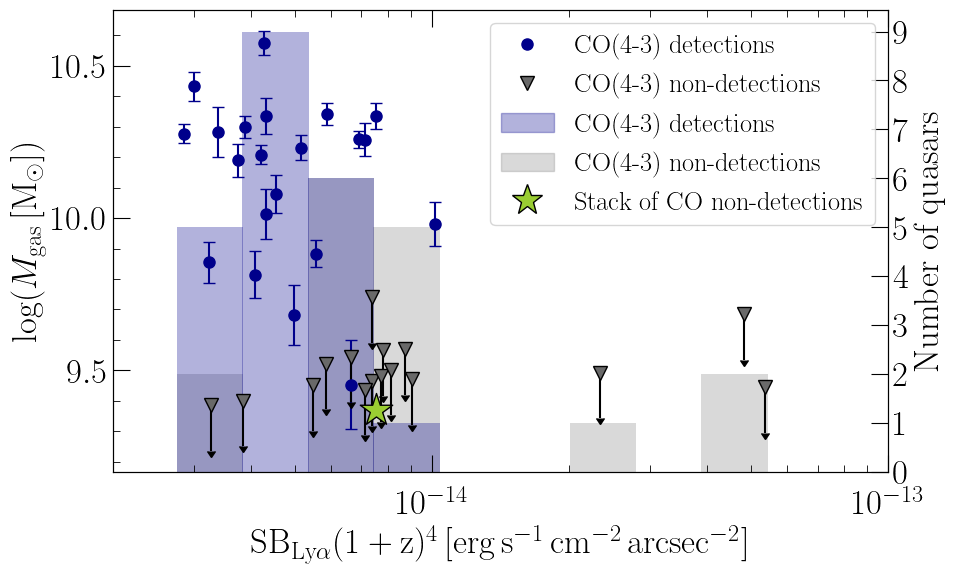}
\caption[]{\textit{Upper panel:} Molecular gas mass as a function of central Ly$\alpha$ surface brightness (within about 12\,kpc) for the CO detections in blue circles and CO non-detections in gray triangles. We additionally show histograms of the detections and non-detections, shaded in blue and gray, respectively. 
The green star indicates the molecular gas mass inferred from stacking of the non-detections at the median Ly$\alpha$ SB of the non-detections.  \textit{Lower panel}: Same as the upper panel, but plotted as a function of the surface brightness averaged over the full extent of each nebula. }
    \label{fig:MH2vsLya}
\end{figure}

We next examine how the molecular gas reservoirs of the quasar hosts relate to the Ly$\alpha$ emission properties of their surrounding nebulae. Figure \ref{fig:MH2vsLya} shows $M_\mathrm{gas}$ as a function of the central Ly$\alpha$ surface brightness (SB), taken from the innermost radial annulus (with radius of $\sim12\,\mathrm{kpc}$) of the MUSE surface brightness profiles presented by \citet{GonzalezLobos2025}. The noisy normalization region of 1~arcsec$^2$ at the quasar location, which is affected by the residuals of the quasar PSF subtraction, was masked to exclude the unresolved nuclear emission when computing these surface brightness profiles. 

We find that quasars with lower central Ly$\alpha$ SB tend to exhibit higher CO detection rates and larger molecular gas masses compared to quasars with higher Ly$\alpha$ SB. We primarily focus on the central Ly$\alpha$ SB, as it probes gas near the quasar host galaxy and is therefore more directly linked to the molecular gas reservoirs, while emission at larger radii has marked differences in morphologies and extent between different targets. 
However, a similar trend is observed when considering the Ly$\alpha$ SB averaged over the full extent of each nebulae (see lower panel of Figure \ref{fig:MH2vsLya} and also Figure \ref{fig:SB_vel_disp_profiles}).

To quantify these relations, we computed the Kendall rank correlation coefficient while accounting for censoring (i.e., upper limits). For this purpose, we used the package by \citet{Flury_kendall, Flury2022}, which is based on the formalism developed by \citet{Akritas1996}. The resulting correlation coefficients are listed in Table \ref{tab:corr_mol_lya}. We also tested for possible relations with the CO effective radius for the subset of spatially resolved sources (Section \ref{sec:gas_fractions}). Performing the Kendall test between the molecular gas mass and the central Ly$\alpha$ surface brightness reveals a clear correlation, indicated by a low $p$ value of $\sim 10^{-4}$. A similar, though weaker, correlation is also obtained when considering the Ly$\alpha$ surface brightness averaged over the full nebula with a $p$ value of $\sim 4\times10^{-3}$.  Furthermore, we find that below a central Ly$\alpha$ surface brightness of $1.3\times10^{-14}\ \mathrm{erg\,s^{-1}\,cm^{-2}\,arcsec^{-2}}$, all quasars are detected in CO, whereas above this threshold about 70\% of the sources are non-detections. A Mann-Whitney U test comparing the Ly$\alpha$ surface brightness distributions of detections and non-detections gives a $p$ value of $\sim6\times10^{-5}$, indicating a significant difference between the two samples.

\begin{table*}[htbp]

\centering
\begin{threeparttable}
\small
\setlength{\tabcolsep}{4pt}
\renewcommand{\arraystretch}{1.1}
\caption{Kendall rank correlations accounting for upper limits between molecular gas and Ly$\alpha$ nebula properties.}
\label{tab:corr_mol_lya}
\begin{tabular}{lcccc}
\hline
 & $\mathrm{SB_{Ly\alpha}(1+\mathrm{z})^4}$ \tnote{a} & $A_{\rm Ly\alpha}$ \tnote{b}  & ${\rm FWHM}_{\rm Ly\alpha}$ \tnote{c} & ${\sigma}_{\rm Ly\alpha}$ \tnote{d}\\
\hline
$M_{\rm gas}$              &\cellcolor{blue!10} $\tau = -0.44,\ p = 1.1 \times 10^{-4}$ (N=37) & $\tau = -0.14,\ p=0.2$  (N=37) & $\tau = -0.01,\ p=0.9$ (N=37) & $\tau = -0.04,\ p=0.7$ (N=37)\\
$I_{\rm CO}$              & \cellcolor{blue!10} $\tau = -0.44,\ p = 1.3 \times 10^{-4}$ (N=37) & $\tau = -0.14,\ p=0.2$  (N=37) & $\tau = -0.002,\ p=1.0$ (N=37) & $\tau = -0.05,\ p=0.7$ (N=37) \\
${\rm FWHM}_{\rm CO}$  \tnote{c}    & $\tau = -0.05,\ p=0.8$   (N=21) & $\tau = -0.1,\ p=0.6$ (N=21) & $\tau = 0.03,\ p=0.9$  (N=21) & $\tau = -0.07,\ p=0.7$ (N=21)\\
${R}_{\rm CO}$  \tnote{e}    & $\tau = 0.33,\ p=0.5$   (N=6) & $\tau = 0.20,\ p=0.7$ (N=6) & $\tau = -0.20,\ p=0.7$  (N=6) & $\tau = -0.07,\ p=1.0$ (N=6)\\
\hline
\end{tabular}
\begin{tablenotes}
\footnotesize
\item \textit{Notes.} We report Kendall's $\tau$, two-sided $p$ values, and the number of objects N used in each test. The properties that show correlations are highlighted in blue.
\item[a] Central Ly$\alpha$ surface brightness corrected for cosmological dimming. 
\item[b] Ly$\alpha$ emission area within the 2$\sigma$ isophotes from \citet{GonzalezLobos2025}. 
\item[c] FWHM from a second-moment analysis.
\item[d] Ly$\alpha$ velocity dispersion within the inner $\sim12\,\mathrm{kpc}$ of the nebulae. 
\item[e] CO effective radius for resolved sources. 
\end{tablenotes}
\end{threeparttable}
\end{table*}

A tentative correlation between Ly$\alpha$ SB and molecular gas mass was previously reported by \citet{MunozElgueta2022} based on an APEX survey targeting CO(6–5), CO(7–6), and [\ion{C}{i}](2-1) in nine quasars, also part of the QSO MUSEUM sample. They found that quasars with more substantial molecular gas reservoirs tend to be surrounded by nebulae with lower Ly$\alpha$ SB. Our analysis confirms this trend, reinforcing the connection between extended Ly$\alpha$ emission and the molecular gas content of quasar host galaxies.

\citet{MunozElgueta2022} also report a tentative trend between CO brightness and physical area of the Ly$\alpha$ nebulae, but we are unable to confirm this trend with our data (see Table \ref{tab:corr_mol_lya}). Given the limitations of the APEX observations (see also Section \ref{sec:APEX}), those results are subject to significant uncertainties. The Kendall rank test indicates that only the molecular gas mass or the CO flux are statistically dependent on the central Ly$\alpha$ surface brightness. All the other tested quantities seem independent as their $p$ values are close to one. 

We find velocity shifts between the nebular Ly$\alpha$ peak redshifts (from \citealt{GonzalezLobos2025}) and molecular gas redshifts with Ly$\alpha$ appearing both redshifted ($\sim6$ to $1400\,\mathrm{km}\,\mathrm{s^{-1}}$) and blueshifted ($\sim -2000$ to $-12\,\mathrm{km}\,\mathrm{s^{-1}}$) relative to CO. Such offsets can arise from resonant scattering of Ly$\alpha$ photons through bulk gas motions, where outflowing gas preferentially produces redshifted emission and inflowing gas blueshifted emission \citep[e.g.,][]{Verhamme2006, Dijkstra2014}. The mean shift corresponds to $\sim-80\,\mathrm{km}\,\mathrm{s^{-1}}$, which could suggest that for most quasars in our sample the halo-gas is inflow-dominated. As the CO redshifts provide a precise measure of the host galaxy systemic velocity, they allow us to investigate the kinematics of the Ly$\alpha$ nebulae in detail. This will be addressed in an upcoming paper.

\subsection{Gas fractions}
To calculate the gas fractions of the quasars, defined as $f_\mathrm{gas}=M_\mathrm{gas}/M_*$, where $M_*$ is the stellar mass, we used the dynamical mass estimate derived in \citet{Chen2021}, which links gas fractions to CO line width and luminosity. This methods assumes a typical dynamical mass estimate, $M_\mathrm{dyn}$, for rotational disks and adopts an average inclination correction of $<\mathrm{sin}^2(i)>=2/3$ \citep{Tacconi2008}. The equation for the dynamical mass estimate then reads
\begin{multline}\label{eq: gas_fractions}
\mathrm{log}\left(\frac{\mathrm{FWHM}_\mathrm{CO}^2 r_\mathrm{e}}{\alpha_\mathrm{CO}}\right) = \\
    \mathrm{log}(L'_\mathrm{CO(1-0)})-\mathrm{log}\left(\frac{3.1\cdot10^5f_\mathrm{gas}(1-f_\mathrm{DM})}{1+f_\mathrm{gas}}\right),
\end{multline}
with a DM fraction of $f_\mathrm{DM}=0.12$ \citep{Genzel2020,Chen2021} and an effective radius, $r_\mathrm{e}$. We assumed that $\alpha_\mathrm{CO}=0.8\mathrm{\, M_\odot}\,(\mathrm{K\,km\,s^{-1}\,pc^2})^{-1}$.

The effective radii were determined from size measurements of the spatially resolved quasars. To assess whether a source is resolved, we adopted the criterion proposed by \citet{Decarli2018}. Specifically, we fit the CO emission in the MFS images (see Section \ref{sec:ALMA_observations}) with a 2D Gaussian using the CASA task \textsc{imfit}. A source was considered resolved if the fitted emission size (convolved with the beam) was at least 1.3 times the beam size and the S/N of the emission exceeds 10. With this method, we find that, out of the 21 CO-detected quasars, six are resolved. They are ID 13, 29, 32, 43, 45, and 58. The task \textsc{imfit} returns the FWHM of the emission deconvolved from the beam and we adopted half of the major axis as the effective radius for the resolved sources. For unresolved sources, we compared the beam sizes to the smallest emission size found among the resolved quasars, which corresponds to an effective radius of about $4\,\mathrm{kpc}$. Since all beam sizes of the unresolved targets exceed this value, we adopted $4\,\mathrm{kpc}$ for the unresolved sources. Adopting an effective radius of 4 kpc is also consistent with recent high-redshift quasar studies (see e.g., \citealt{Damato2020, ArrigoniBattaia2022, Molyneux2025}). For ID3 and ID49 the dynamical mass estimates derived from Equation \ref{eq: gas_fractions} break down under the assumption of a typical disk inclination of $<\mathrm{sin}^2(i)>=2/3$, as these sources exhibit unusually narrow CO line widths. Since both targets are known blazars and therefore likely viewed close to face-on, we instead adopted a lower inclination angle of $i = 20^\circ$. The gas fractions reported for these two sources are thus computed using this revised inclination, yielding values consistent with the remainder of the sample. We obtain on average low gas fractions (see Table \ref{tab:gas_fractions}) with a median value of $f_\mathrm{gas} = 0.10$. The gas fractions are thus generally lower than for star-forming galaxies at a similar redshift range (compare e.g., \citealt{Tacconi2020,Sanders2023}).

\begin{figure}[!ht]
    \centering
    \begin{adjustbox}{width=1.0\linewidth,center}
        \includegraphics{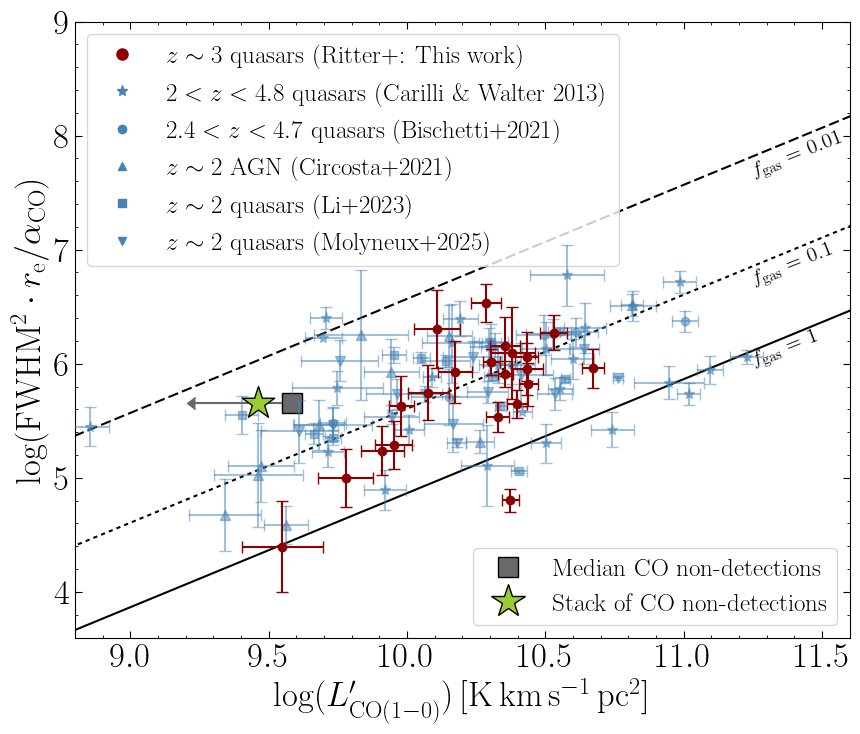}
    \end{adjustbox}
\caption[]{Visualization of gas fractions using a dynamical mass estimate from \citet{Chen2021}. The diagonal lines represent
gas mass fractions of 1\%, 10\%, and 100\%. The CO detections of this work are shown in red. The literature values for QSOs are shown in light blue and are taken from \citet{2013CarilliWalter}, \citet{Bischetti2021}, \citet{Circosta2021}, \citet{Li2023} and \citet{Molyneux2025}. The gray square indicates the median upper limit of the CO non-detections in our sample. The green star shows the value inferred from stacking these non-detections (see Figure \ref{fig:stacked_spectrum}). Both values assume a FWHM of $300\,\mathrm{km\,s^{-1}}$.} 
    \label{fig:Gas_fractions}
\end{figure}

Figure \ref{fig:Gas_fractions} shows the correlation between $\mathrm{log}\left(\mathrm{FWHM}_\mathrm{CO}^2 r_\mathrm{e}/\alpha_\mathrm{CO}\right)$ and $\mathrm{log}(L'_\mathrm{CO(1-0)})$, where different gas fractions are indicated by diagonal dashed lines. We added quasars and AGN from the literature \citep{2013CarilliWalter, Bischetti2021, Circosta2021, Li2023, Molyneux2025} for which we adopted the CO FWHM and $L
'_\mathrm{CO}$ values from the published works. The CO luminosities were converted to $L'_\mathrm{CO(1-0)}$ using the conversion factors compiled by \citet{2013CarilliWalter}. The gas fractions were calculated using Equation \ref{eq: gas_fractions} and assuming a uniform effective radius of 4 kpc for all sources for consistency with our analysis. We also included the non-detections from our sample assuming FWHM$=300$~km~s$^{-1}$ and $r_e=4$~kpc. These are shown both as a median upper limit on $L'_\mathrm{CO(1-0)}$ and as the $L'_\mathrm{CO(1-0)}$ inferred from stacking the non-detections. These estimates imply that, if detected individually, the non-detections would lie at even lower gas fractions. Overall, our measurements are consistent with the ranges found in previous studies.

\renewcommand{\arraystretch}{1.1}
\begin{table}[htbp]
\centering
\small
\setlength{\tabcolsep}{10pt}
\renewcommand{\arraystretch}{1.3}
\begin{threeparttable}
\caption{Gas fractions of the resolved and unresolved CO-detected quasars with the assumed effective radius, $r_\mathrm{e}$, as explained in the text.}
\label{tab:gas_fractions}
\begin{tabular}{cccc}
\toprule
ID & Resolved? & $r_\mathrm{e}$  & $f_\mathrm{gas}$   \\
 &  &  [kpc] &  \\
\midrule
2 &  & 4 & \asym{0.26}{0.14}{0.08}  \\
3 & &4 &  \asym{0.10}{0.22}{0.07} \tnote{*} \\
4 &  & 4 &   \asym{0.02}{0.03}{0.01}  \\
6 & & 4 & \asym{0.06}{0.06}{0.03}  \\
11 &  & 4 & \asym{0.20}{0.19}{0.09}  \\
13 & \checkmark & $8.4\pm2.6$ & \asym{0.02}{0.01}{0.01} \\
14 &  & 4 & \asym{0.11}{0.05}{0.03}    \\
17 && 4 & \asym{0.08}{0.03}{0.02}  \\
18 &  & 4 & \asym{0.09}{0.08}{0.04} \\
20 &  & 4 & \asym{0.09}{0.07}{0.04} \\
27 &  & 4 & \asym{0.18}{0.14}{0.07}  \\
29 & \checkmark & $7.1\pm2.0$ & \asym{0.07}{0.07}{0.03}  \\
32 & \checkmark & $6.3\pm1.4$ & \asym{0.07}{0.04}{0.02} \\
34 &  & 4 & \asym{0.08}{0.14}{0.05} \\
36 &  & 4 & \asym{0.21}{0.21}{0.10} \\
43 & \checkmark & $5.4\pm1.3$ & \asym{0.12}{0.11}{0.06} \\
44 &  & 4  & \asym{0.30}{0.17}{0.10}  \\
45 & \checkmark & $4.0\pm1.1$ & \asym{0.23}{0.16}{0.09}  \\
46 & & 4 & \asym{0.28}{0.41}{0.15}  \\
49 &  & 4 & \asym{0.32}{0.14}{0.09} \tnote{*} \\
58 & \checkmark & $7.6\pm4.0$ & \asym{0.09}{0.09}{0.04} \\
\bottomrule
\end{tabular}
\begin{tablenotes}
\footnotesize
\item[*] For ID3 and ID49 (both blazars, exhibiting narrow line widths), we adopted an inclination of $i=20^\circ$, since the dynamical mass estimate breaks down at higher inclinations.
\end{tablenotes}
\end{threeparttable}
\end{table}

\subsection{Comparison to APEX observations} \label{sec:APEX}
\citet{MunozElgueta2022} conducted a study of nine quasars of the QSO MUSEUM survey with the APEX telescope, targeting CO(6-5), CO(7-6) and [\ion{C}{I}](2-1). All of these sources are also part of our ALMA sample (ID3, ID7, ID8, ID13, ID18, ID29, ID39, ID43, and ID50), allowing for a comparison of the single-dish and interferometric observations. 
We note that the APEX line candidates are mostly low-significance detections and, without precise systemic redshifts, the uncertain line frequencies make baseline and rms estimates more challenging. Additionally, the APEX beam has a FWHM of about 30$\,\arcsec$, corresponding to large physical scales of approximately 240$\,\mathrm{kpc}$ at $z\sim3$, which means that emission from nearby sources within the field might contribute to the measured signals. Therefore, the ALMA detections can generally be considered more robust.

When comparing the reported detections from ALMA and APEX, we find that the APEX non-detections and upper limits are overall consistent with our results. However, the reported CO and [\ion{C}{I}] detections are difficult to reconcile with the CO(4--3) luminosities and upper limits measured from ALMA, since the corresponding higher-$J$ CO and [\ion{C}{I}]
emission are generally expected to fall below the APEX detection threshold. This suggests that the reported single-dish detections may not be directly associated with the quasars and could correspond to spurious detections, interlopers or line emission from nearby sources. A more in-depth comparison of the APEX and ALMA observations, beyond the scope of this work, would be required to assess the origin of each reported line detection in the single-dish data.

\section{Discussion} \label{sec:Discussion}
In this section we discuss our results, starting with the connection between the large-scale Ly$\alpha$ emission and the molecular gas content of the host galaxies. We find that quasars with dimmer Ly$\alpha$ nebulae exhibit higher molecular gas masses, and we explore several possible scenarios that could explain this trend. We place our findings into the broader context of previous observational studies and simulations. We then discuss the quasars' gas fractions in terms of effects of AGN activity. Lastly, we assess any evidence of our quasars residing in overdense environments and investigate effects of the quasar environments on potential companion galaxies.

\subsection{Molecular gas reservoirs and Ly$\alpha$ nebulae}
As described in Section \ref{sec:COvsLyalpha}, we find a relation between the molecular gas masses estimated from CO(4-3) and the central surface brightness of the surrounding Ly$\alpha$ nebulae, with brighter Ly$\alpha$ emission associated with quasars with lower molecular gas masses. This trend is supported by the Kendall rank correlation test (Table~\ref{tab:corr_mol_lya}). We primarily focus on the central Ly$\alpha$ surface brightness, as it probes gas in the immediate vicinity of the quasar host galaxy, and therefore provides a more direct probe of the connection between the Ly$\alpha$ emission and the molecular gas reservoirs.

As Ly$\alpha$ is a resonant line, its radiative transfer depends strongly on the kinematics and distribution of the surrounding gas. The absorption and re-emission of Ly$\alpha$ photons by hydrogen atoms occurs on very short timescales (typically $\sim10^{-9}\,\mathrm{s}$, e.g., \citealt{OsterbrockFerland2006, Dijkstra2014}) causing them to undergo numerous scatterings before escaping dense environments such as the ISM and cool clouds in the CGM \citep[e.g.,][]{OsterbrockFerland2006, Dijkstra2017}. The origin of the bright, extended Ly$\alpha$ nebulae observed around quasars is still debated, but is thought to involve multiple contributing mechanisms. \citet{GonzalezLobos2025} find that the Ly$\alpha$ nebulae around quasars in the QSO MUSEUM survey are likely powered by a combination of quasar photoionization and subsequent recombination in the CGM; resonant scattering of photons from the quasar BLR within the CGM; and collisional excitation in shocks. These powering mechanisms are also supported by simulations \citep[e.g.,][]{GronkeBird2017, Costa2022}. 

The observed connection between molecular gas masses and Ly$\alpha$ SB should therefore be interpreted within the broader context of Ly$\alpha$ radiative transfer and the mechanisms powering extended nebulae. On the one hand, quasars with more substantial molecular gas reservoirs also tend to have a higher dust content (typical gas-to-dust ratios are measured to be $\sim 70 -100$ \citep{Wang2016,Davies2017, Venemans2017, Bischetti2021}). A dusty host medium can absorb Ly$\alpha$ photons originating from the BLR, preventing them from reaching the CGM. Moreover, dust can obscure the central quasar, reducing the flux of ionizing photons that excite hydrogen in the halos and produce the large-scale Ly$\alpha$ emission. In this framework, the observed anti-correlation between molecular gas masses and Ly$\alpha$ SB may be driven, at least in part, by increased dust attenuation in more gas-rich systems. This is consistent with recent findings by \citet{Hall2026}, who report a negative correlation between the integrated Ly$\alpha$ luminosity and the far-infrared dust emission in a sample of four quasars hosted by dusty starbursts. These systems also exhibit fainter and shallower Ly$\alpha$ surface brightness profiles compared to quasars with lower levels of dust obscuration of similar bolometric luminosity. Furthermore, \citet{GonzalezLobos2023} find that the ratio of the nebula to quasar Ly$\alpha$ luminosity decreases with increasing dust mass.

At the same time, we find tentative evidence that the CO detection rate depends on quasar properties. In our sample, quasars hosting more massive black holes and exhibiting lower Eddington ratios are more likely to be detected in CO and to have more massive CO-inferred molecular gas reservoirs (Section \ref{sec:COvsQSOproperties}). On the other hand, the high-Eddington ratio quasars in our sample are comparatively gas-poor, while exhibiting brighter Ly$\alpha$ nebulae. This could indicate that strong radiative feedback in the high-Eddington systems reduces the gas and dust content, allowing for a higher escape fraction of ionizing and Ly$\alpha$ photons. However, we note that our sample is biased as the quasars with higher Eddington ratios also tend to host lower-mass black holes (see Figure \ref{fig:LbolvsMBH_MH2}). This raises the question of whether the observed trend could simply reflect underlying scaling relations, given that molecular gas content, stellar mass, and black hole mass are interrelated. To test this, we derive an expected relation between molecular gas mass and black hole mass using the $M_\mathrm{gas}/M_*$ relation from \citet{Tacconi2020} (for MS galaxies since we do not have estimates of SFR for our quasar sample) and the $M_\mathrm{BH}$-$M_*$ relation at $z\sim3$ from \citet{Decarli2010}. For the median black hole mass and gas mass of our sample, the slope of the derived scaling relation is significantly flatter than the observed trend and fails to account for the low gas masses of the non-detections at lower $M_{\mathrm{BH}}$.

\begin{figure*}
\centering

\begin{minipage}[c]{0.38\textwidth}
    \centering
    \includegraphics[
        width=\linewidth]{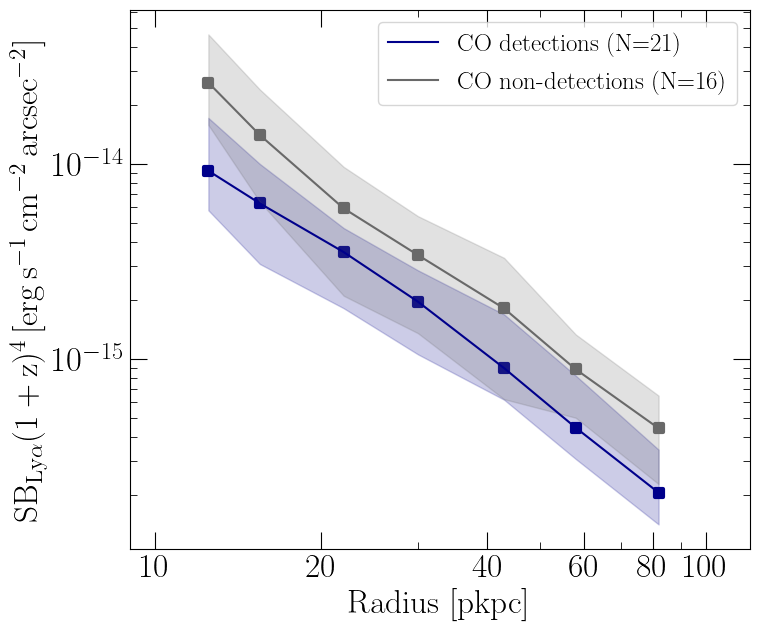}
\end{minipage}
\hfill
\begin{minipage}[c]{0.38\textwidth}
    \centering
    \includegraphics[
        width=\linewidth]{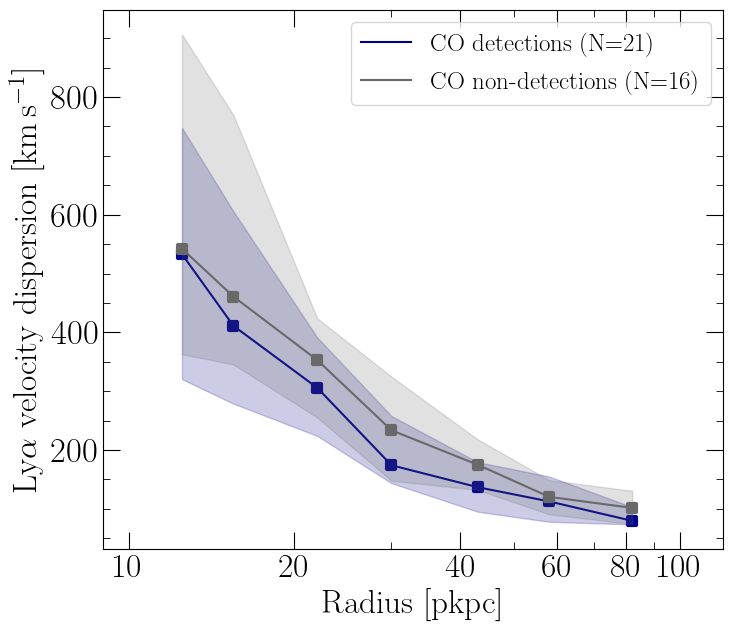}
\end{minipage}
\hfill
\begin{minipage}[c]{0.20\textwidth}
    \caption{
    \textit{Left}: Ly$\alpha$ surface-brightness profiles, corrected
    for cosmological dimming, as a function of projected distance.
    \textit{Right}: Ly$\alpha$ velocity-dispersion profiles as a
    function of projected distance. Squares show the median profiles
    for CO detections (blue) and non-detections (gray). Shaded regions
    indicate the $16^\mathrm{th}$--$84^\mathrm{th}$ percentile ranges.
    }
    \label{fig:SB_vel_disp_profiles}
\end{minipage}

\end{figure*}

This suggests that the observed trends are likely not driven solely by scaling relations, but may instead reflect differences in quasar feedback and the impact on the surrounding gas. Similarly, \citet{GonzalezLobos2025} find that within the full QSO MUSEUM sample, quasars with higher bolometric luminosity and with higher Eddington ratios (at fixed black hole mass) show brighter Lyman alpha nebulae with larger velocity dispersions in the inner $\sim40$~kpc. Taken together, these observations highlight the key role of ionizing photons from the quasar in affecting the cool CGM gas (extended Ly$\alpha$ emission), and of more violent AGN feedback in shaping the host galaxy (CO) and inner CGM scales (Ly$\alpha$ large velocity dispersion). In Figure~\ref{fig:SB_vel_disp_profiles}, we present the median stacked Ly$\alpha$ surface-brightness and velocity-dispersion profiles for the CO-detected and CO-undetected quasars. Consistent with the trends discussed above, the CO non-detections, corresponding to quasars with lower molecular gas masses and higher median Eddington ratios, exhibit a higher Ly$\alpha$ surface brightness at a fixed projected radius. We note, however, that the 16th–84th percentile ranges start to overlap beyond $\sim12$ kpc, indicating that the difference becomes less significant at larger radii. This is consistent with the lower panel of Figure~\ref{fig:MH2vsLya}, where the trend with Ly$\alpha$ surface brightness averaged over the full nebula is still present, but less pronounced. This is likely because Ly$\alpha$ emission at larger radii from the quasar is increasingly shaped by the extended CGM environment, and therefore less directly coupled to the host galaxy. In contrast, we do not find a statistically significant difference between the velocity-dispersion profiles of the CO detections and non-detections (see also Section \ref{sec:COvsLyalpha} and Table \ref{tab:corr_mol_lya}). 

Cosmological, radiation-hydrodynamic simulations by \citet{Costa2022} have shown that quasar-driven outflows are required to reproduce the bright, extended Ly$\alpha$ nebulae observed around high-redshift quasars. In particular, these outflows lead to low-column density channels in hydrogen and dust, enhancing the escape fraction of both ionizing and BLR Ly$\alpha$ photons by lowering absorption and scattering, while affecting the isotropy of escaping flux at the same time. In a recent paper, \citet{Rueda-Vargas2026} find tentative evidence for a correlation between the outflow power as measured from [\ion{O}{iii}] and the Ly$\alpha$ nebula luminosity in 6 $z\sim2-3$ quasars. Taken together with our observations, this suggests that quasars with higher Eddington ratios may drive stronger outflows, which can both reduce the molecular gas content of the hosts and enhance the reprocessed Ly$\alpha$ emission. 

To further investigate the link between gas depletion and AGN feedback, we examine the relation between molecular gas fraction and Eddington ratio. For the CO-detected quasars we do not find a significant correlation. This holds even when considering only spatially resolved sources (which have more accurate size measurements) or including additional resolved quasars from \citet{Bischetti2021} and \citet{Molyneux2025}. However, the CO non-detections are expected to have systematically lower gas fractions (see Figure~\ref{fig:Gas_fractions}). Including these sources could therefore reveal a dependence of molecular gas fraction on Eddington ratio similar to that seen for the total molecular gas content. Our analysis, however, covers only a limited dynamic range in Eddington ratios, and the gas fractions themselves are subject to considerable uncertainty due to the assumptions involved in their derivation (see also Section 
\ref{sec:gas_fractions}).

Despite the possible connection between quasar-driven outflows, gas depletion, and Ly$\alpha$ surface brightness, we do not detect signatures of molecular outflows in our sample. Cold, molecular outflows have been detected in CO on kiloparsec scales \citep[e.g.,][]{Feruglio2010, Fiore2017, Brusa2018, Herrera-Camus2019, Chartas2020, Vayner2021, Speranza2024}. However, their detection generally requires a high S/N and spatial resolution. Observational studies of outflows in multiphase tracers have shown that the cold, molecular phase likely constitutes the bulk of the outflow mass budget \citep[e.g.,][]{Feruglio2015,Tombesi2015, Rupke2017, Fluetsch2021, Speranza2024}. On the other hand, energy injection from AGNs or stellar feedback could heat the cold molecular gas, producing a warmer phase that could dominate the outflow \citep{Davies2024, Dan2025, Barfety2025}. Moreover, simulations by \citet{Costa2018a, Costa2018b}, which model outflows driven by AGN radiation pressure on dust, show that warm, dense gas ($T\sim10^5\,\mathrm{K}$, $n_\mathrm{H}>1\,\mathrm{cm^{-3}}$ ) carries a large amount of outflow mass on kiloparsec scales, even when not explicitly tracking molecular gas. Additional quasar ionization could make this phase observable in emission lines such as [\ion{O}{iii}] \citep{Costa2022}. Even if the molecular gas does not directly participate in the outflows, it may still be impacted indirectly, for instance, through heating or through photodissociation driven by quasar radiation.

In addition to affecting the surface brightness, quasar-driven outflows also lead to more extended nebulae in the simulations by \citet{Costa2022}. This occurs because low-column-density channels reduce scattering in the central few kiloparsecs, allowing Ly$\alpha$ photons to remain in resonance with halo gas and scatter out to larger radii. However, the same simulations also show that more luminous and longer-lived quasars can reduce the overall hydrogen optical depth, resulting in less efficient Ly$\alpha$ scattering and more compact nebulae. \citet{MunozElgueta2022} reported a tentative trend between molecular gas mass and Ly$\alpha$ emission area, with quasars hosting more massive molecular gas reservoirs displaying less extended nebulae. Given the limitations of the APEX data (see Section \ref{sec:APEX}), we revisit this relation using our dataset by examining the relation between molecular gas mass and Ly$\alpha$ emission area (measured within the 2$\,\sigma$ isophotes). We do not find a significant correlation, even when considering variations in Eddington ratio or bolometric luminosity (see also Table \ref{tab:corr_mol_lya}). This lack of correlation may reflect the discussed degeneracies and requires further investigation. 

Another aspect that could affect the brightness and extent of the Ly$\alpha$ nebulae is the orientation of the systems. Systems with host galaxies that are viewed edge-on are more likely to exhibit nebulae with an asymmetric geometry and lower surface brightness \citep{MunozElgueta2022, Costa2022}. For a disk galaxy, the Ly$\alpha$ flux escapes primarily along the rotation axis, as it encounters lower column densities along this direction. However, the observed trend with molecular gas is unlikely to be driven by orientation alone, as it would require that all quasars with larger molecular gas reservoirs happen to be viewed more edge-on by chance. Moreover, the tentative correlation with Eddington ratio supports the idea that intrinsic quasar properties, rather than orientation, play a key role for our findings. We also do not find that the two quasars classified as blazars (ID3 and ID49) -- whose accretion disks are expected to be observed more face-on -- show any peculiar Ly$\alpha$ properties (surface brightness or Ly$\alpha$ emission area) compared to the rest of the sample. However, the escape and scattering of Ly$\alpha$ photons primarily depend on the orientation of the host galaxy, which can be misaligned with the axis of the AGN accretion disk or jet \citep[e.g.,][]{Kinney2000, Schmitt2002}. In order to firmly test the hypothesis of orientation being a major contribution to the observed geometry and brightness of Ly$\alpha$ nebulae, we would require higher-resolution data to properly determine the galaxy morphology and kinematics.

An alternative explanation for the observed connection between Ly$\alpha$ surface brightness and molecular gas reservoirs involves the halo mass. Quasars with more massive molecular gas reservoirs may preferentially reside in more massive DM halos, with deeper potential wells to retain baryonic matter. This could be relevant for our sample, since the CO detections tend to have higher black-hole masses, albeit with substantial uncertainties, which may statistically point to higher-mass halos. In turn, a higher halo mass would indicate a system more dominated by hot accretion. Indeed, according to simulations by \citet{DekelBirnboim2006}, whether gas from the IGM shock-heats or gets accreted in cold form depends sensitively on the halo mass. In lower mass halos, a larger fraction of the accreted gas remains in a cold phase, boosting the reservoir of cool CGM gas available to emit Ly$\alpha$ and ultimately resulting in brighter nebulae. Such a halo-mass dependence would complicate a direct interpretation of the observed trend solely in terms of quasar-driven outflows. 

An equally important question is whether molecular gas could form within the cool halo gas traced by the Ly$\alpha$ nebulae, provided it reaches sufficiently high densities \citep[e.g.,][]{Decarli2021, MunozElgueta2022}. Cloud-crushing simulations by, for example, \citet{McCourt2018}, \citet{GronkeOh2018}, or \citet{Kanjilal2021}, show that cold gas can be entrained in a hot halo environment. However, extended CO emission on circumgalactic scales has only been detected in a handful of systems \citep{Emonts2019, Cicone2021, Jones2023, Li2023}, whereas several other studies of quasars do not find evidence for molecular gas beyond the central galaxy \citep{Riechers2006, Decarli2021}. In our sample, the majority of CO emission remains unresolved, indicating that it is concentrated within the inner $\sim 4\,\mathrm{kpc}$. Six targets are resolved in the 12\,m data according to the criterion of \citet{Decarli2018}, revealing larger spatial extents with radii up to  $\sim 8.4\,\mathrm{kpc}$. These sizes are comparable with other reports of extended molecular gas around $z\sim2-3$ quasars \citep[e.g.,][]{Scholtz2023} and consistent with predicted extents of the ISM in quasar host galaxies at similar redshifts from cosmological simulations of up to $r\sim20\,\mathrm{kpc}$ \citep[e.g.,][]{Obreja2024}. This suggests that the detected emission may still trace the outer regions of the ISM rather than the CGM. We caution, however, that size estimates from \textsc{imfit} are subject to significant uncertainties, in particular considering the low spatial resolution and S/N of our data and we therefore do not interpret this as evidence for extended molecular gas reservoirs.

In summary, the observed relation between Ly$\alpha$ surface brightness and molecular gas mass likely reflects a combination of radiative transfer effects, dust attenuation, and underlying connections with quasar properties. Quasar-driven outflows may enhance the escape of Ly$\alpha$ photons while at the same time reducing the molecular gas content, but alternative explanations related to halo mass and host galaxy orientation cannot be excluded.

\subsection{Low gas fractions in quasar host galaxies} \label{sec:gas_fractions}
We find that most quasars in our sample exhibit relatively low gas fractions, with a median of $ f_\mathrm{gas} = 0.10$. The CO non-detections (which we generally find to exhibit high Eddington ratios and high Ly$\alpha$ surface brightness, see previous section) are expected to exhibit even lower gas fractions (see Figure \ref{fig:Gas_fractions}). We note, however, that the estimates of gas fractions are highly uncertain due to the assumptions involved, such as the adopted value of the $\alpha_\mathrm{CO}$, the inclination of the individual galaxies and the assumed effective radius. Nevertheless, the inferred gas fractions of the quasars in our sample are systematically lower compared with inactive galaxies on the star formation main sequence (MS) at the same redshifts. Relative to the \citet{Tacconi2020} relation for MS galaxies with similar stellar mass $\sim10^{11}\,\mathrm{M}_\odot$ and redshift ($z\sim3.2$), the median gas fractions of our quasars are lower by about 1 dex. Comparably low gas fractions have been found in high-redshift quasar studies, such as the hyperluminous quasars ($z\sim2.4-4.6$) of the WISSH survey \citep{Bischetti2021}, the luminous quasars ($z\sim2$) in \citet{Molyneux2025} or X-ray selected lower luminosity AGN from \citet{Circosta2021} (see also Figure \ref{fig:Gas_fractions}). These results suggest that gas depletion may be a common feature during certain phases of quasar evolution. 

Consistent with this picture, dust-obscured quasars have shown evidence for strong gas depletion \citep[e.g.,][]{Brusa2015, Perna2018, Sun2024}. These systems are often hypothesized to represent a transitional stage in the evolutionary sequence from a starburst galaxy to an unobscured quasar \citep[e.g.,][]{Hopkins2008, Banerij2012, Temple2019, CalistroRivera2021}. This transition, commonly referred to as the “blow-out phase,” is thought to be driven by AGN (and supernova) feedback, which disperses the residual gas through powerful outflows and eventually gives rise to unobscured, blue quasars with depleted gas reservoirs. This interpretation aligns with models in which AGN activity and in particular energetic outflows (such as driven by radiation pressure \citealt{Costa2018a, Costa2018b}) can significantly reduce the molecular gas content in quasars, shaping their star formation and evolutionary path.

In summary, we find that the quasars in our sample show strong evidence for significant gas depletion, consistent with results from previous AGN studies. This further supports a scenario in which AGN feedback plays a key role in shaping the gas reservoirs in the host galaxies of these systems.

\subsection{Companion galaxies in quasar fields} \label{sec:companions}
As mentioned in Section \ref{sec:Linedetections_comp}, we detect 14 potential companion galaxies within the spectral windows of the quasars' CO emission that appear to be physically associated with the quasars. This allows us to assess whether the quasars in our sample reside in overdense environments and to quantify the quasar-galaxy clustering. Previous works on LAEs in the environments of $z\sim3$ quasars with extended Ly$\alpha$ nebulae have reported strong overdensities \citep{Fossati2021}. In addition, clustering analyses indicate that such environments correspond to DM halos with masses of order $10^{12.5}\,\mathrm{M_\odot}$ \citep{Shen2007, Timlin2018, Fossati2021}.

To assess the likelihood that the detected lines arise from lower-redshift interlopers, we computed the probability that any of the detected CO line emitters correspond to lower-redshift galaxies emitting in CO(1-0) at $z\sim0.04$, CO(2-1) at $z\sim1.1$, or CO(3-2) at $z\sim2.1$. We used the CO luminosity functions for these transitions from \citet{Saintonge2017} and \citet{Decarli2019} at the corresponding redshifts. To account for the decreasing sensitivity with increasing radius, we modeled the primary beam response as a 2D Gaussian with a FWHM of $\sim53''$, corresponding to the HPBW of the primary beam for our observations, following \citet{GarciaVergara2022}. For each CO transition, the limiting luminosity at the phase center was determined from the median line rms of our sample (see Table~\ref{tab:observing_summary}), assuming a representative line width of $300\,\mathrm{km\,s^{-1}}$ and requiring a $5\,\sigma$ detection. We then scaled the limiting luminosity with radius according to the primary-beam response, yielding $L_{\mathrm{lim}}(R)$. At each radial integration step, $L_{\mathrm{lim}}(R)$ was used to compute the number density of CO emitters brighter than this limit. The radius-dependent number densities were then integrated over the effective cylindrical companion-search volume of each field, defined by a projected radius of $\sim38''$ from the phase center and a line-of-sight redshift window corresponding to $\pm2000\,\mathrm{km\,s^{-1}}$. The probability that a given field contains a lower-redshift CO(1-0), CO(2-1), or CO(3-2) contaminant was computed to be $\sim0.2\%$, $\sim7\%$, and $\sim3\%$, respectively. To further assess possible lower-redshift contamination, we inspected the MUSE broad-band images for continuum counterparts at the positions of the CO line candidates. Strong continuum emission in the MUSE spectral range would likely indicate a lower-redshift galaxy, whereas sources at $z\sim3$ are expected to be very faint or undetected in the continuum. In the case of ID43, we identify faint continuum emission that is spatially coinciding with the CO companion, but no spectral features which could be indicative of a lower redshift interloper \footnote{At a projected separation of about $1.6\,\arcsec$ ($12\,\mathrm{kpc}$) from the emitter in the field of ID43, we identify an unrelated foreground galaxy at $z\sim0.54$, which is, however, unlikely to be associated with the detected line. }. Applying the same checks for all line candidates (in both the continuum images and the MUSE spectra) yields no evidence for lower-redshift contamination. Using synthetic SDSS r-band images created from the MUSE datacubes, we estimate a 3\,$\sigma$ continuum depth of $m_r\sim 26$\,AB for a $\sim1"$ aperture, indicating that our companions must be fainter in continuum than this limit. This is in agreement with templates of $z\sim3$ SMGs (e.g.,  \citealt{daCunha2015}). Our quasar-galaxy sample is thus determined to consist of 14 potential companion galaxies detected in CO(4-3).

\begin{figure}[!ht]
    \centering

    \includegraphics[width=\linewidth]{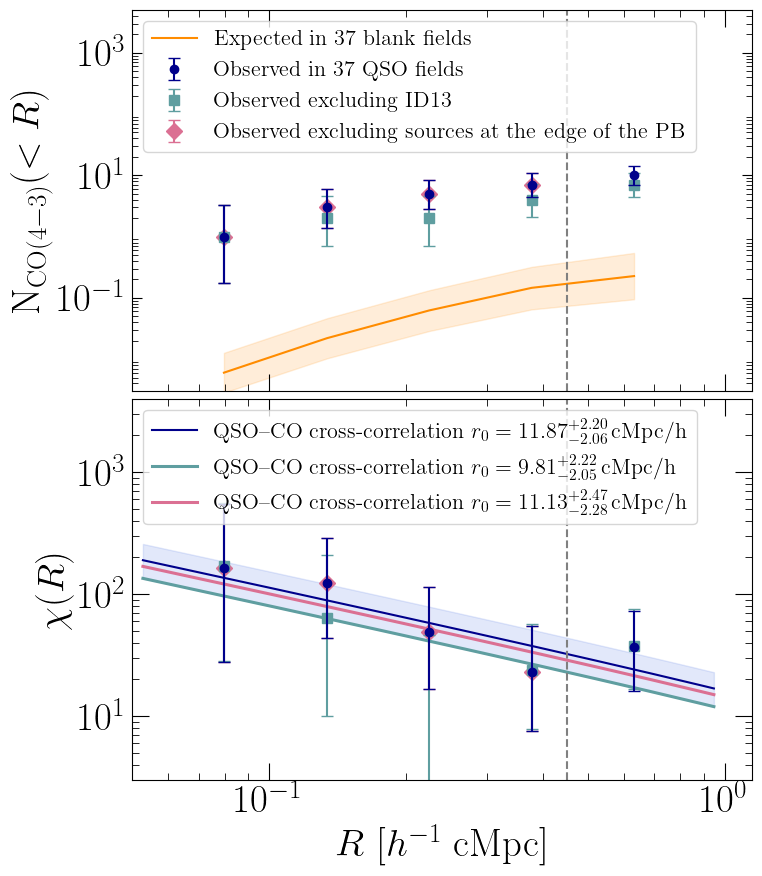}

    \caption[]{\textit{Upper panel}: Cumulative number counts of CO(4-3) lines observed in the quasar fields, including all targets (blue), excluding ID13 (teal) and excluding all sources at the edge of the primary beam (magenta). The magenta sample differs from the full sample only in the outermost radial bin. At smaller radii, the cumulative counts are unchanged and therefore overlap with the blue points. We show the number counts with asymmetric Poisson error bars computed from Garwood confidence intervals \citep{Garwood1936}. The orange curve shows the expected cumulative counts from 37 blank fields, with the shaded region indicating the associated uncertainty. The dashed gray line marks the projected radius beyond which candidate detections become less reliable, as they approach
    the primary-beam edge. \textit{Lower panel}: Quasar–CO cross-correlation function, $\chi(R)$, with Poisson error bars. The solid dark blue line shows the best-fit model (clustering length $r_0$) for the full sample, while the magenta and teal lines show the best-fit models when excluding ID13 and sources near the primary-beam edge, respectively. The shaded blue area indicates the uncertainty on the best-fit model, which is similar for all fits but is only shown for the full sample for clarity. }
    \label{fig:clustering_analysis}
\end{figure}

For the clustering analysis, we further restricted companion galaxies to lie within a redshift window of $\pm1000\,\mathrm{km\,s^{-1}}$ (corresponding to $\Delta z \sim 0.03$), while all galaxies considered already have a fidelity of $\geq80\%$, consistent with the CO(4–3) clustering study at $z\sim4$ by \citet{GarciaVergara2022}. The redshift was measured relative to the CO(4-3) redshift of the associated quasar or, if the quasar is undetected, to the peak Ly$\alpha$ redshift of the nebula. As the companions of ID8, ID45, ID46 and ID49 fall outside this redshift window, we excluded them from the following analysis, leaving a total of 10 companion galaxy candidates.

We first compared the number of companion galaxies detected in the 37 quasar fields to the number expected in blank fields within the primary beam of our observations. The latter was estimated from the CO(4-3) luminosity function of \citet{Decarli2019}, using the same sensitivity model as described above. At the center of our ALMA cubes, the representative limiting luminosity is $L_{\mathrm{lim}} = 6.4 \times 10^{9}\,\mathrm{K\,km\,s^{-1}\,pc^{2}}$, evaluated at the median redshift of the sample ($z\sim3.16$) and derived from the median line rms assuming a $5\,\sigma$ detection and a line width of $300\,\mathrm{km\,s^{-1}}$. In this calculation, the luminosity function was integrated over the comoving volume defined by the projected area of the 37 primary beams, adopting the HPBW of $\sim53''$ as the diameter of each field, and a redshift interval of $\Delta z \sim 0.03$. The expected number of SMGs across 37 blank fields is thus estimated to be about 0.19. In contrast, within the same volume, we detect 7 high-fidelity companion galaxy candidates, indicating a clear excess of galaxies. The total CO(4-3) line overdensity is $37^{+7}_{-6}$, considering a one-sided Poisson confidence interval for small number statistics (\citealt{Gehrels1986}). If we exclude the peculiar system ID13 (the Fabulous ELAN; \citealt[e.g.,][]{ArrigoniBattaia2022}), which represents an extreme outlier in terms of its extended Ly$\alpha$ nebula and unusually rich companion population, the inferred CO(4-3) line overdensity in the remaining 36 quasar fields decreases to $22^{+6}_{-5}$.

Using our data, we can furthermore determine the real-space cross-correlation length $r_{0,\mathrm{QG}}$, following a similar approach to \citet{GarciaVergara2017, GarciaVergara2019, GarciaVergara2022}. To this end, we modeled the expected number of quasar-galaxy pairs in logarithmically spaced radial bins using a power-law cross-correlation function of the form $\xi=(r/r_0)^\gamma$, integrated over the cylindrical volume of each bin (with a redshift window of $\pm1000\,\mathrm{km\,s^{-1}}$), while accounting for the radius-dependent limiting luminosity. We adopted a fixed slope, $\gamma=1.8$, consistent with previous studies \citep[e.g.,][]{Ouchi2004, GarciaVergara2017, GarciaVergara2019, Fossati2021, GarciaVergara2022}. The best-fit $r_{0,\mathrm{QG}}$ was then obtained by maximizing the log-likelihood of the observed pair counts given the model. For this calculation we also took into account potential companion galaxies detected outside the primary beam half-power radius. We obtain a quasar-galaxy cross-correlation length of $11.87^{+2.20}_{-2.06}\,h^{-1}\mathrm{cMpc}$. Excluding again the system ID13, we obtain a quasar-galaxy cross-correlation length of $9.81^{+2.22}_{-2.05}\,h^{-1}\mathrm{cMpc}$. Excluding sources located near the edge of the primary beam (beyond a radius of $\sim20\,\arcsec$), where detections are expected to be less reliable due to reduced sensitivity, yields a cross-correlation length of $11.13^{+2.47}_{-2.28}\,h^{-1}\mathrm{cMpc}$. Figure \ref{fig:clustering_analysis} shows the cumulative number counts of CO(4-3) lines observed in our quasar fields as well as the cross-correlation function between quasars and CO(4-3) line emitters, given by $\chi(R)=\frac{\langle QG(R) \rangle}{\langle QR(R) \rangle} -1 $, with $\langle QG(R) \rangle$ the observed number of quasar-galaxy pairs and $\langle QR(R) \rangle$ the expected number of quasar-galaxy pairs assuming the blank field number density of CO(4-3) emitters \citep[see also][]{GarciaVergara2022}.

The derived values for the cross-correlation length are consistent within uncertainties of the value obtained for CO(4-3) emitters at $z\sim 4$ in \citet{GarciaVergara2022} of $8.37^{+2.42}_{-2.04}\,h^{-1}\mathrm{cMpc}$.
On the other hand, LAEs around high-redshift quasars exhibit weaker clustering, with correlation lengths of $2.78^{+1.16}_{-1.05}\,h^{-1}\mathrm{cMpc}$ at $z\sim4$ \citep{GarciaVergara2019} or $1.95^{+0.22}_{-0.23}\,h^{-1}\mathrm{cMpc}$ at $z=3-4.5$ \citep{Fossati2021}. \citet{ArrigoniBattaia2019} present a search for LAEs in the fields of the bright QSO MUSEUM sample, finding similar number counts as expected for blank fields. This suggests that compared to SMGs, LAEs might be more affected by the extreme environments within the halos of massive active galaxies.

Nonetheless, overall only about 30\% of the quasars host CO line emitting companions within their fields, compared to the much higher fraction of $\sim$80\% reported for hyper-luminous quasars in the WISSH survey \citep{Bischetti2021} and $\sim$70\% found for ELANe environments in \citet{Li2023}. While, on average, the quasars in our sample are found in overdense environments, many individual fields do not show evidence for such overdensities. We note, however, that some faint companions may remain undetected in our data due to sensitivity limits.

\begin{figure}[!ht]
    \centering

    \includegraphics[width=\linewidth]{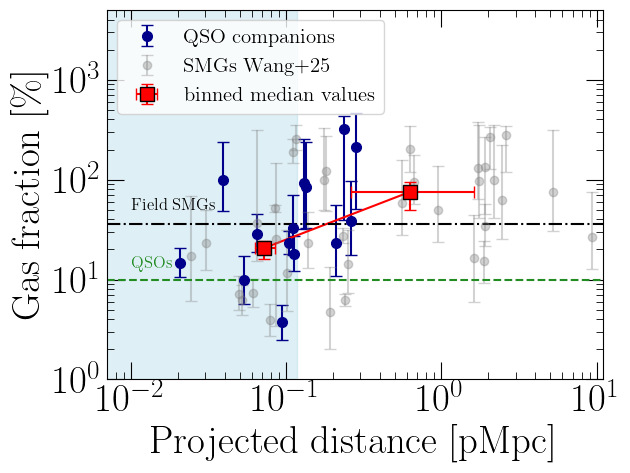}

    \caption[]{Gas fraction of companion galaxies as a function of projected distance between quasar and companion. The blue points show the potential companion galaxies identified in our survey and the gray points SMGs compiled in \citet{Wang2025}. The shaded blue region marks the extent of the virial radius corresponding to a halo mass of $M_\mathrm{halo} = 10^{12.5}\,M_\odot$ at the median redshift of our quasar sample. The upper dashed line shows the median gas fraction of field SMGs, and the lower dashed line the median value of this work's quasar sample, which is consistent with median gas fractions reported for quasars at similar redshifts in the literature. Errors on the median values (in red squares) are obtained from bootstrapping, by resampling the data within each bin and recomputing the medians. The figure is adapted from \citet{Wang2025}. }
    \label{fig:fgas_companions}
\end{figure}

To assess the impact of quasars on nearby galaxies, we analyzed the gas fractions of detected companion galaxies as a function of projected separation from the quasar. We followed the analysis in \citet{Wang2025}, which showed that SMGs around quasars at $z\sim 2-3$ show a trend for lower gas fractions within the quasar virial radius. We calculated the gas fractions using Equation \ref{eq: gas_fractions} with $\alpha_\mathrm{CO}=0.8\, \mathrm{M_\odot}\,(\mathrm{K\,km\,s^{-1}\,pc^2})^{-1}$ and a ratio of $r_{41}=0.34$ as used in \citet{Wang2025} for the SMGs in the field (and $r_{41}=0.87$ for the companions AGN1 and QSO2 of ID13). We compared our data to the sample of SMGs around quasars compiled in \citet{Wang2025}. To ensure a fair comparison, we scaled their gas fractions, originally computed with $\alpha_\mathrm{CO}=1.0$, to match our adopted $\alpha_\mathrm{CO}$. Figure \ref{fig:fgas_companions} shows the combined data, with our companion measurements overplotted on the SMG sample from \citet{Wang2025}. We calculated the median gas fraction inside and outside the virial radius (for a halo mass of $M_\mathrm{halo} = 10^{12.5}\,M_\odot$ at the median redshift of our quasar sample) for the combined dataset, using both our companion measurements and the SMGs from \citet{Wang2025} (Figure \ref{fig:fgas_companions}). We find a median $f_\mathrm{gas}$ inside the virial radius of $0.21^{+0.04}_{-0.05}$ and of $0.76^{+0.18}_{-0.26}$ outside. A Kendall rank correlation test gives $\tau=0.32$ and $p=1\times10^{-3}$, which hints at a significant trend between companion gas fraction and projected distance to the quasar. Including our sources together with the SMGs compiled by \citet{Wang2025} decreases the $p$ value by a factor of about 5, corresponding to an increase in statistical significance. The companion of ID43, which is the second-closest companion found in our sample at a distance of about $40\,\mathrm{kpc}$, appears to be an outlier, showing an unusually high gas fraction given its location within the virial radius. The overall observed decrease in gas fractions within the virial radius is consistent with the trends reported by \citet{Wang2025} and indicate that our data reinforce the picture of quasar environments influencing the gas content of satellites, potentially through ram-pressure stripping or strangulation.

\section{Conclusions} \label{sec:Conclusions}
We present CO(4–3) observations of a sample of 37 quasars at $z\sim3$. As part of the QSO MUSEUM survey, these quasars are known to host bright, extended Ly$\alpha$ nebulae previously observed with VLT/MUSE. With our new ALMA data, we investigated the molecular gas content of the quasar host galaxies and explored their connections to the surrounding cool Ly$\alpha$-emitting halo gas and effects of AGN activity.
The main results of our work can be summarized as follows:
\begin{itemize}
  \item Of the 37 targeted quasars, 21 are detected in CO(4-3), corresponding to a detection rate of about $57\%$. We find CO line widths of $\sim 70 - 640\,\mathrm{km\,s^{-1}}$ and derive
  molecular gas masses of the detected quasars of $M_\mathrm{gas}=(3-40) \times10^9\,\mathrm{M_\odot}$. For the quasars undetected in CO, we derive $3\,\sigma$ upper limits on the molecular gas masses, with a median value of $\sim 3.1 \times 10^9\,\mathrm{M_\odot}$.
  \item We explored correlations between Ly$\alpha$ properties and molecular gas properties and find that the quasars with the dimmest Ly$\alpha$ nebulae are associated with the most massive molecular gas reservoirs. In particular quasars that host nebulae with central Ly$\alpha$ SB corrected for cosmological dimming above $\sim1.3\times10^{-14}\,\mathrm{erg\,s^{-1}\,{cm}^{-2}\,arcsec^{-2}}$ are mostly not detected in CO. This indicates that host galaxy properties play a pivotal role in shaping the characteristics of the surrounding Ly$\alpha$ nebulae. In particular, strong obscuration in the quasar hosts by gas and dust could reduce both the escape fraction of Ly$\alpha$ photons from the BLR and the ionizing photon flux, thereby reducing the halo emission. 
  \item We find that in our sample, quasars with lower Eddington ratios (below $\sim 0.8$) exhibit more massive molecular gas reservoirs. By contrast, quasars with higher Eddington ratios remain mostly undetected in CO. This trend can be interpreted as evidence that strongly accreting quasars drive more powerful outflows, which deplete the gas and dust content of their host galaxies and thereby enhance the reprocessed Ly$\alpha$ emission. This interpretation is consistent with simulations by \citet{Costa2022} that have shown that quasar-driven outflows can produce brighter, more extended Ly$\alpha$ nebulae. 
  \item By using a dynamical mass estimate, we find that the detected quasars in our sample exhibit on average low gas fractions (median of $M_\mathrm{gas}/M_* \sim 0.10$) compared to inactive, star-forming galaxies at similar stellar masses and redshifts. This is consistent with findings in the literature and supports the idea that AGN activity may heat and deplete molecular gas reservoirs. We also investigate a potential trend with Eddington ratio but find no significant correlation.
  \item Six of the detected quasars are spatially resolved in CO and we derive host galaxy sizes with effective radii of up to $\sim8\,\mathrm{kpc}$. For the remaining unresolved sources, half the beam major-axis FWHM corresponds to $>4\,\mathrm{kpc}$, consistent with the CO emission in most quasars in our sample being concentrated within $\sim 4\,\mathrm{kpc}$. 
  \item We detect 14 potential companion galaxies in the 37 fields, implying an overall overdensity with respect to blank fields. A clustering analysis for the detected CO line emitters yields a quasar-galaxy cross-correlation length of $9.81^{+2.22}_{-2.05}\,h^{-1}\mathrm{cMpc}$, compared to $8.37^{+2.42}_{-2.04}\,h^{-1}\mathrm{cMpc}$ for $z\sim 4$ CO(4-3) emitters in \citet{GarciaVergara2022}. However, on an individual basis, only 30\% of the fields show evidence of companion galaxies in the submillimeter.
\end{itemize}

Our results highlight the importance of using multiple gas tracers across different physical scales to gain a comprehensive understanding of the processes shaping both the quasar host galaxies and their surrounding CGM. By combining interferometric observations with ALMA to determine host galaxy properties, such as systemic redshifts, molecular gas content, and geometry, with MUSE observations of the extended Ly$\alpha$ emission, we will be able to constrain Ly$\alpha$ radiative transfer effects in future studies and better understand the processes powering these large-scale gas reservoirs. Moreover, expanding the analysis to the fainter QSO MUSEUM targets, characterized by lower black hole masses and Eddington ratios, may provide further confirmation of our trends. Additionally, ionized gas tracers such as [\ion{O}{iii}] or H$\alpha$ could provide valuable information on the presence of outflows and help test our interpretation.

\begin{acknowledgements}
We thank the anonymous referee for the valuable comments that helped improve the manuscript. 
J.R. thanks Guinevere Kauffmann, Stephen Molyneux and Aniket Bhagwat for helpful discussions. We thank Yu-Jan Wang and Stephen Molyneux for kindly providing data used in this work.
C.-C.C. acknowledges support from the National Science and Technology Council of Taiwan (NSTC 111-2112-M-001-045-MY3 and 114-2628-M-001-006-MY4), as well as Academia Sinica through the Career Development Award (AS-CDA-112-M02). A.O. is funded by the Carl-Zeiss-Stiftung through the NEXUS Programm. J.G.L. acknowledges funding from the DLR (German Aerospace Agency) via grant 50 OR2401. This paper makes use of the following ALMA data: ADS/JAO.ALMA\#2023.1.00963.S. ALMA is a partnership of ESO
(representing its member states), NSF (USA) and NINS (Japan), together with NRC (Canada), NSTC and ASIAA (Taiwan), and KASI (Republic of Korea), in cooperation with the Republic of Chile. The Joint ALMA Observatory is operated by ESO, AUI/NRAO and NAOJ.
\end{acknowledgements}

\bibliography{references}
\bibliographystyle{aa}

\onecolumn
\begin{appendix}

\section{CO and continuum observations}

\begin{figure*}[htbp]
    \centering

    \begin{minipage}{\linewidth}
        \centering
        \includegraphics[width=0.3\linewidth]{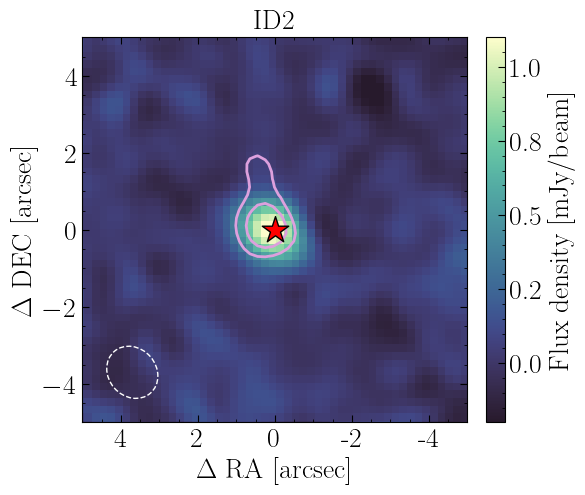}
        \includegraphics[width=0.3\linewidth]{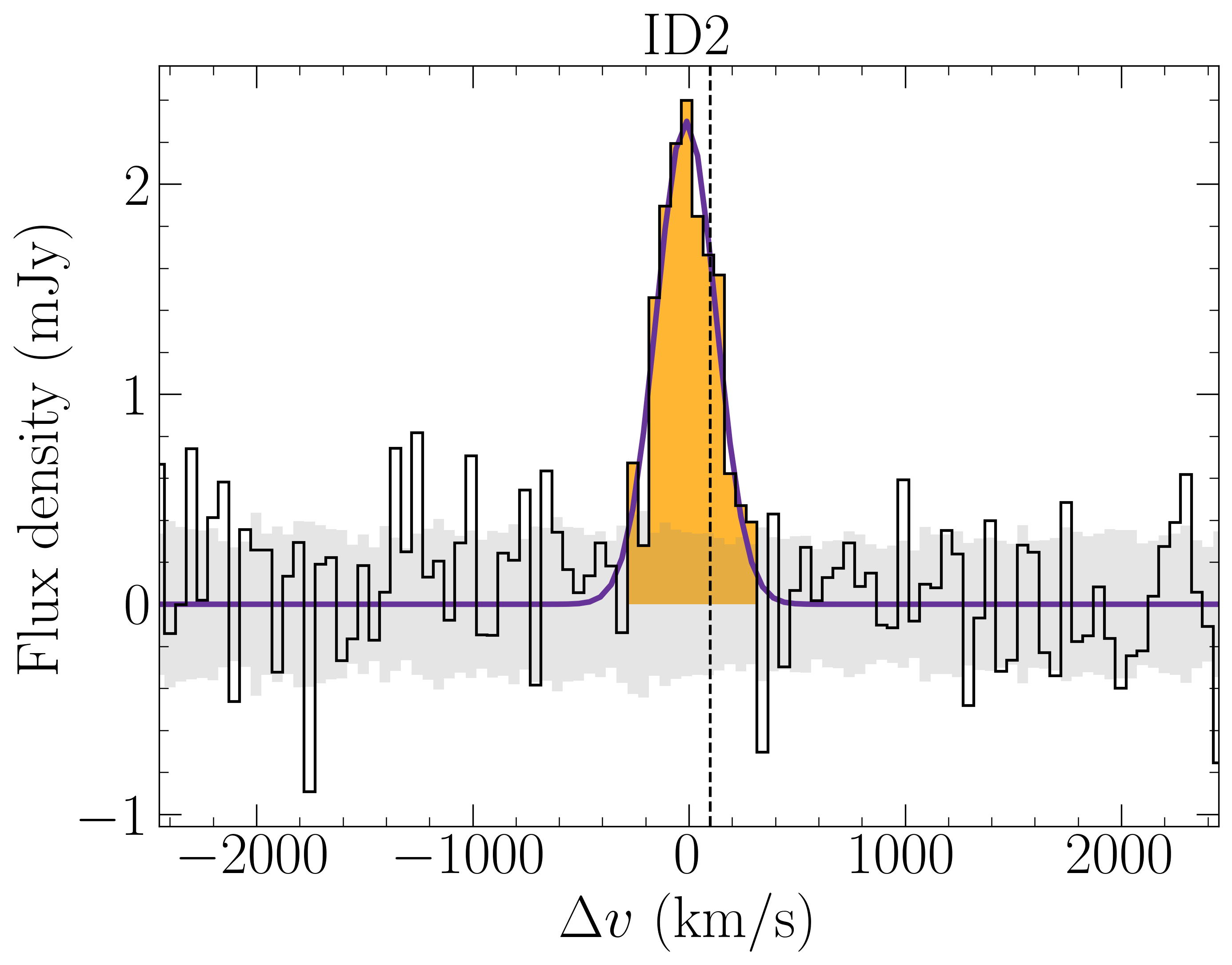}
        \includegraphics[width=0.3\linewidth]{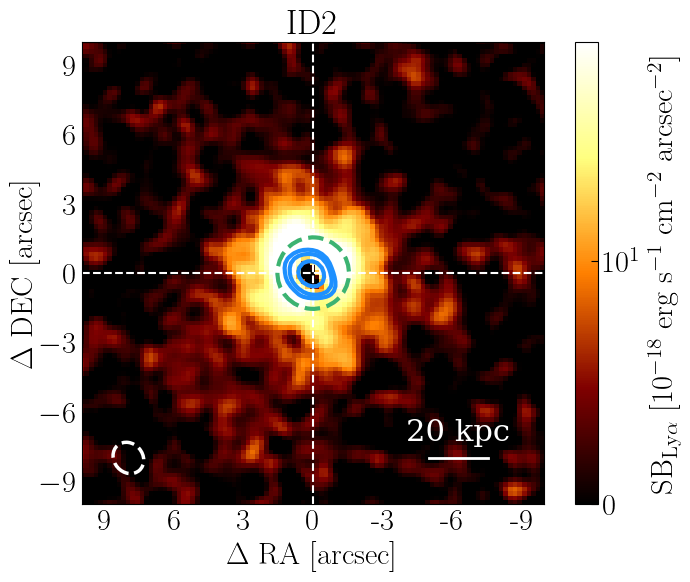}
    \end{minipage}

    \begin{minipage}{\linewidth}
        \centering
        \includegraphics[width=0.3\linewidth]{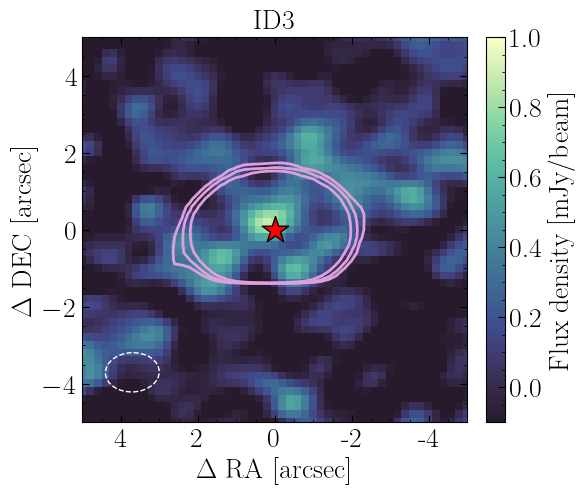}
        \includegraphics[width=0.3\linewidth]{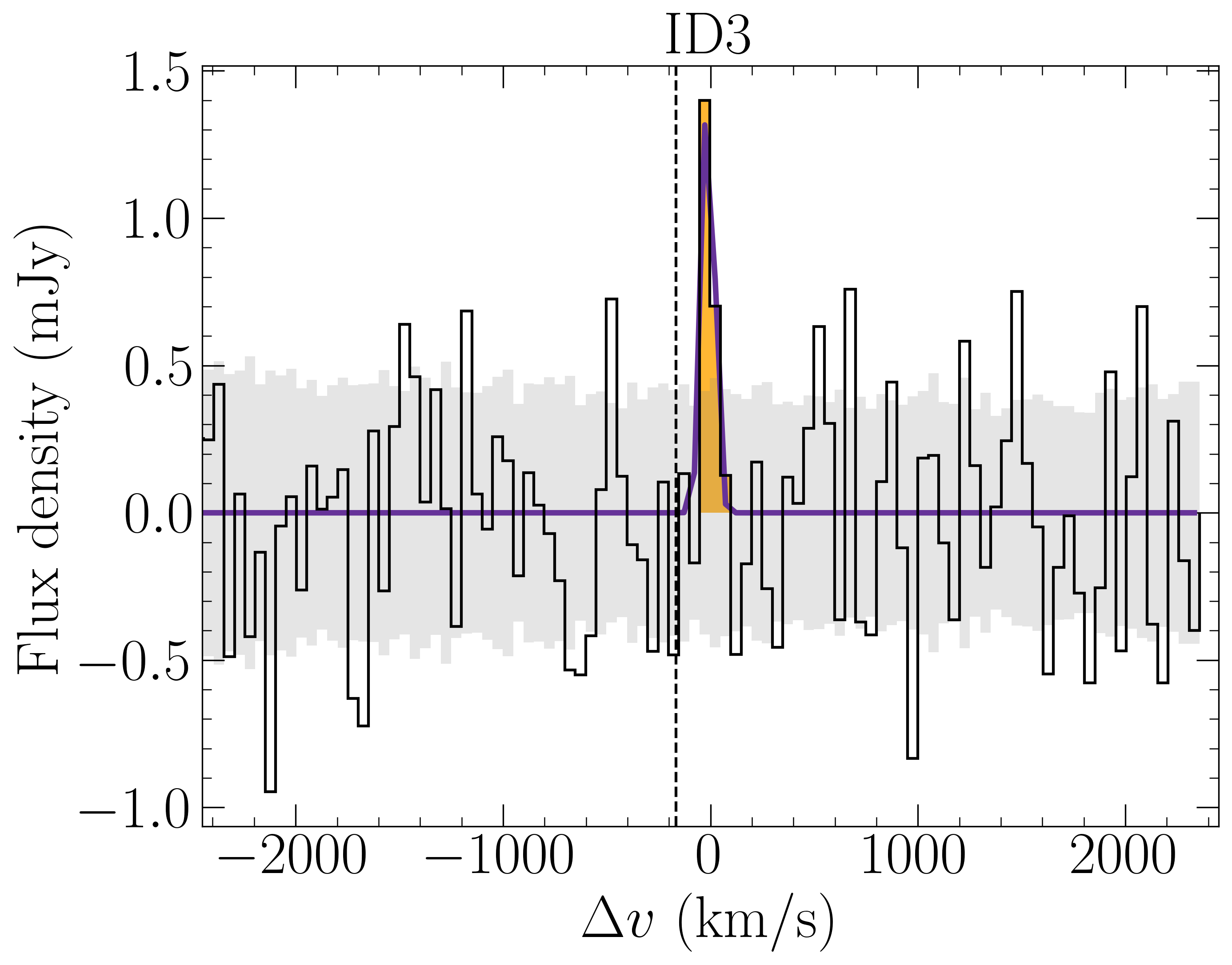}
        \includegraphics[width=0.3\linewidth]{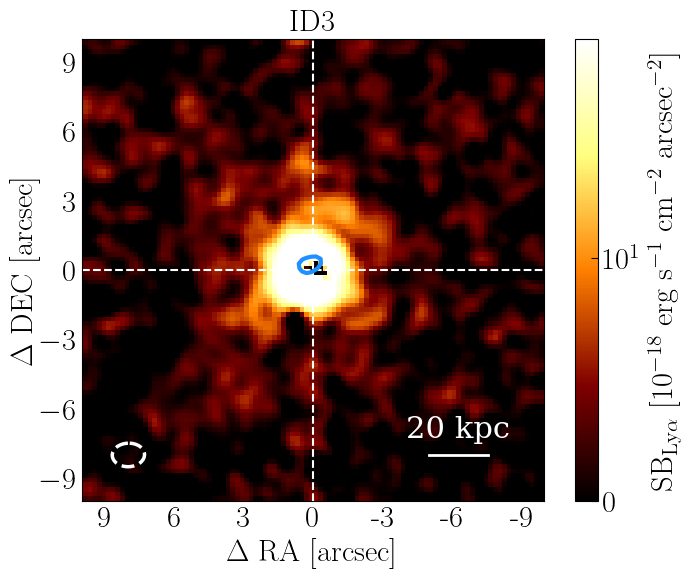}
    \end{minipage}

    \begin{minipage}{\linewidth}
        \centering
        \includegraphics[width=0.3\linewidth]{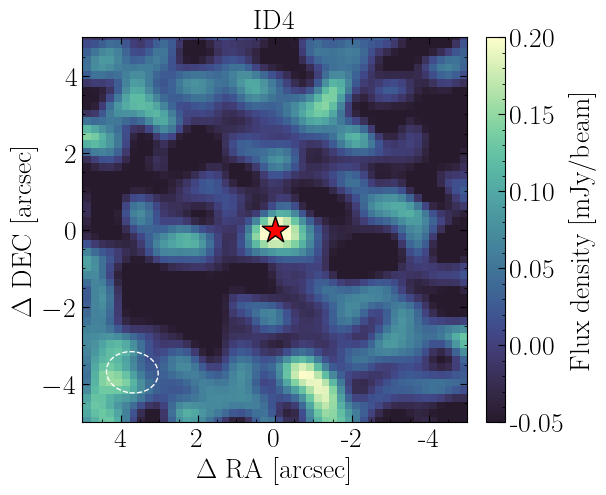}
        \includegraphics[width=0.3\linewidth]{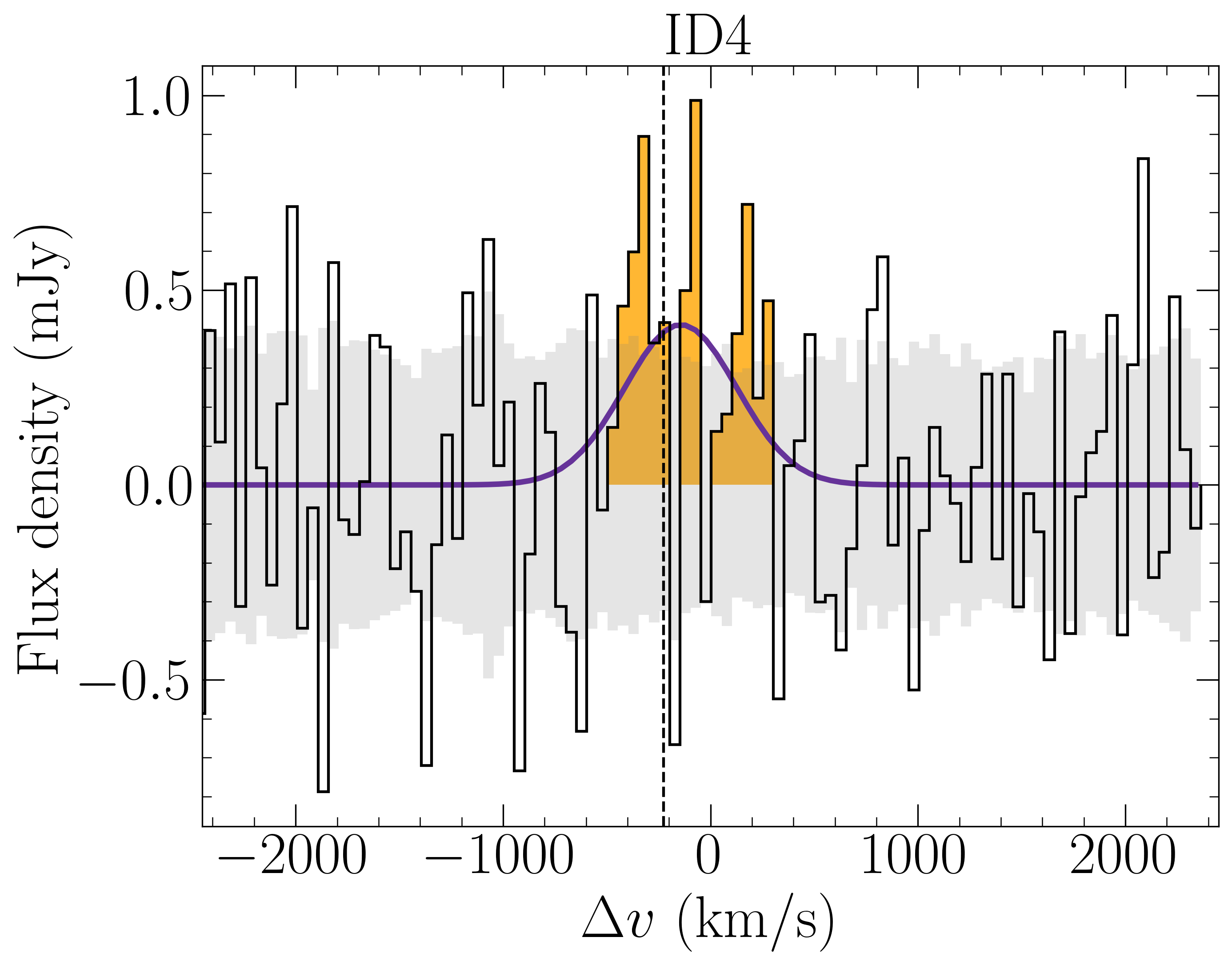}
        \includegraphics[width=0.3\linewidth]{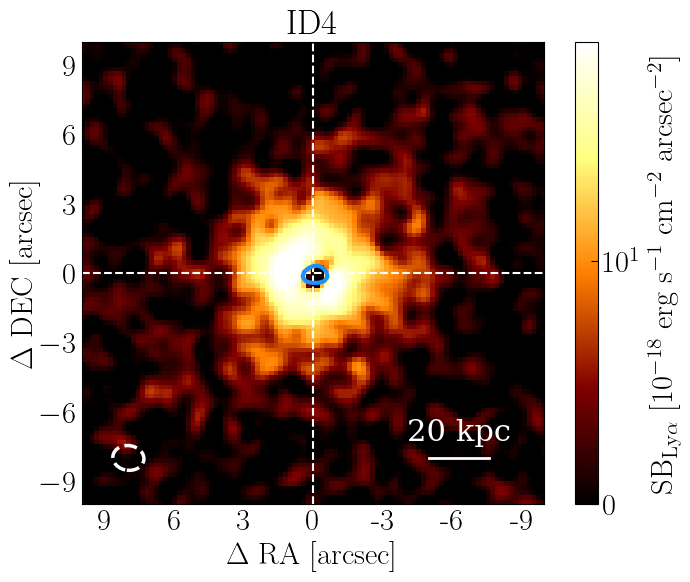}
    \end{minipage}

    \begin{minipage}{\linewidth}
        \centering
        \includegraphics[width=0.3\linewidth]{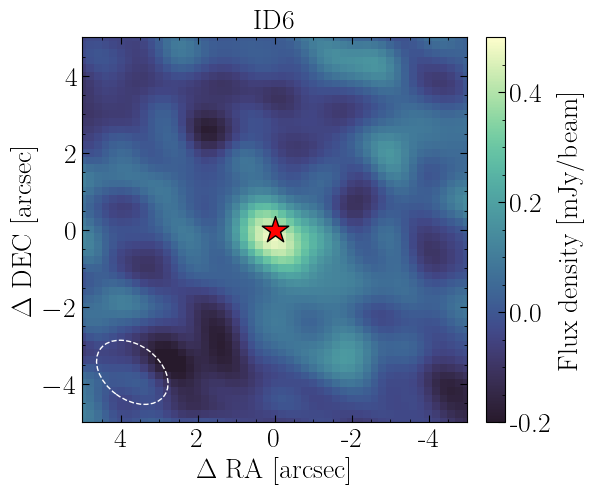}
        \includegraphics[width=0.3\linewidth]{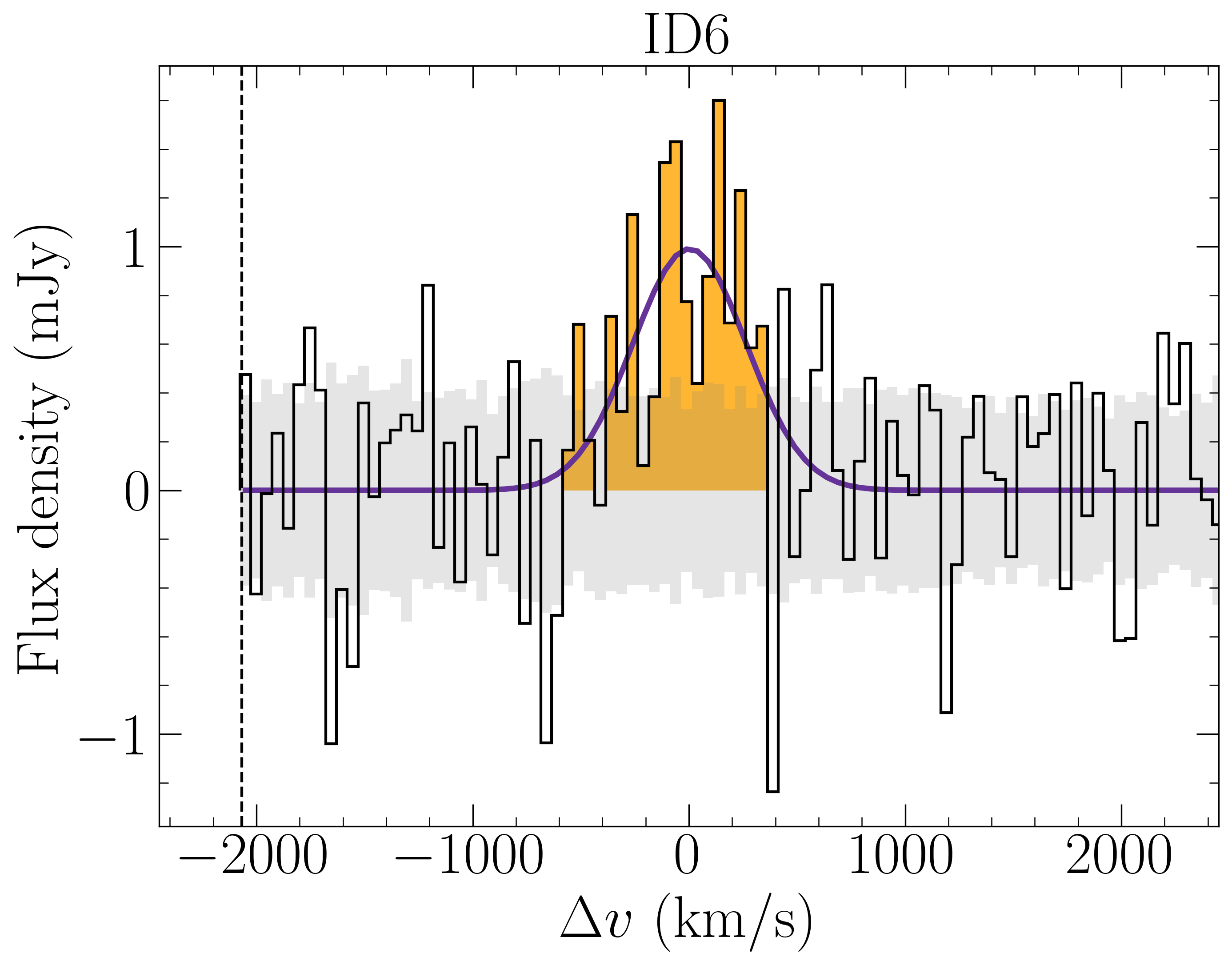}
        \includegraphics[width=0.3\linewidth]{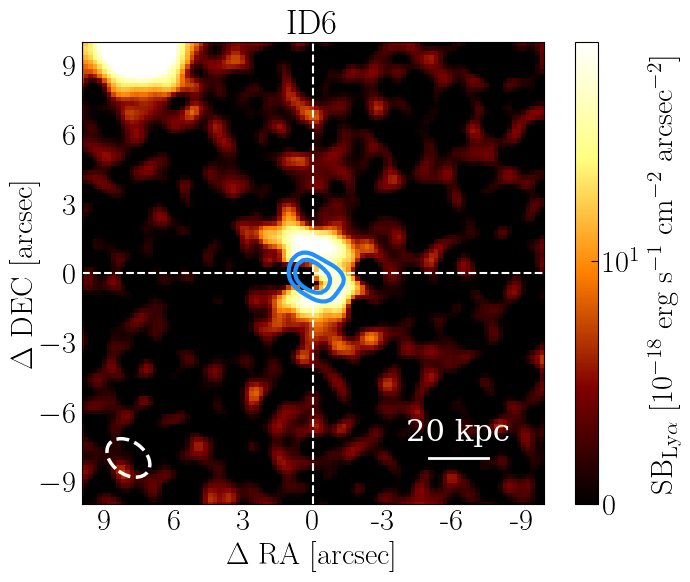}
    \end{minipage}

    \caption{\textit{Left panel}: Multifrequency synthesis (MFS) image of the CO(4-3) emission of the quasars. Contours indicate the continuum emission (at 3, 5 and 10$\,\sigma$), plotted for those quasars with continuum detection. Red stars mark the position of the quasars in the optical. The synthesized beam is shown as an ellipse in the lower left corner. \textit{Middle panel}: Spectrum of the CO(4-3) emission. $0\,\mathrm{km}\mathrm{s}^{-1}$ corresponds to the centroid of the CO(4-3) emission (determined from a first-moment analysis). The purple line indicates the Gaussian fit. The shaded orange regions highlight the line emissions. The shaded gray area shows the rms noise level of the spectrum. The dashed black line shows velocity offset corresponding to the peak Ly$\alpha$ wavelength of the nebula (from \citealt{GonzalezLobos2025}). \textit{Right panel}: CO contours (at 3, 5 and 10$\,\sigma$) on the PSF- and continuum-subtracted Ly$\alpha$ surface brightness maps. The dashed green circle in the first map indicates a radius of 12\,kpc within which the central Ly$\alpha$ surface brightness is calculated after masking the noisy PSF normalization region (see Figure \ref{fig:MH2vsLya}). For clarity it is only shown once per page, as the scales are comparable for all maps.} 
    \label{fig:flux_maps_spectra}
\end{figure*}

\begin{figure*}[htbp]
    \addtocounter{figure}{-1}
    \centering
    \begin{minipage}{\linewidth}
        \centering
        \includegraphics[width=0.3\linewidth]{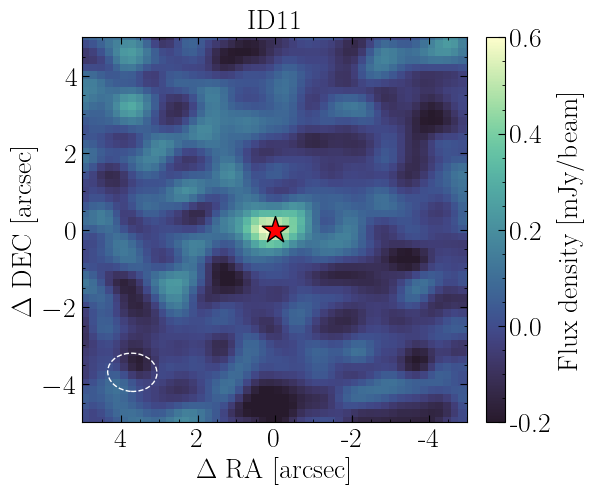}
        \includegraphics[width=0.3\linewidth]{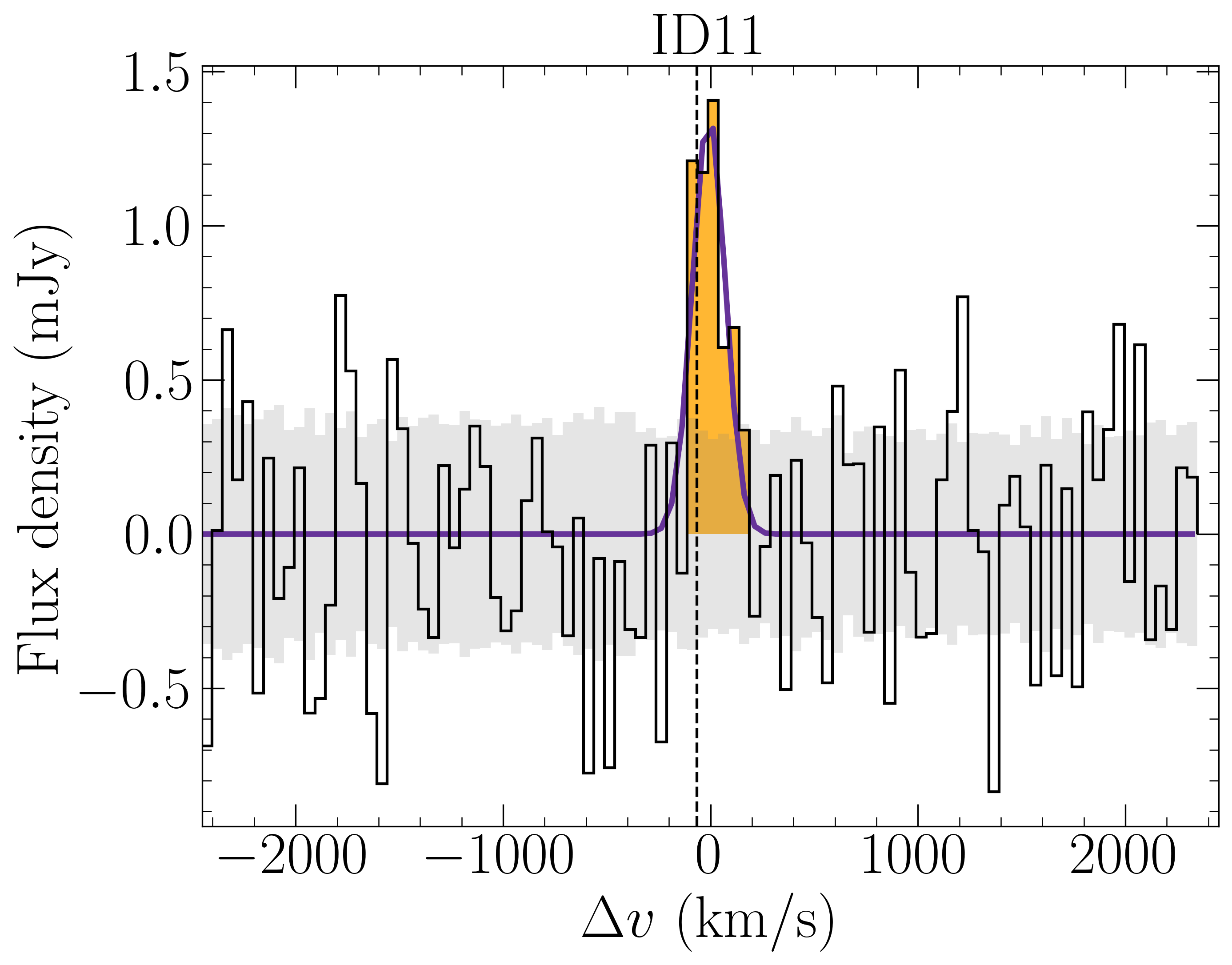}
        \includegraphics[width=0.3\linewidth]{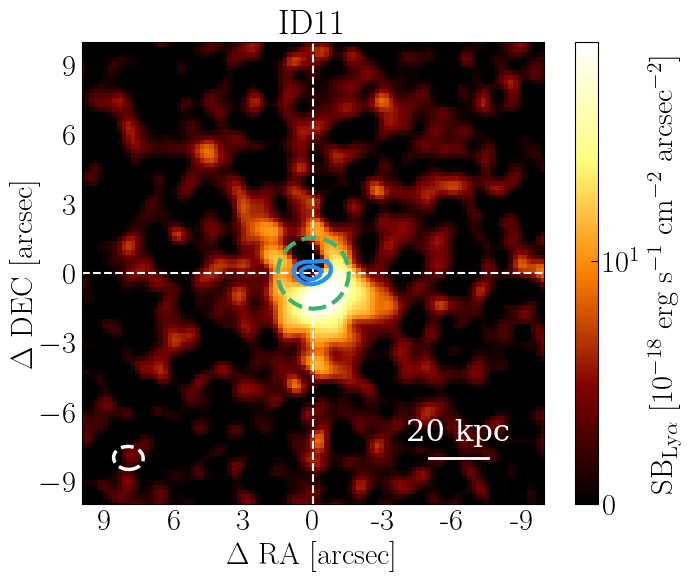}
    \end{minipage}

    \begin{minipage}{\linewidth}
        \centering
        \includegraphics[width=0.3\linewidth]{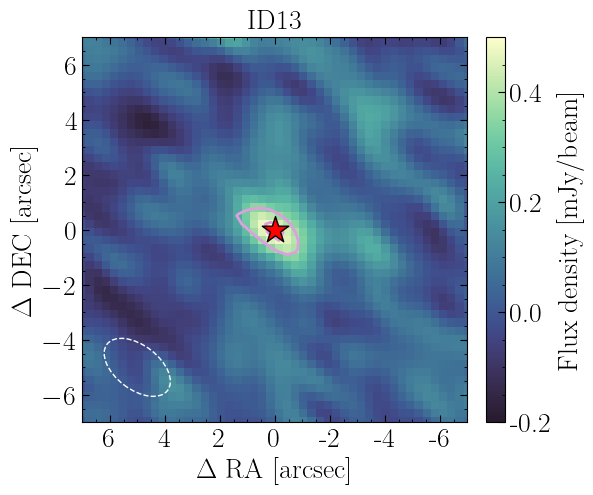}
        \includegraphics[width=0.3\linewidth]{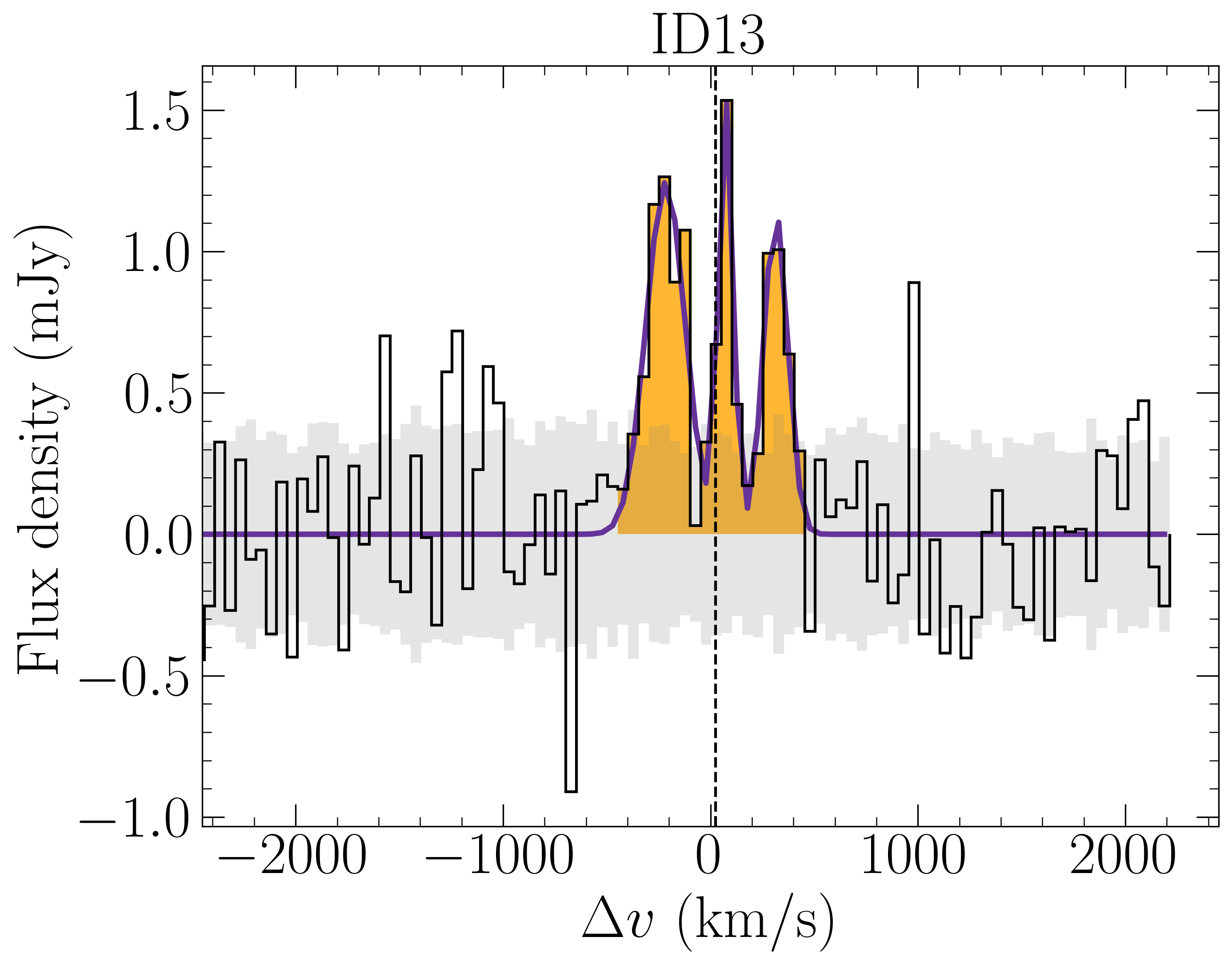}
        \includegraphics[width=0.3\linewidth]{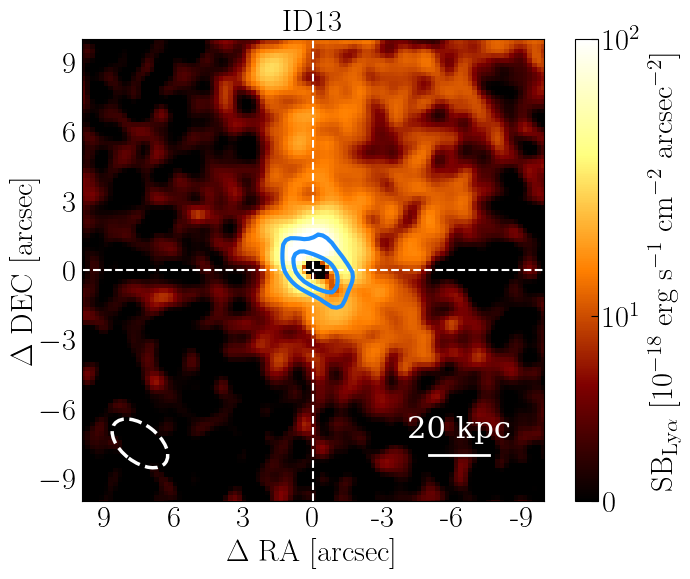}
    \end{minipage}

    \begin{minipage}{\linewidth}
        \centering
        \includegraphics[width=0.3\linewidth]{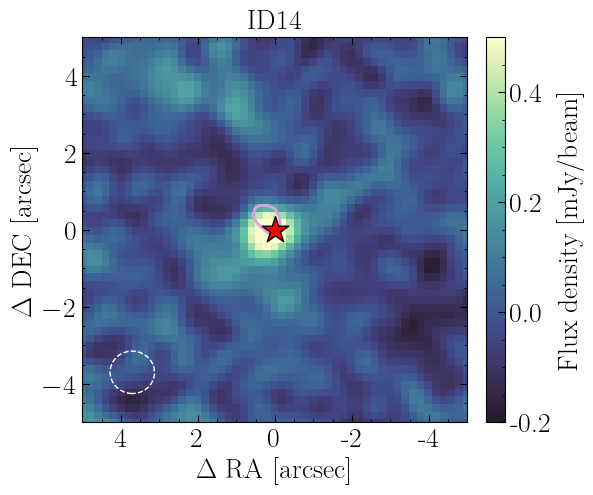}
        \includegraphics[width=0.3\linewidth]{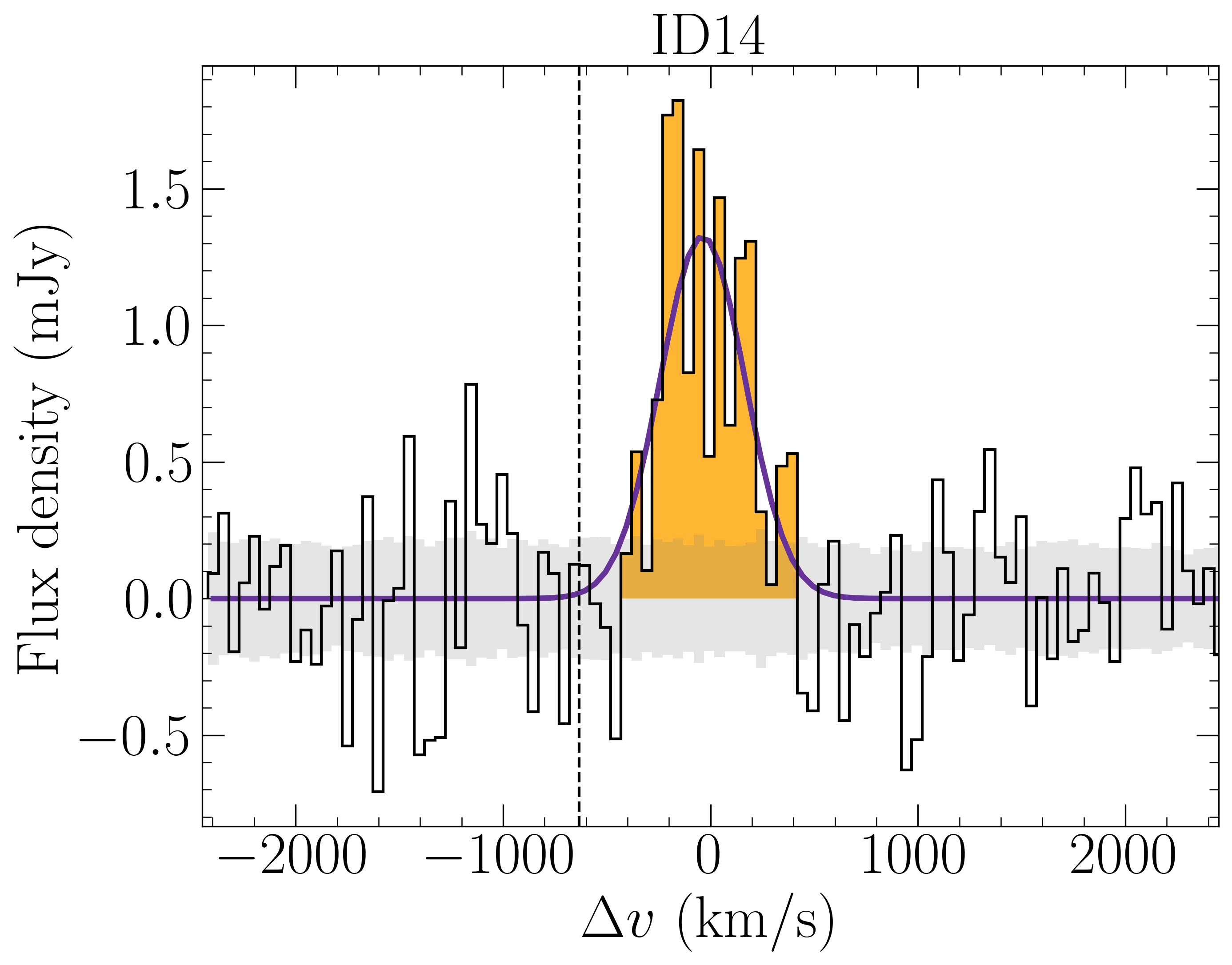}
        \includegraphics[width=0.3\linewidth]{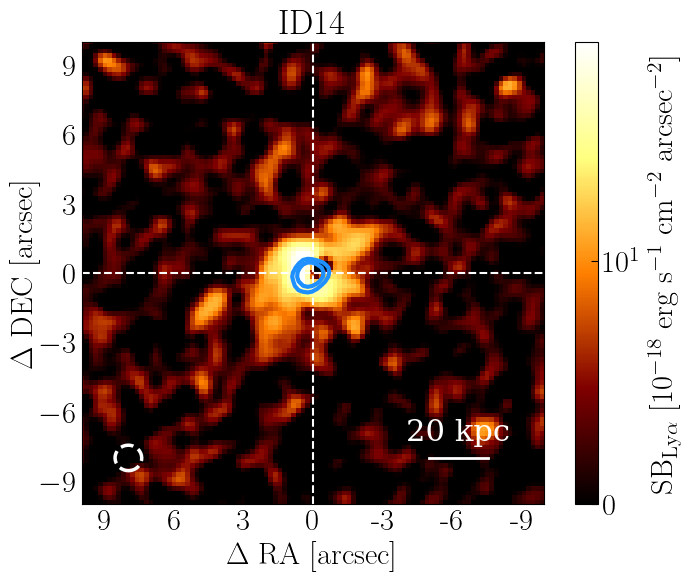}
    \end{minipage}

    \begin{minipage}{\linewidth}
        \centering
        \includegraphics[width=0.3\linewidth]{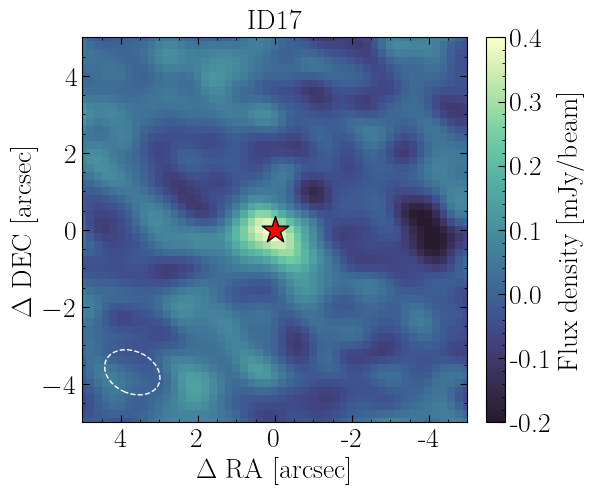}
        \includegraphics[width=0.3\linewidth]{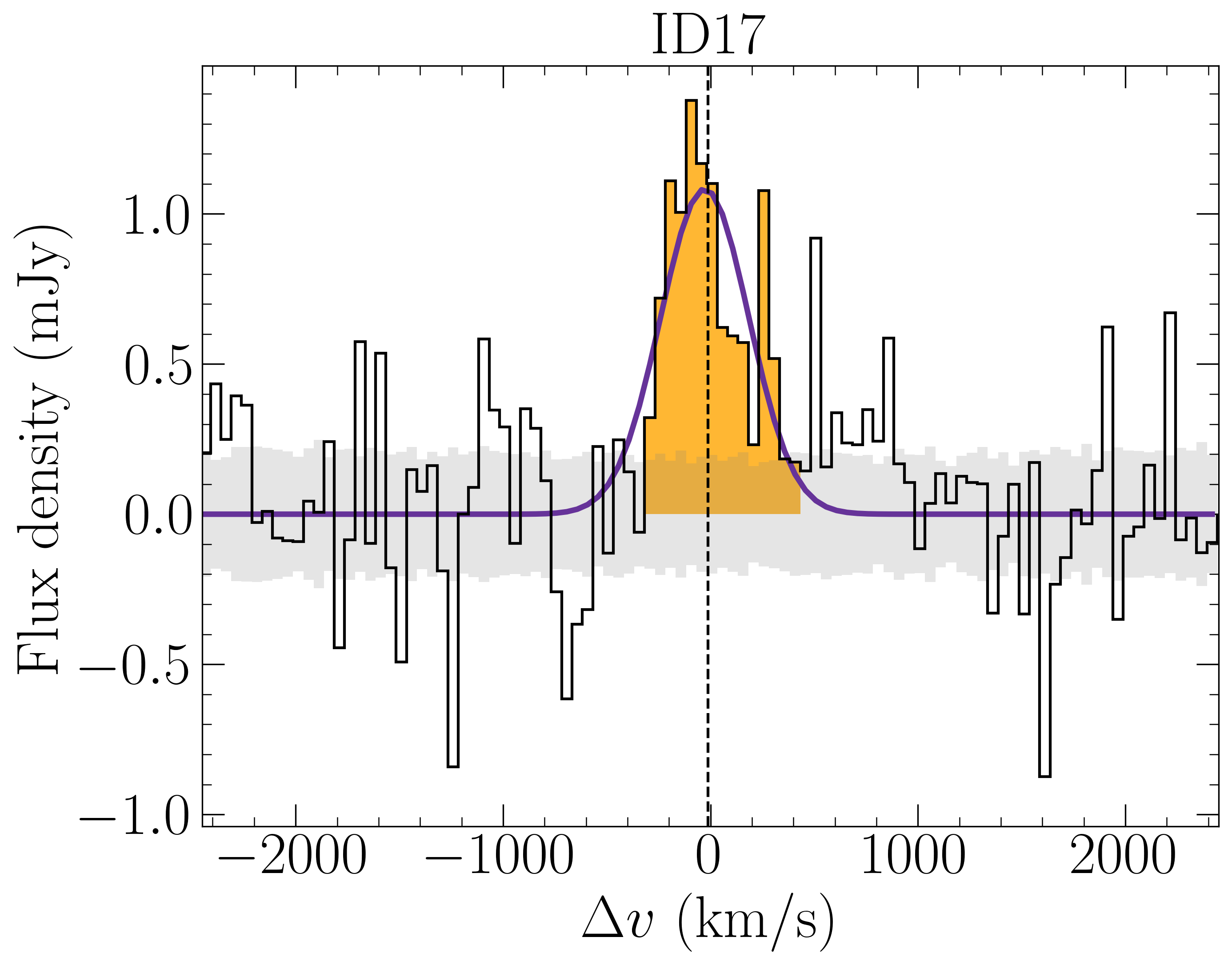}
        \includegraphics[width=0.3\linewidth]{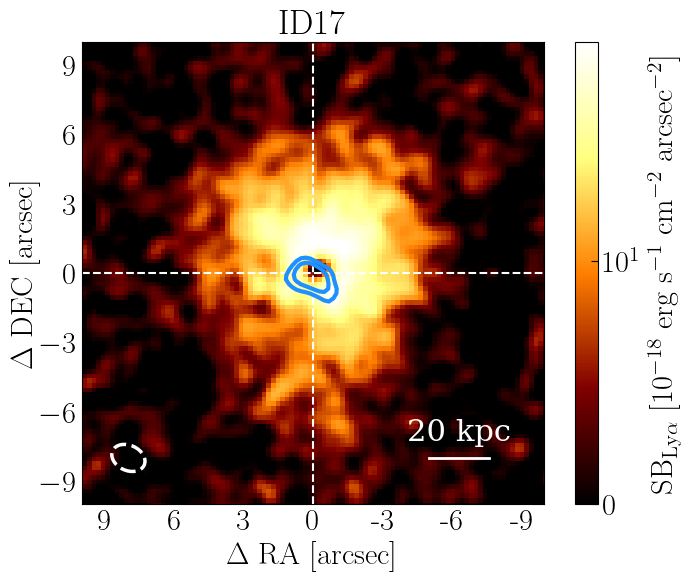}
    \end{minipage}

    \begin{minipage}{\linewidth}
        \centering
        \includegraphics[width=0.3\linewidth]{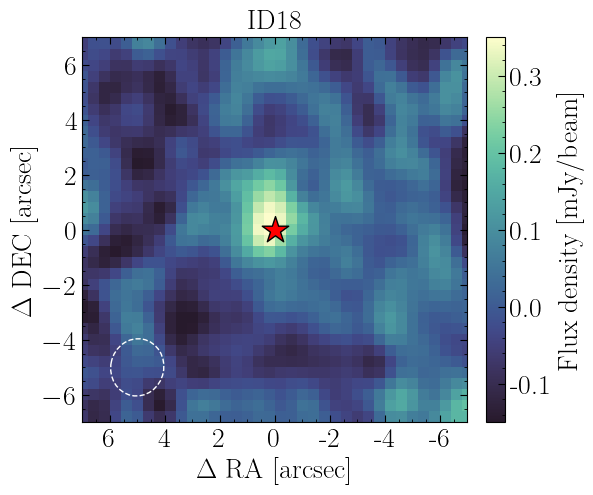}
        \includegraphics[width=0.3\linewidth]{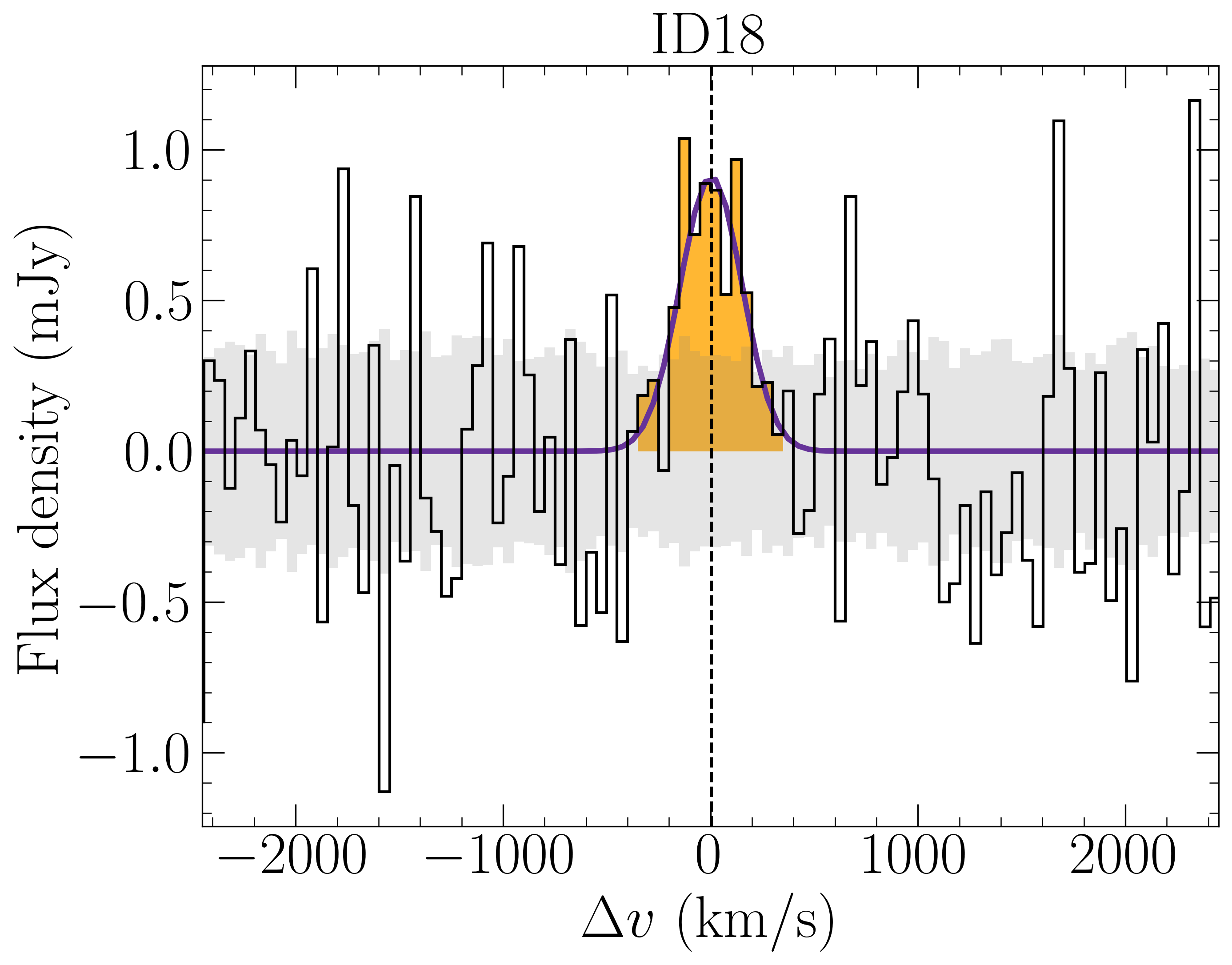}
        \includegraphics[width=0.3\linewidth]{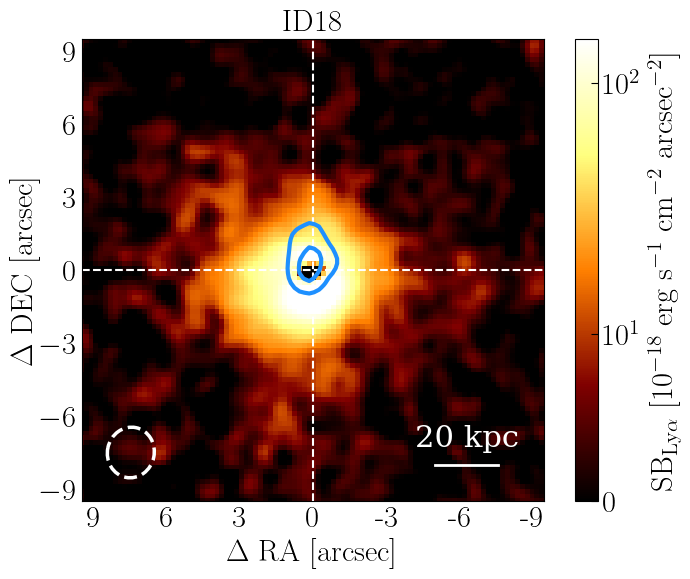}
    \end{minipage}

    \caption{continued.}
    \label{fig:flux_maps_spectra_2}
\end{figure*}

\begin{figure*}[htbp]
    \addtocounter{figure}{-1}
    \centering

    \begin{minipage}{\linewidth}
        \centering
        \includegraphics[width=0.3\linewidth]{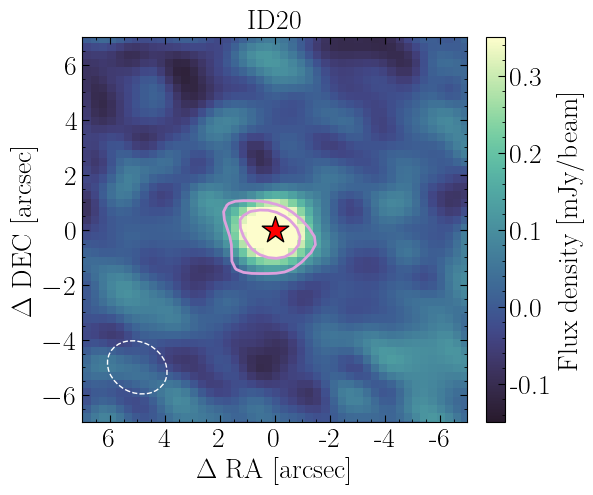}
        \includegraphics[width=0.3\linewidth]{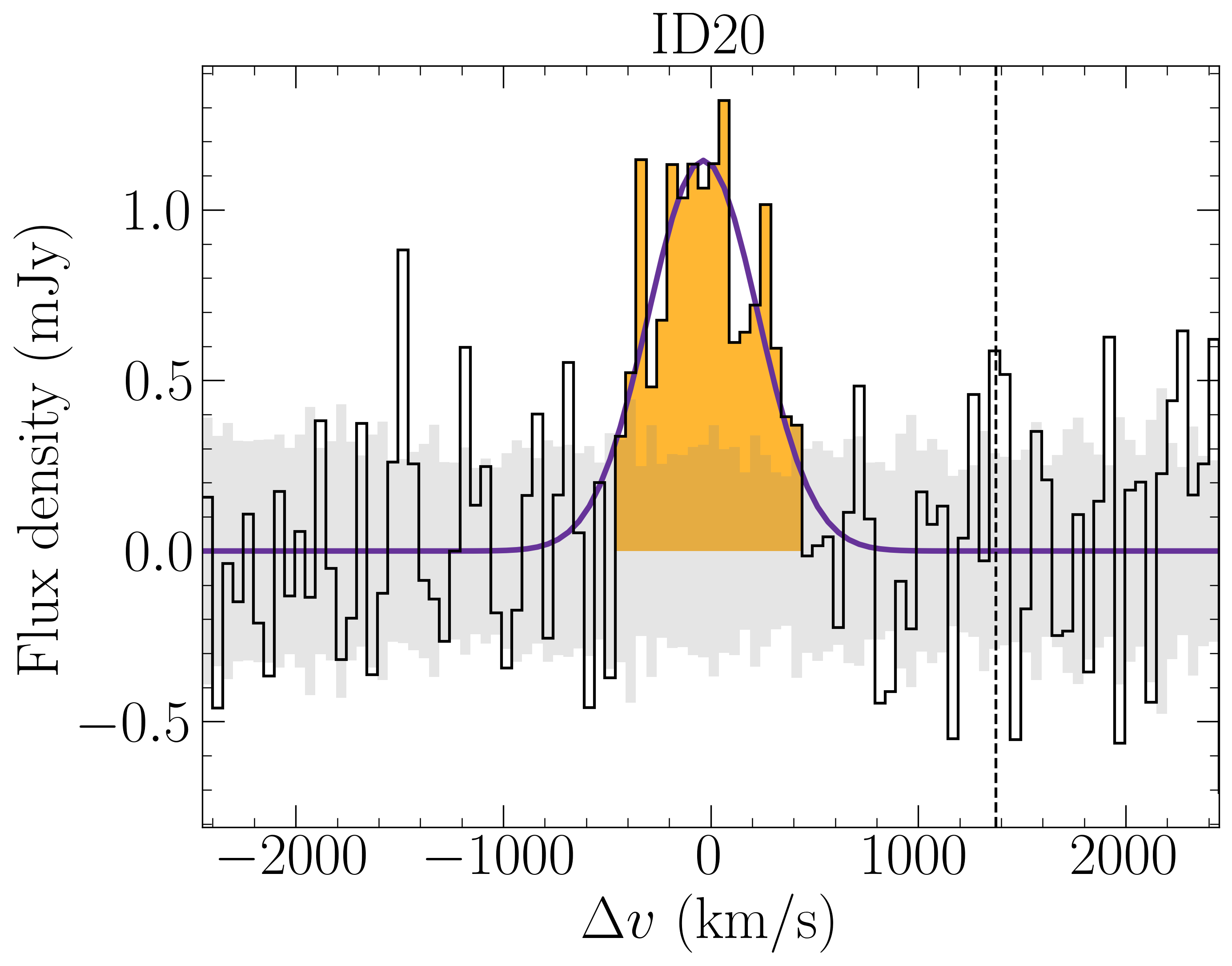}
        \includegraphics[width=0.3\linewidth]{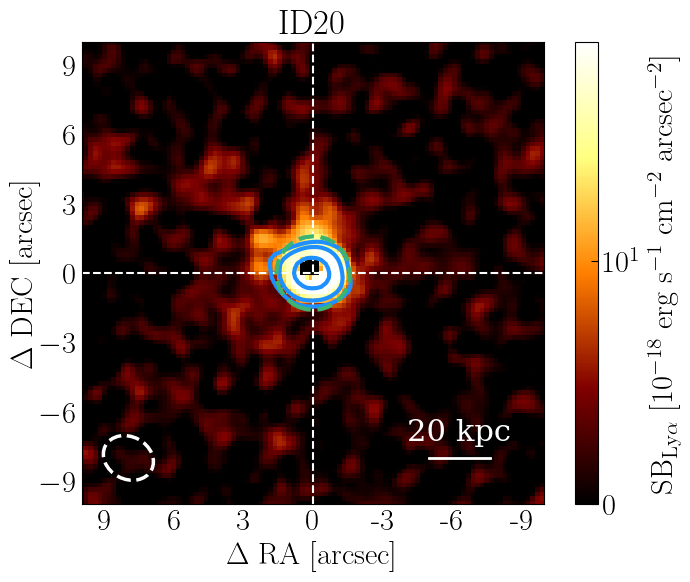}
    \end{minipage}

    \begin{minipage}{\linewidth}
        \centering
        \includegraphics[width=0.3\linewidth]{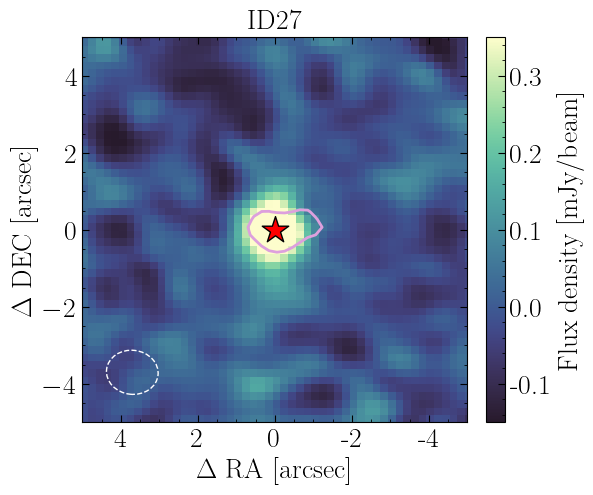}
        \includegraphics[width=0.3\linewidth]{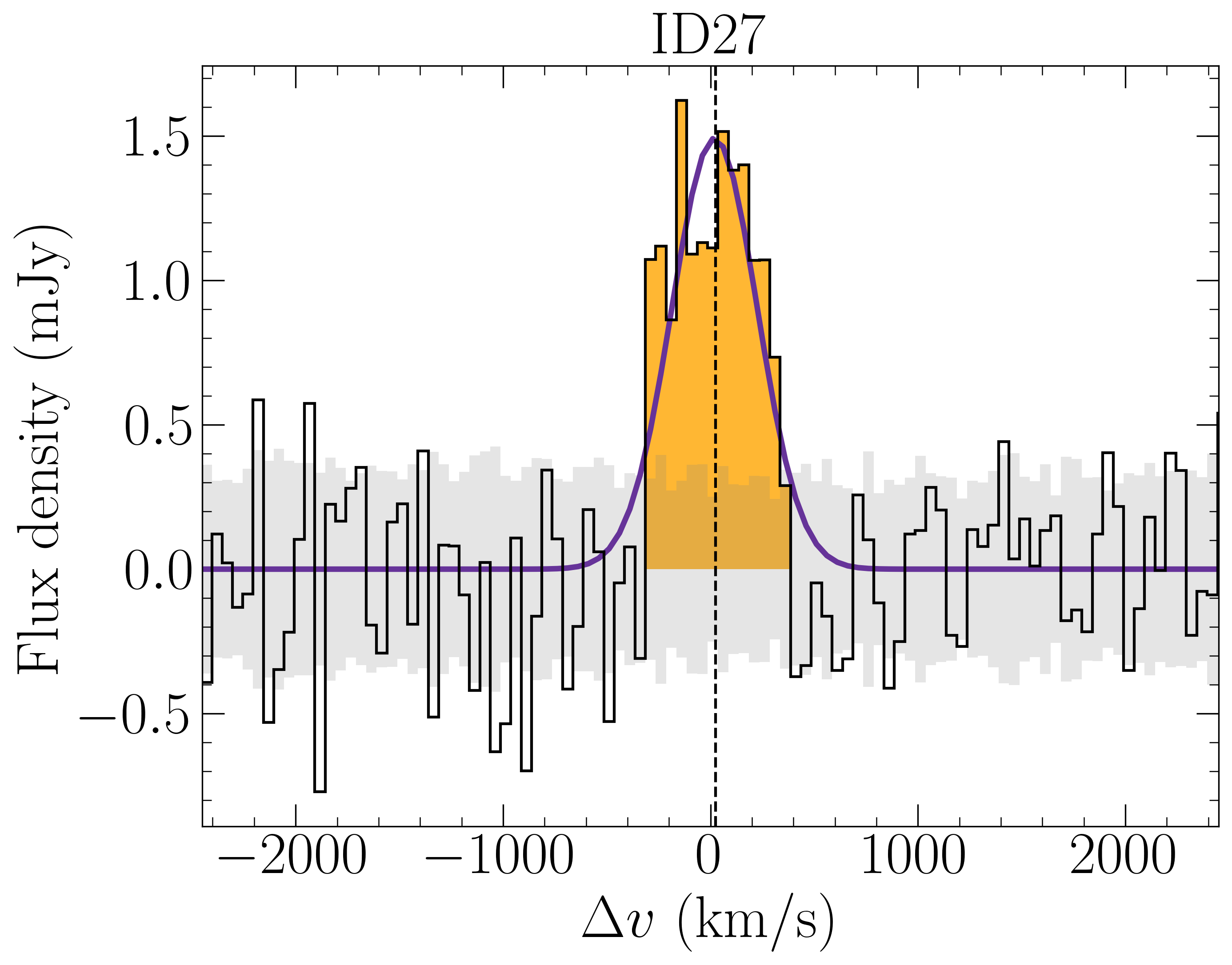}
        \includegraphics[width=0.3\linewidth]{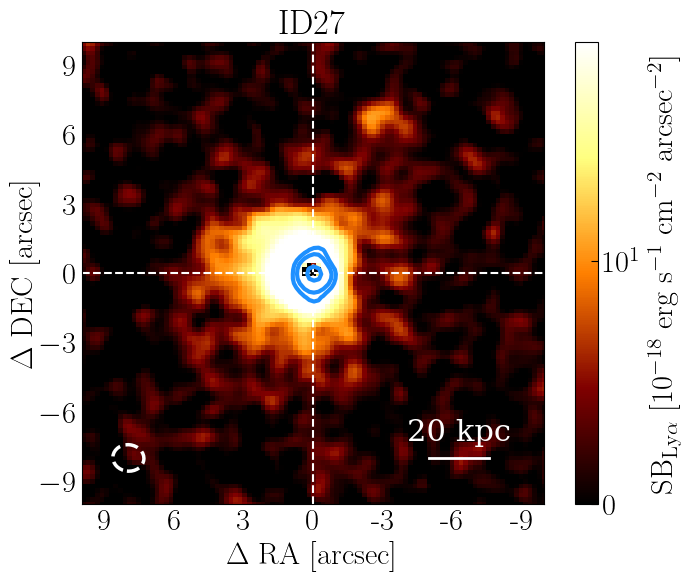}
    \end{minipage}

    \begin{minipage}{\linewidth}
        \centering
        \includegraphics[width=0.3\linewidth]{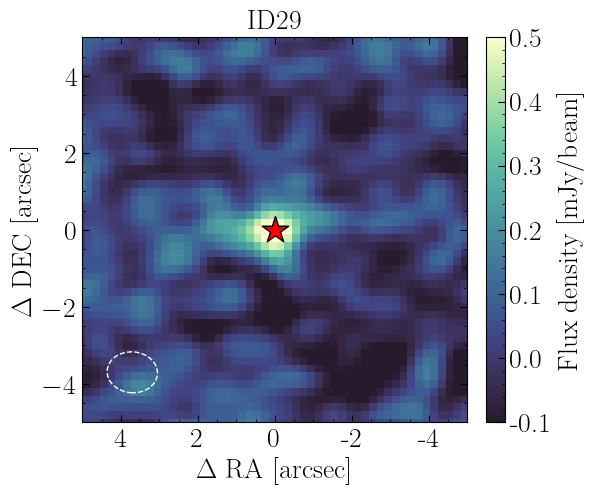}
        \includegraphics[width=0.3\linewidth]{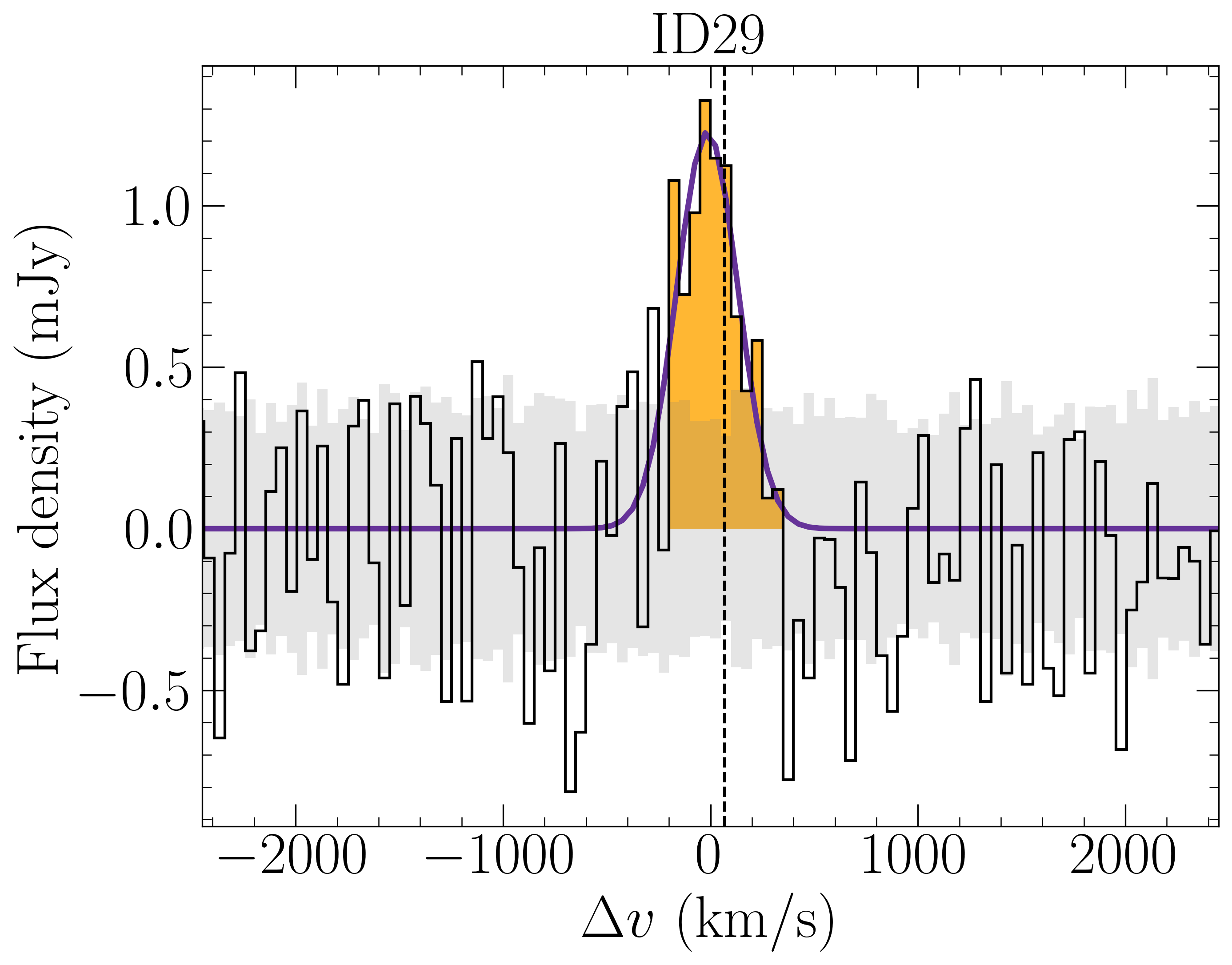}
        \includegraphics[width=0.3\linewidth]{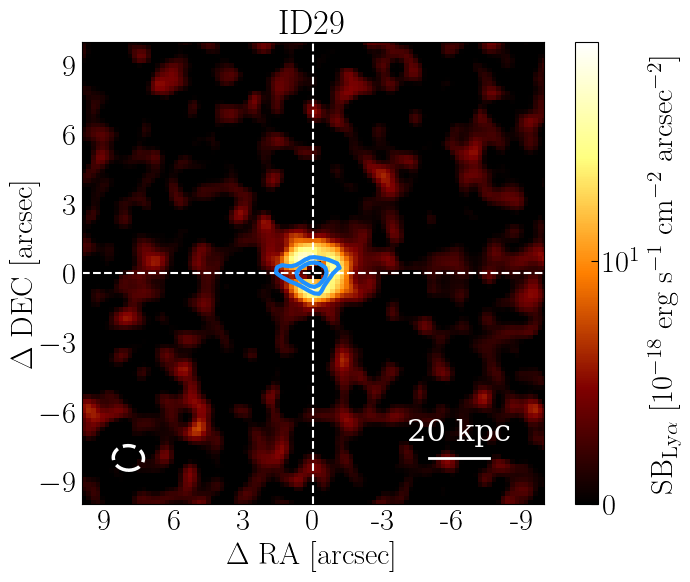}
    \end{minipage}

    \begin{minipage}{\linewidth}
        \centering
        \includegraphics[width=0.3\linewidth]{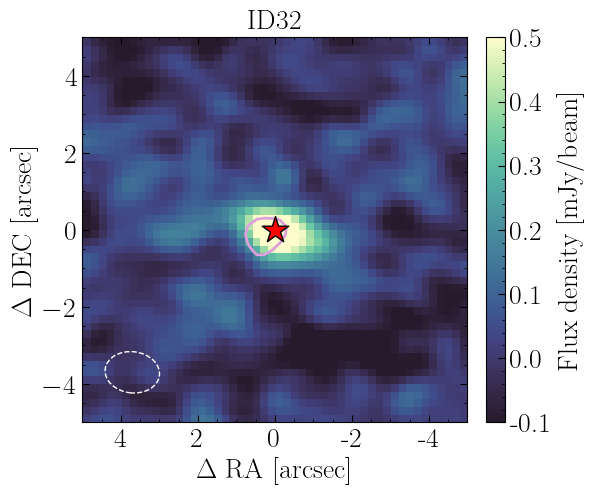}
        \includegraphics[width=0.3\linewidth]{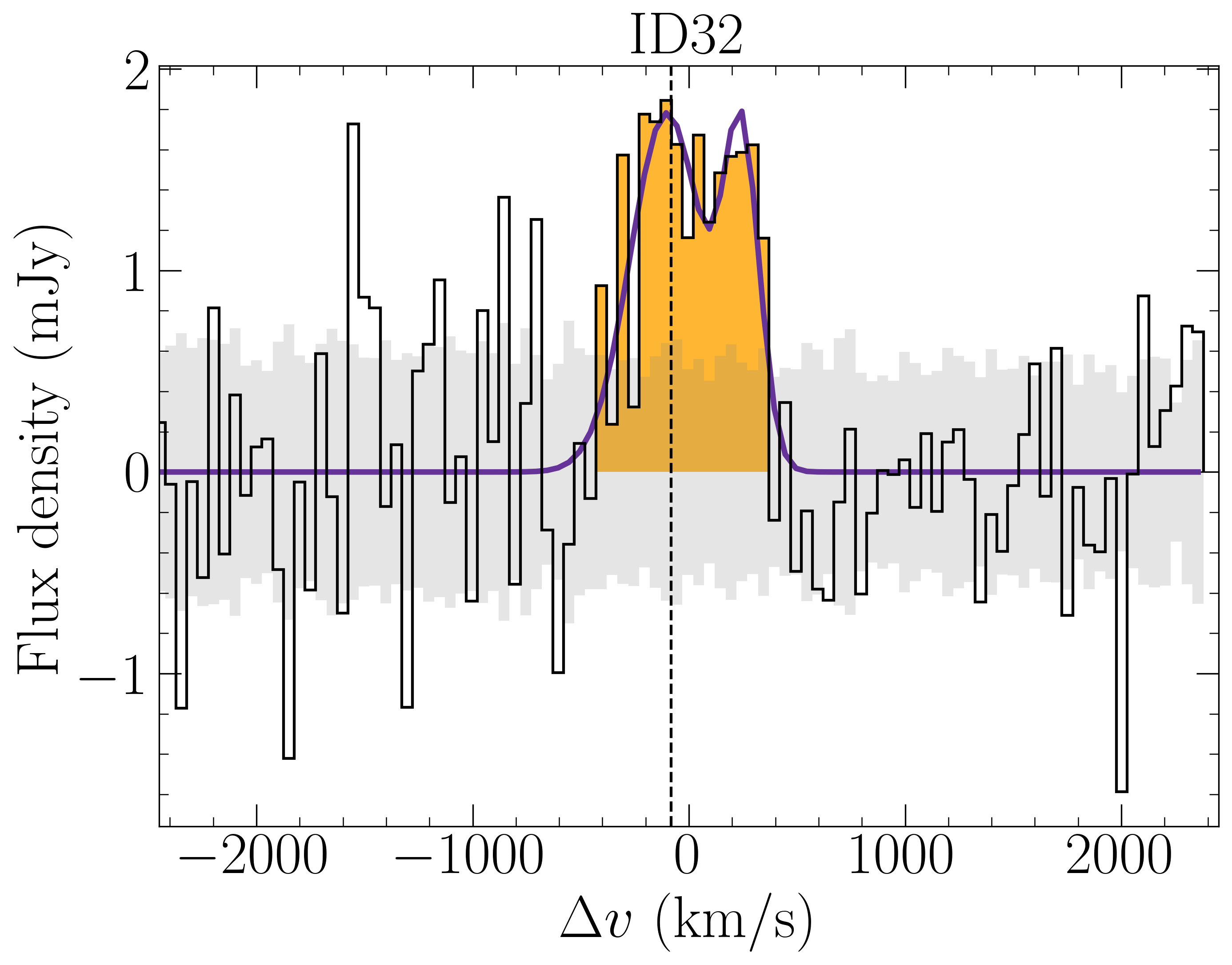}
        \includegraphics[width=0.3\linewidth]{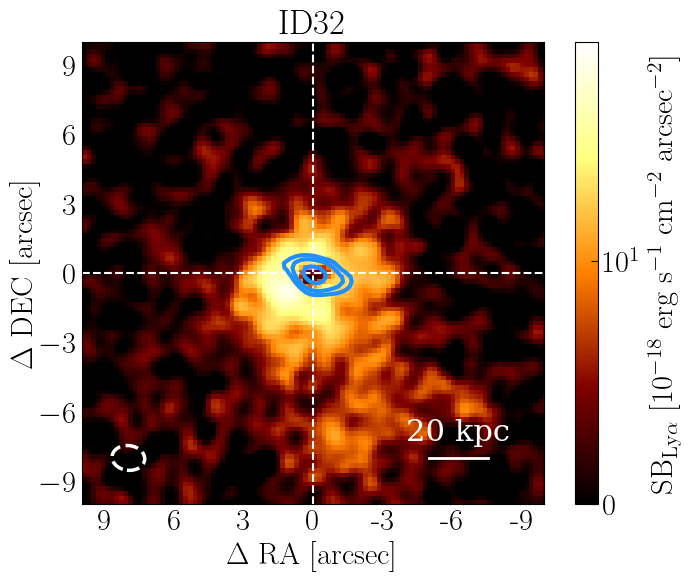}
    \end{minipage}

    \begin{minipage}{\linewidth}
        \centering
        \includegraphics[width=0.3\linewidth]{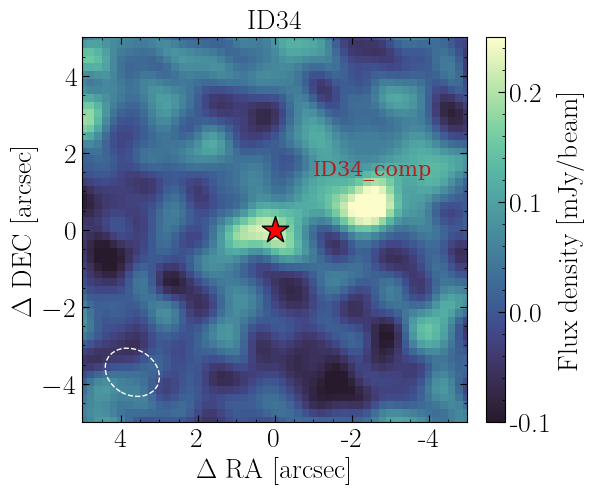}
        \includegraphics[width=0.3\linewidth]{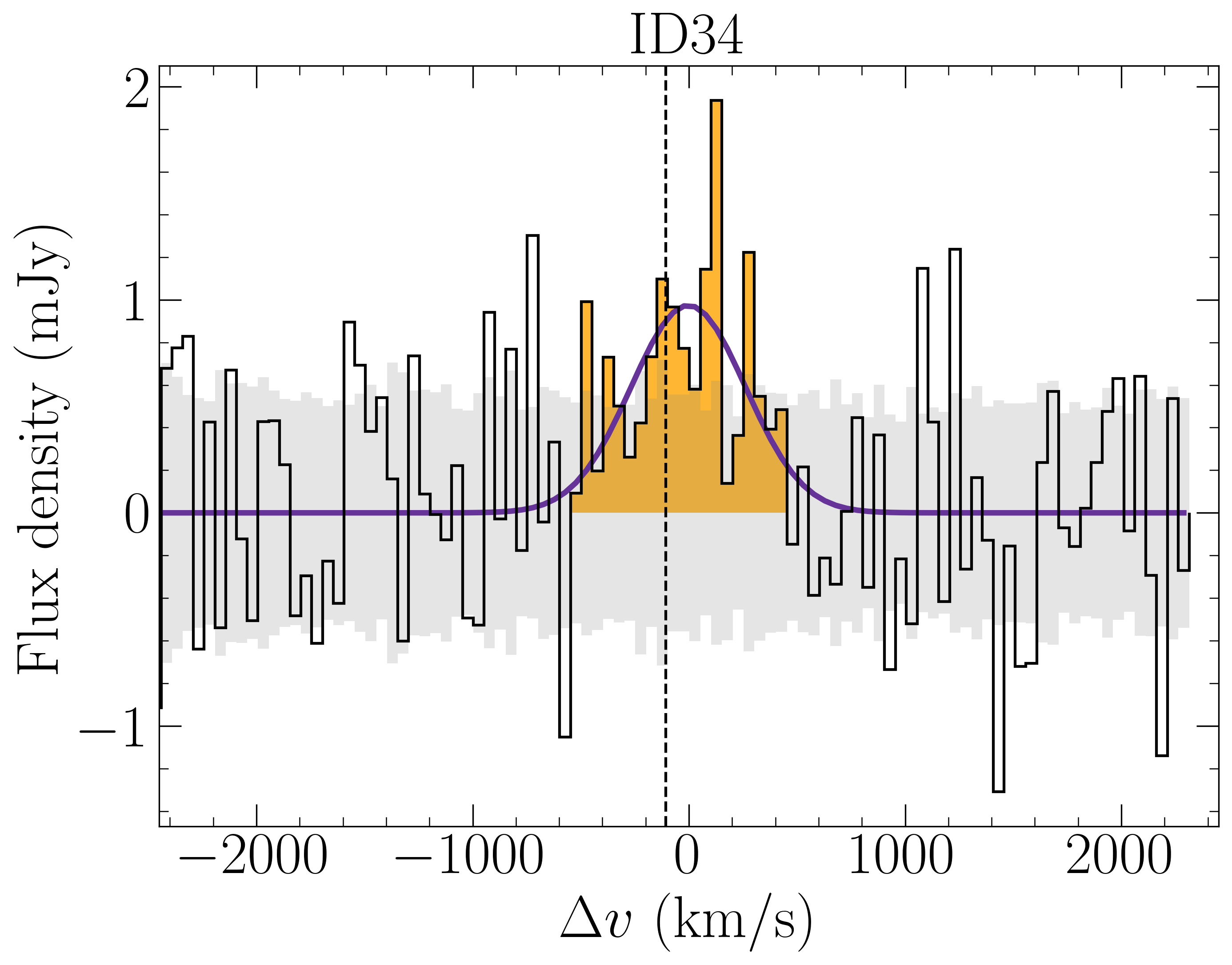}
        \includegraphics[width=0.3\linewidth]{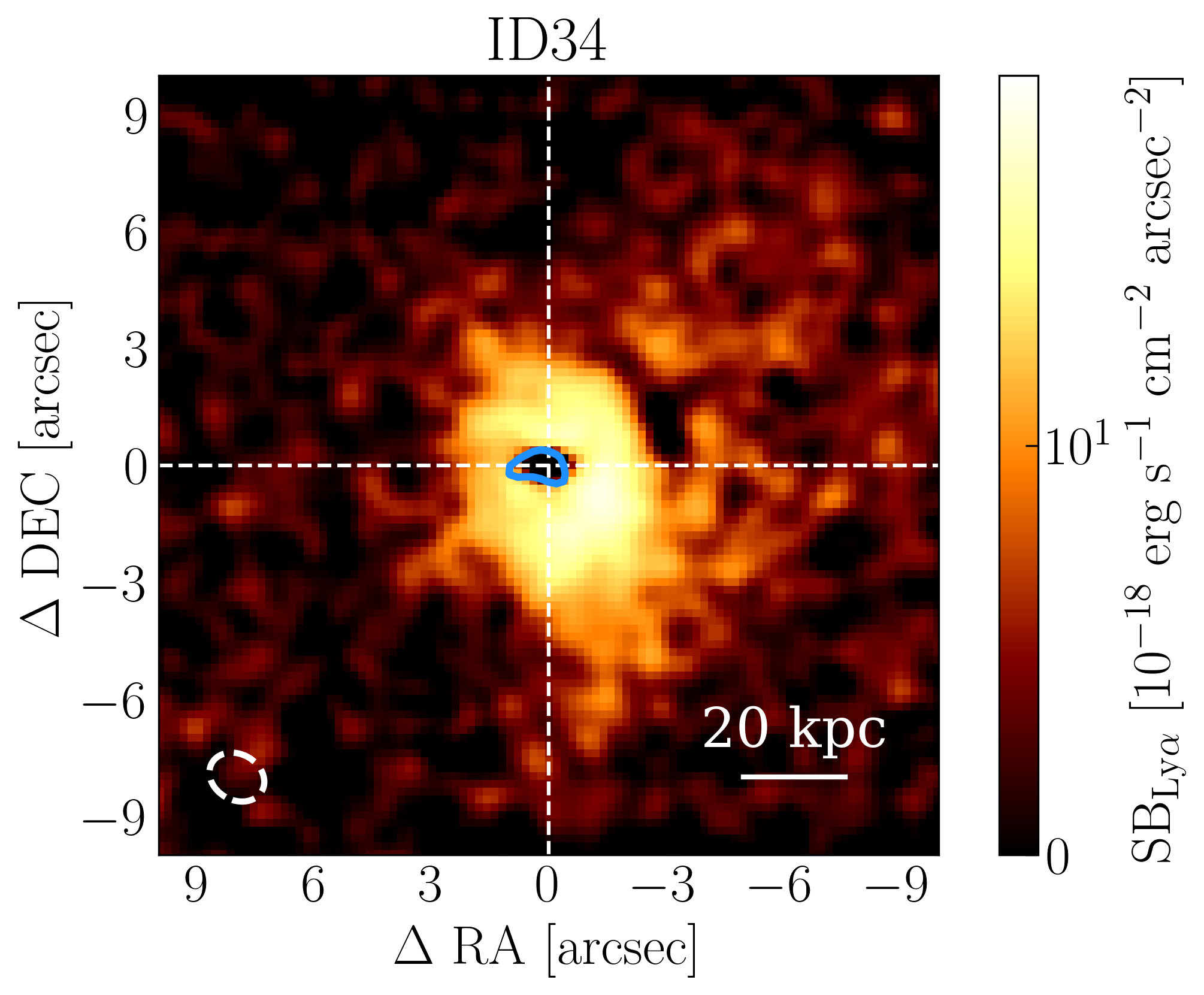}
    \end{minipage}

    \caption{continued.}
    \label{fig:flux_maps_spectra_3}
\end{figure*}

\begin{figure*}[htbp]
    \addtocounter{figure}{-1}
    \centering

    \begin{minipage}{\linewidth}
        \centering
        \includegraphics[width=0.3\linewidth]{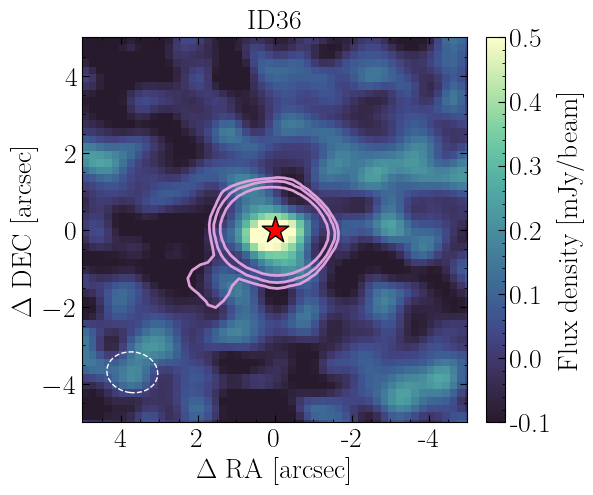}
        \includegraphics[width=0.3\linewidth]{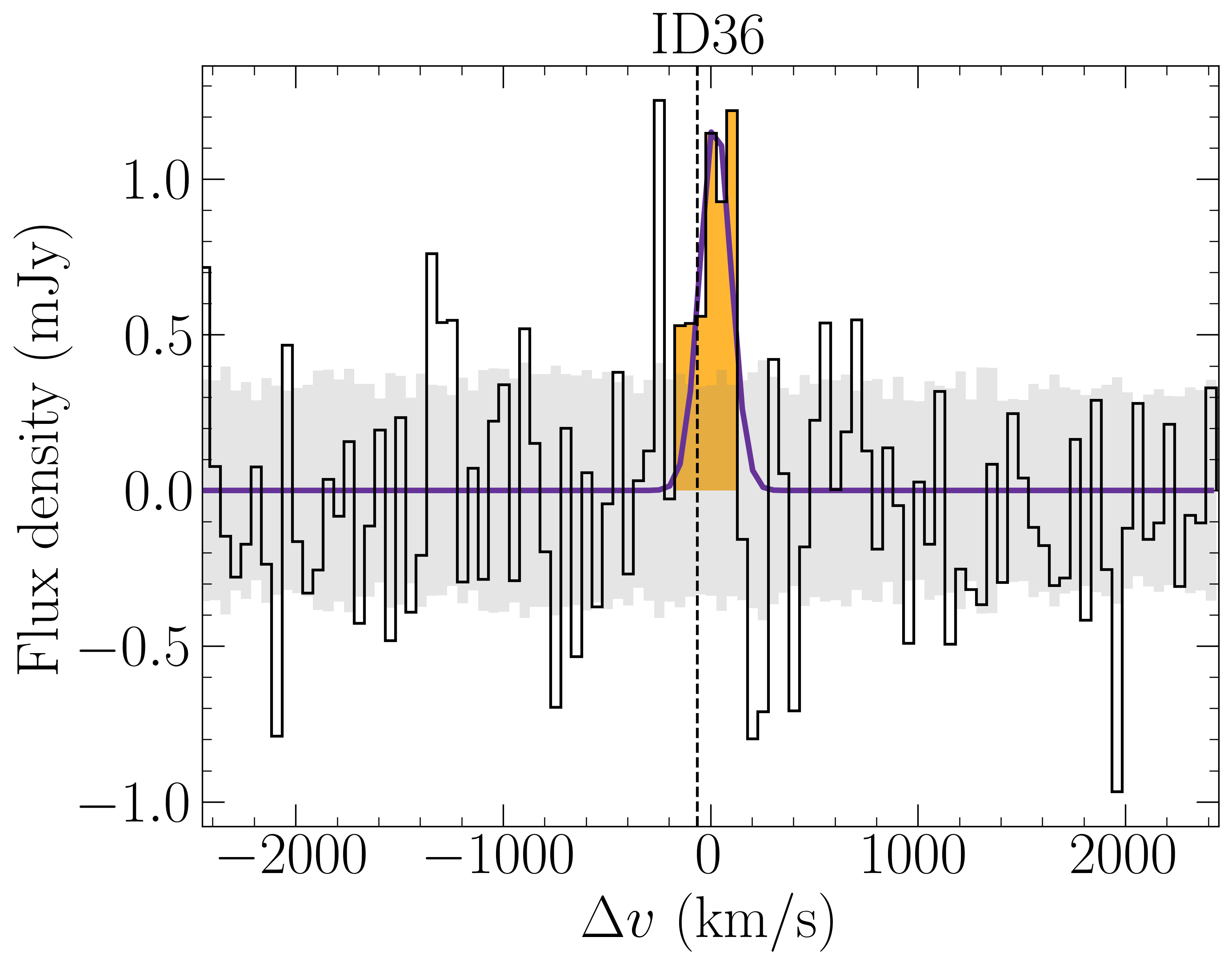}
        \includegraphics[width=0.3\linewidth]{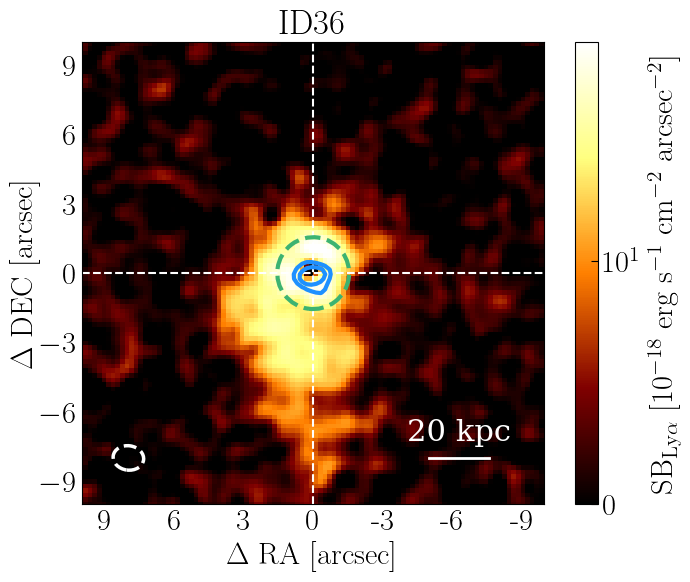}
    \end{minipage}

    \begin{minipage}{\linewidth}
        \centering
        \includegraphics[width=0.3\linewidth]{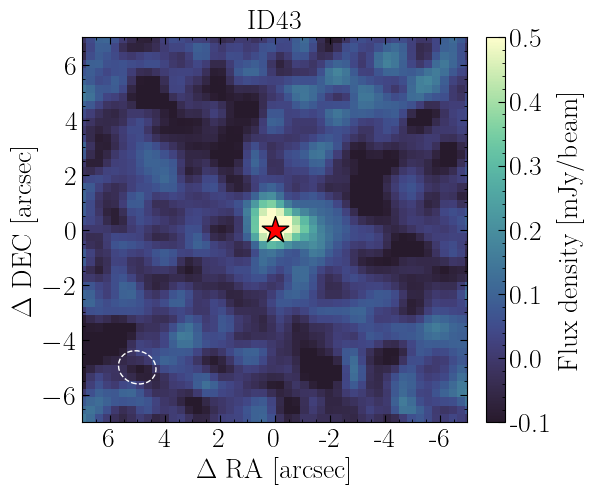}
        \includegraphics[width=0.3\linewidth]{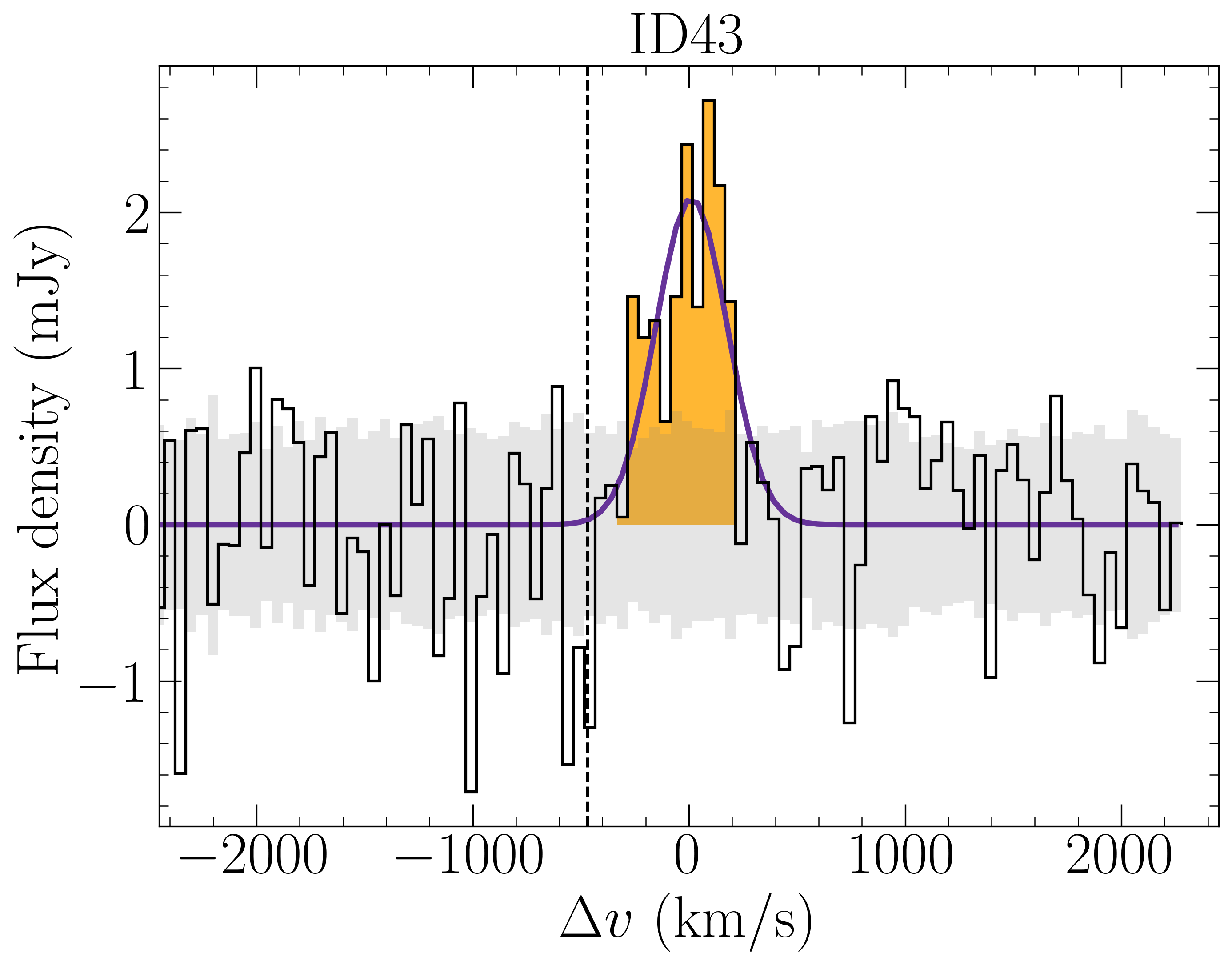}
        \includegraphics[width=0.3\linewidth]{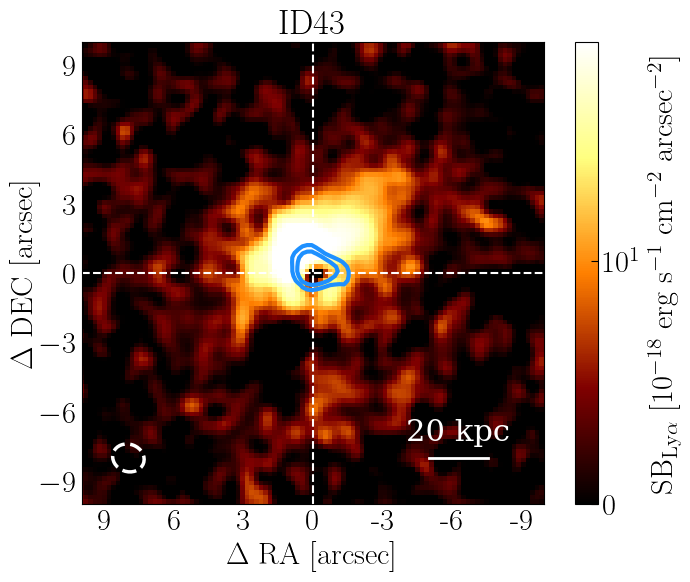}
    \end{minipage}

    \begin{minipage}{\linewidth}
        \centering
        \includegraphics[width=0.3\linewidth]{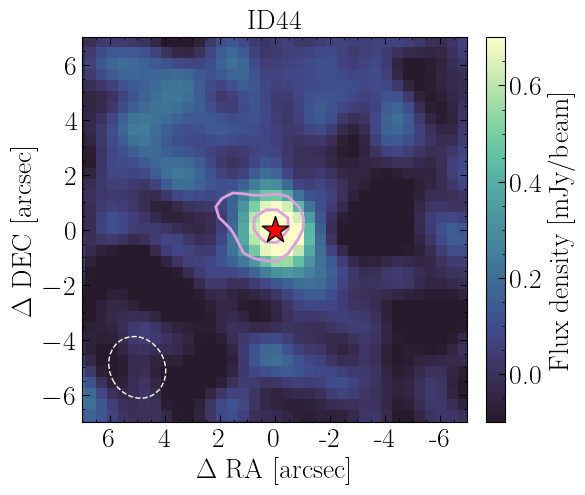}
        \includegraphics[width=0.3\linewidth]{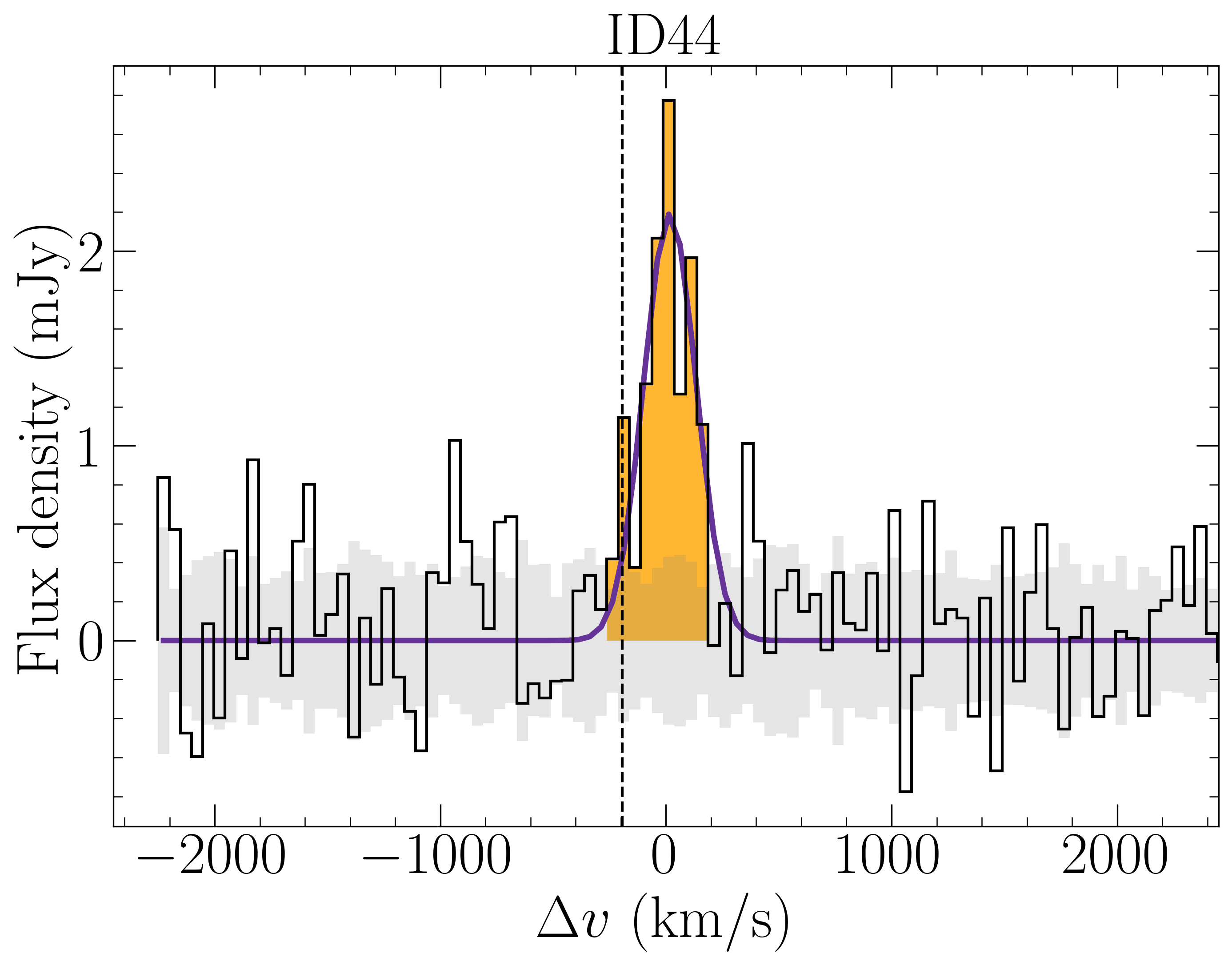}
        \includegraphics[width=0.3\linewidth]{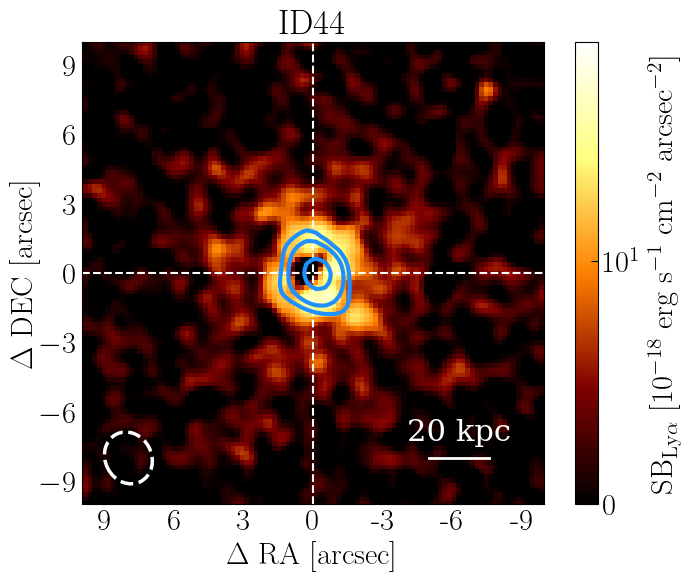}
    \end{minipage}

    \begin{minipage}{\linewidth}
        \centering
        \includegraphics[width=0.3\linewidth]{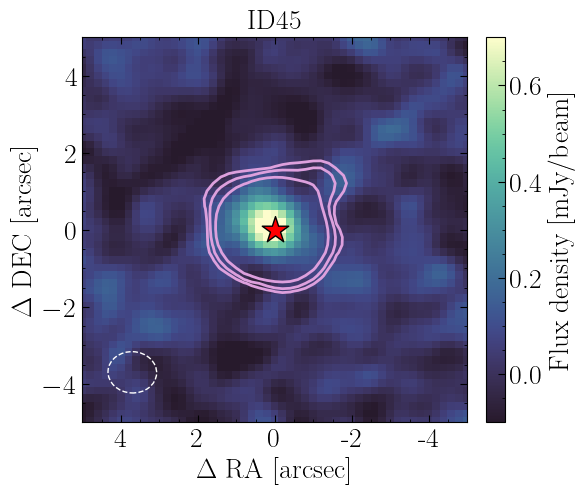}
        \includegraphics[width=0.3\linewidth]{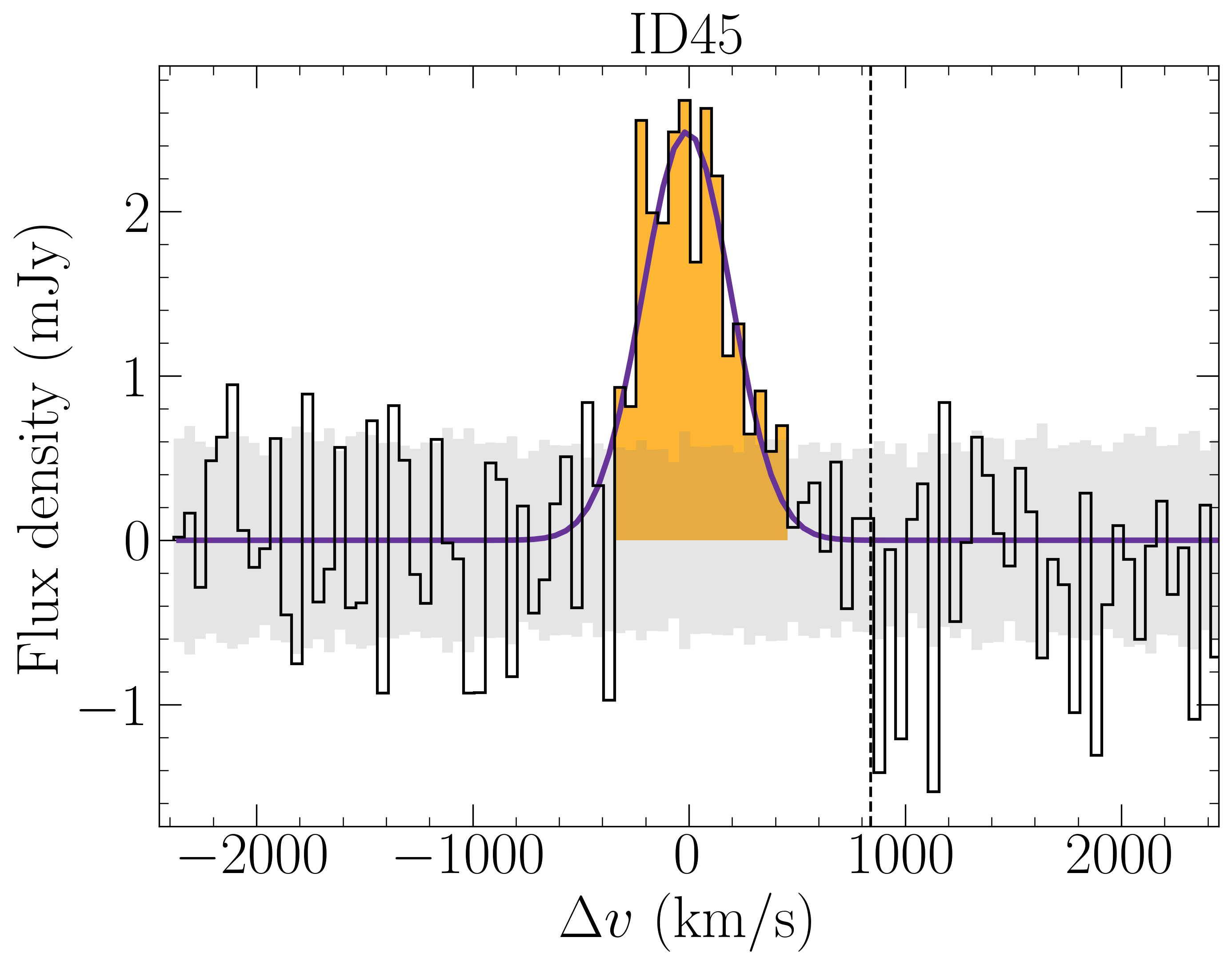}
        \includegraphics[width=0.3\linewidth]{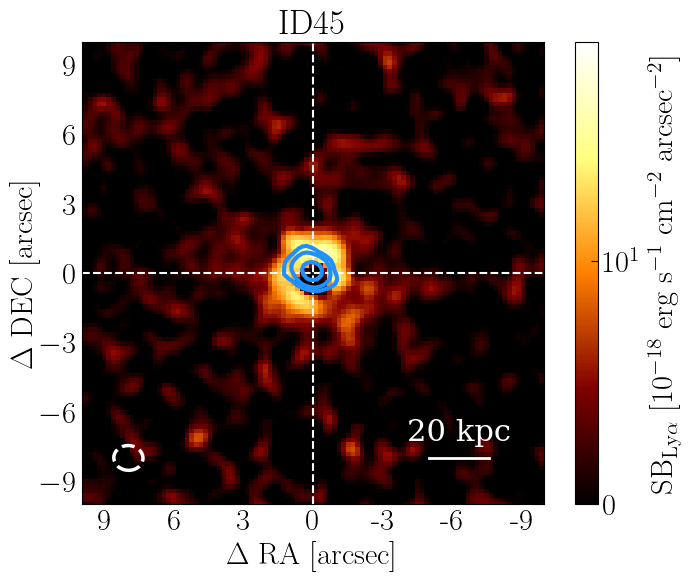}
    \end{minipage}

    \begin{minipage}{\linewidth}
        \centering
        \includegraphics[width=0.3\linewidth]{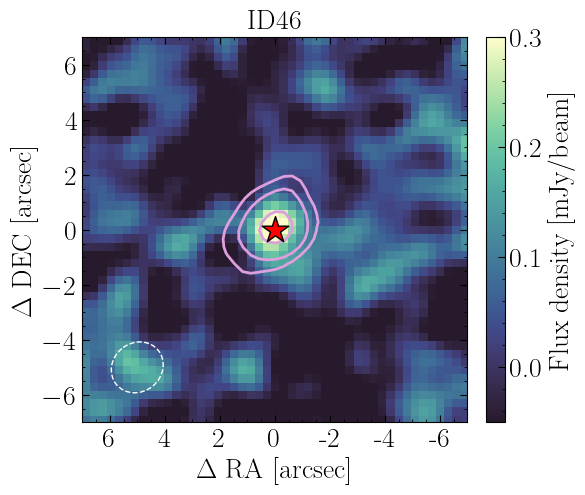}
        \includegraphics[width=0.3\linewidth]{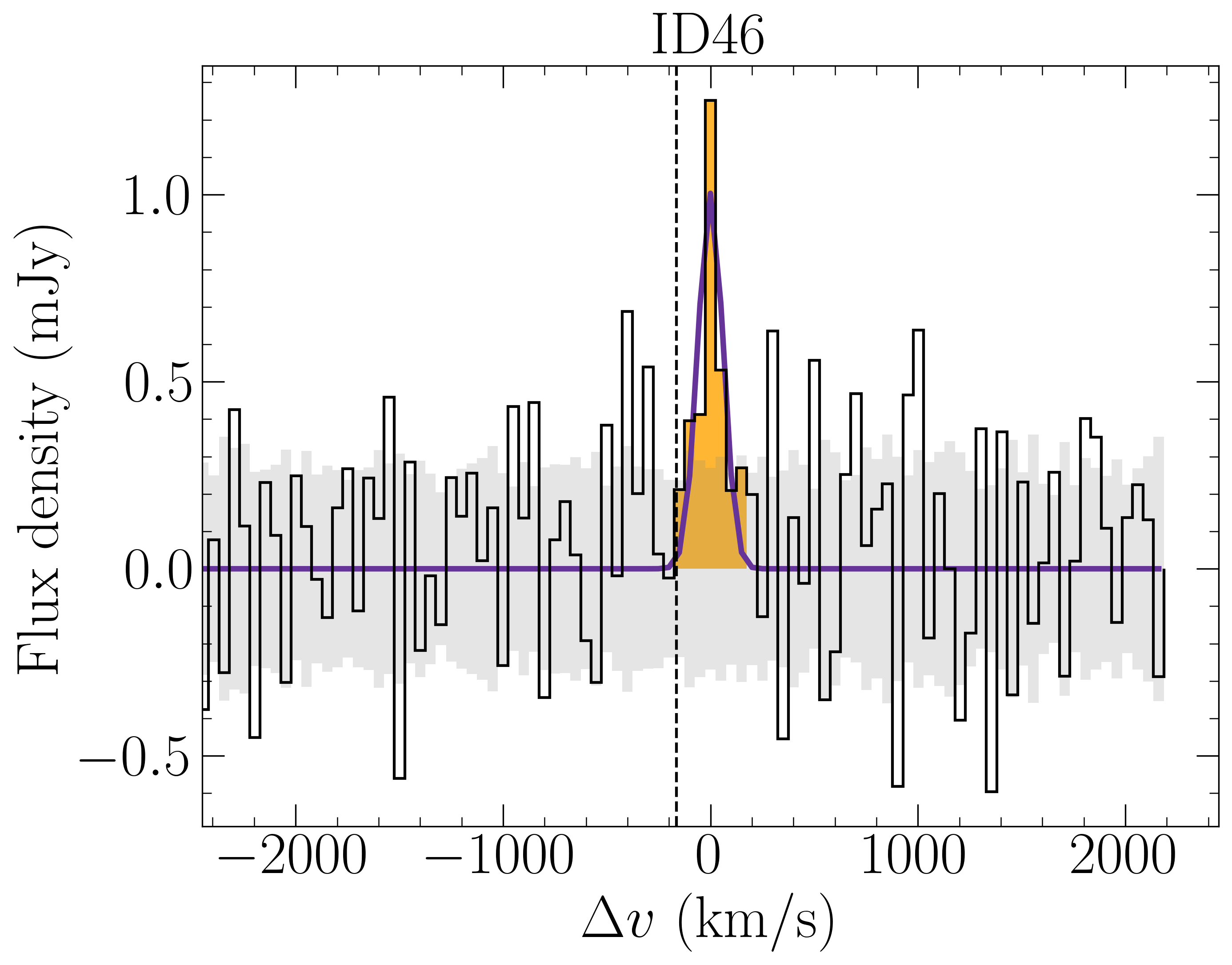}
        \includegraphics[width=0.3\linewidth]{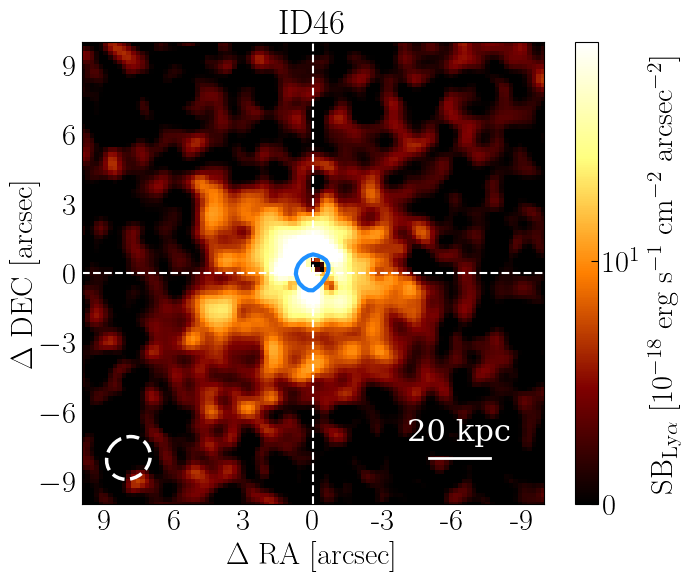}
    \end{minipage}

    \caption{continued.}
    \label{fig:flux_maps_spectra_4}
\end{figure*}

\begin{figure*}[htbp]
    \addtocounter{figure}{-1}
    \centering

    \begin{minipage}{\linewidth}
        \centering
        \includegraphics[width=0.3\linewidth]{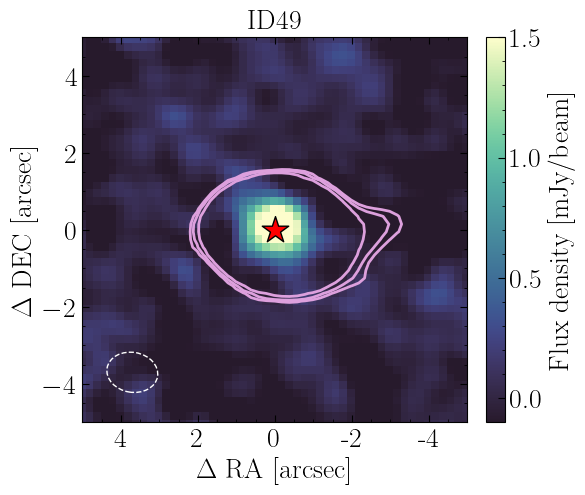}
        \includegraphics[width=0.3\linewidth]{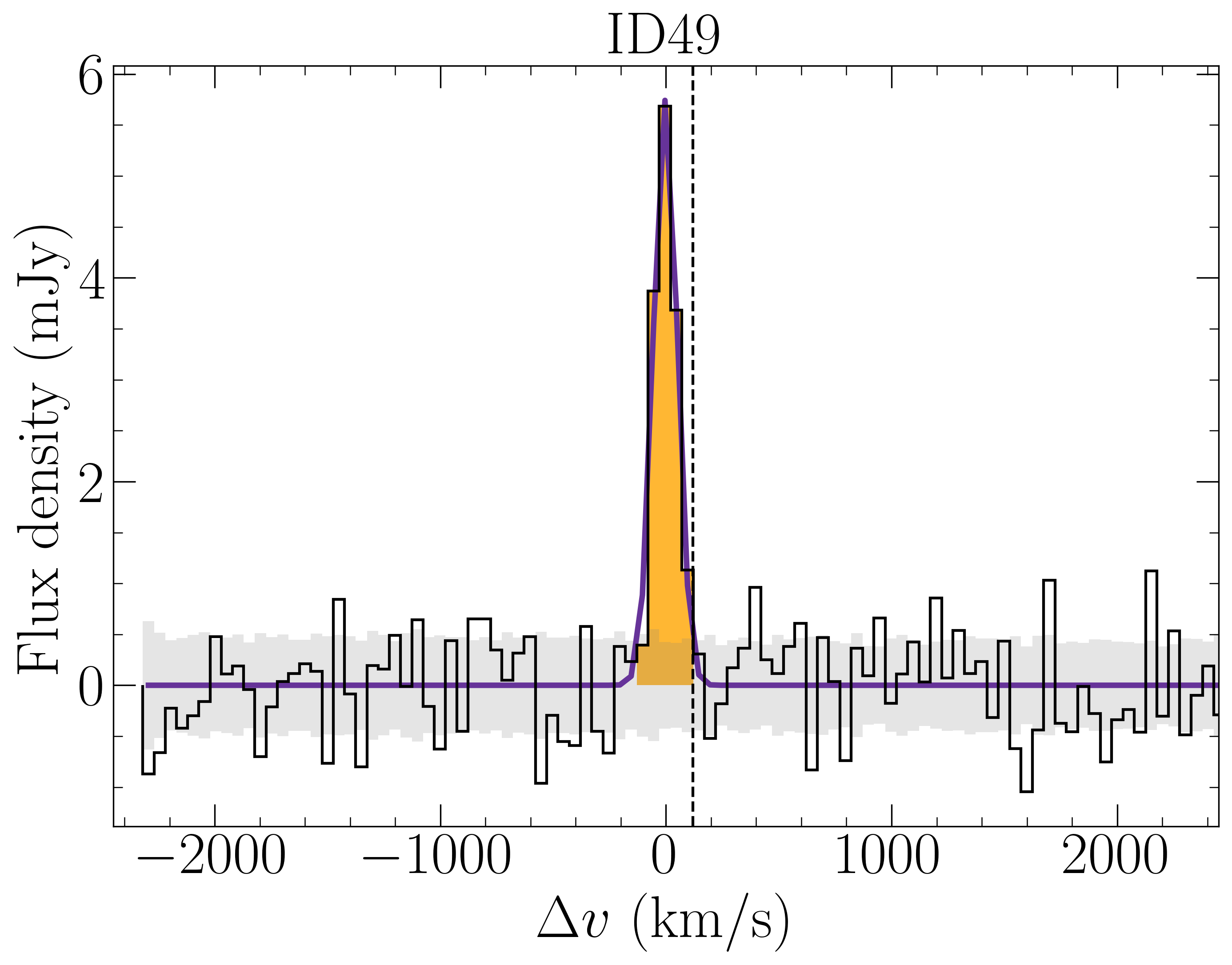}
        \includegraphics[width=0.3\linewidth]{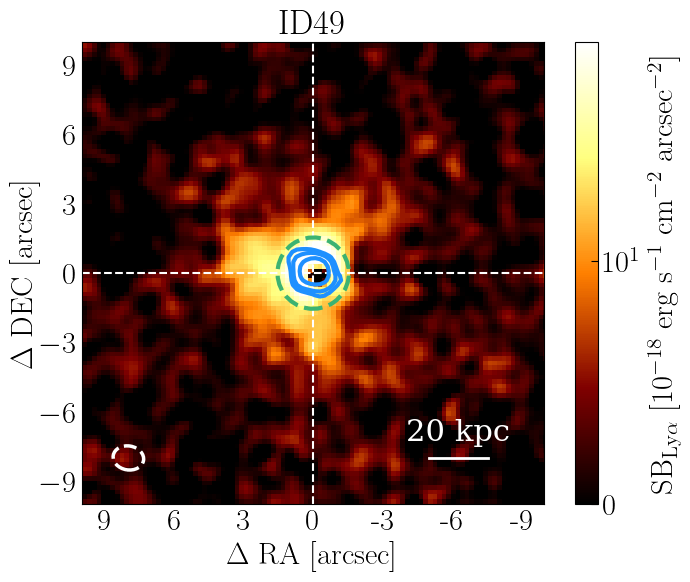}
    \end{minipage}

    \begin{minipage}{\linewidth}
        \centering
        \includegraphics[width=0.3\linewidth]{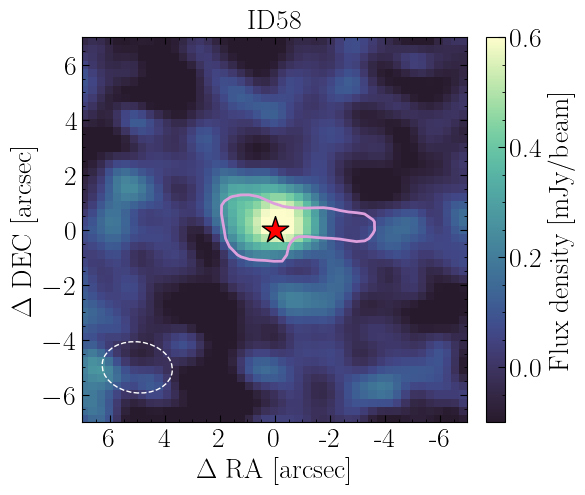}
        \includegraphics[width=0.3\linewidth]{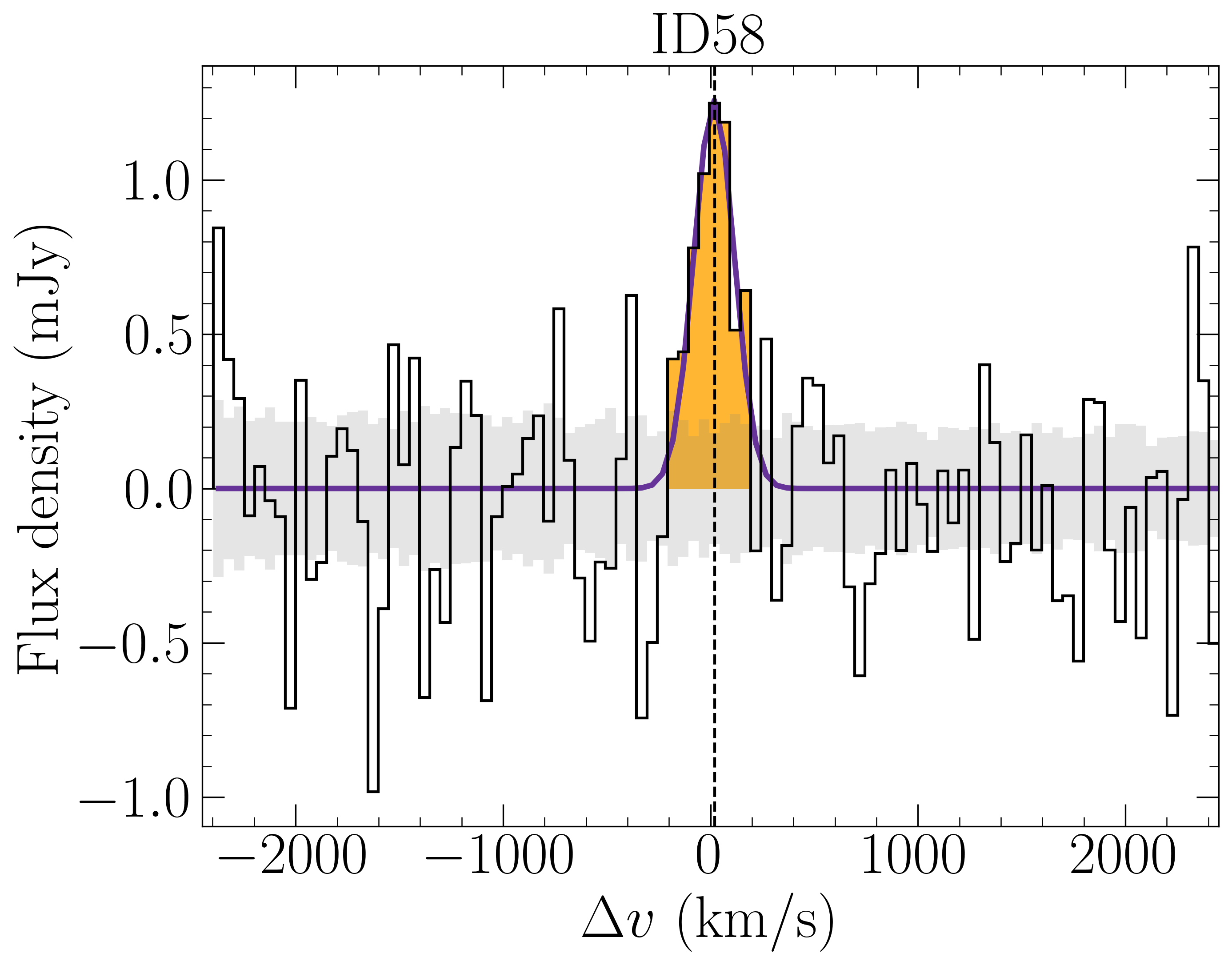}
        \includegraphics[width=0.3\linewidth]{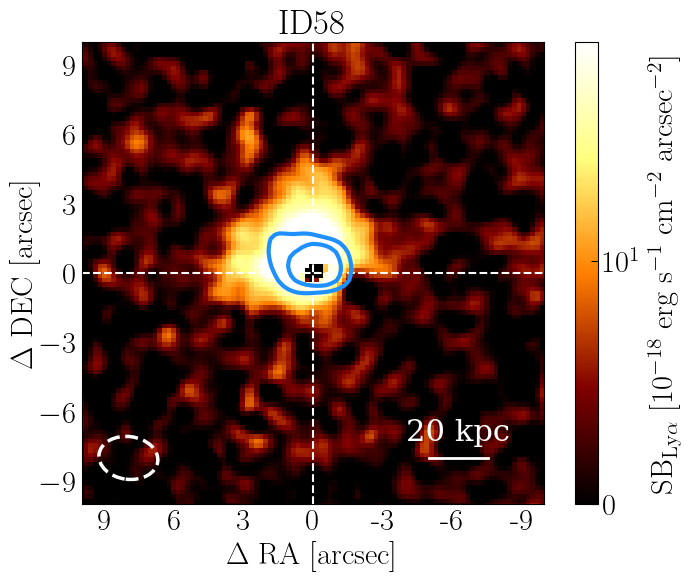}
    \end{minipage}

    \caption{continued.}
    \label{fig:flux_maps_spectra_5}
\end{figure*}

\begin{figure*}[htbp]
    \centering
    \includegraphics[width=0.28\linewidth]{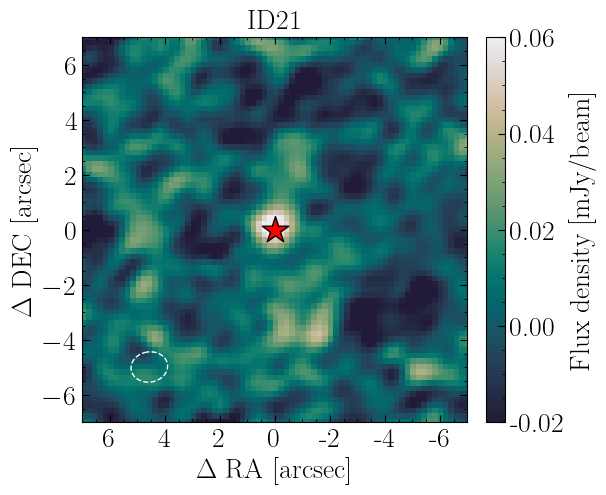}
    \includegraphics[width=0.28\linewidth]{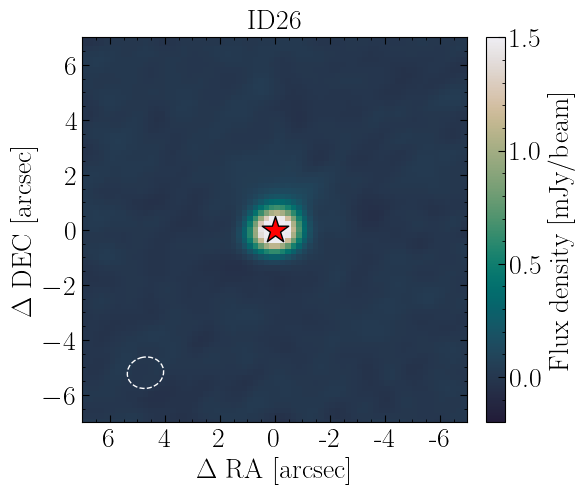}
    \includegraphics[width=0.28\linewidth]{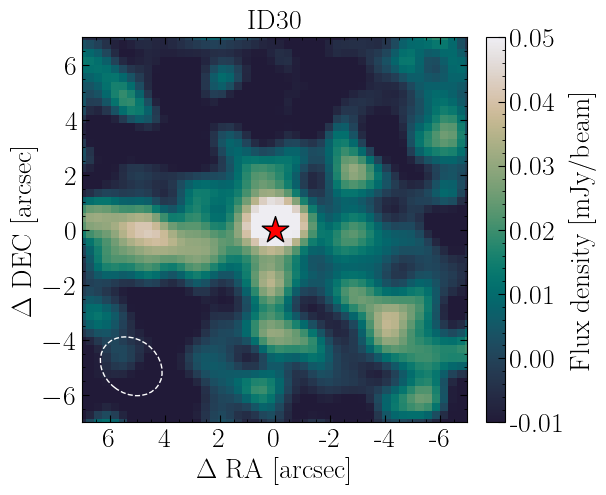}
    \includegraphics[width=0.28\linewidth]{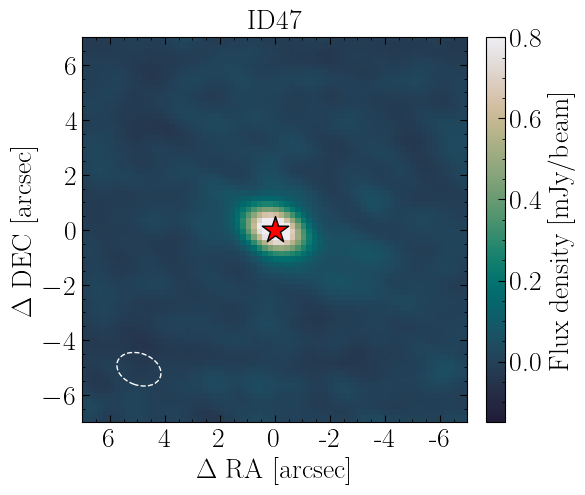}
    \includegraphics[width=0.28\linewidth]{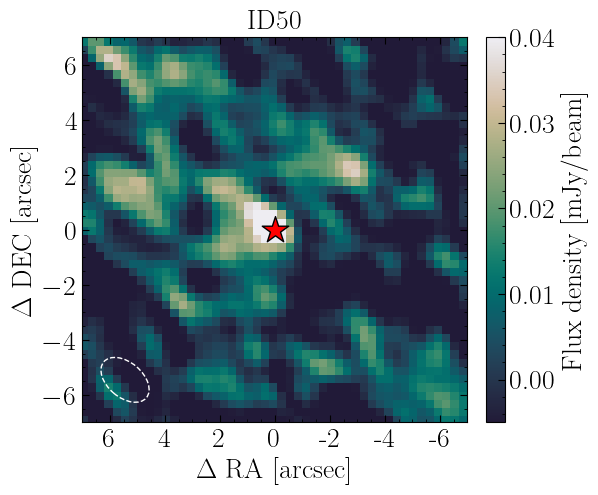}
    \includegraphics[width=0.28\linewidth]{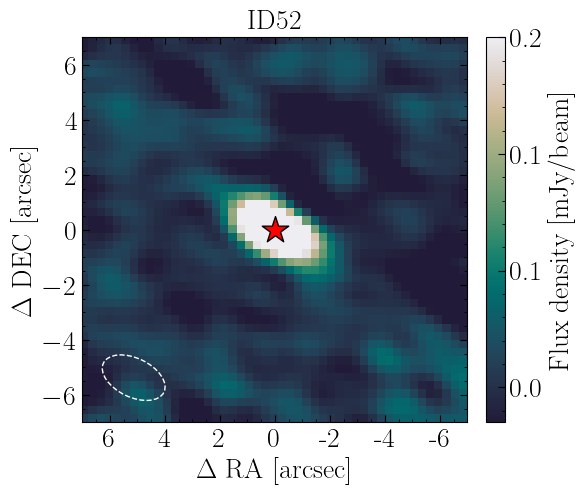}
    \includegraphics[width=0.28\linewidth]{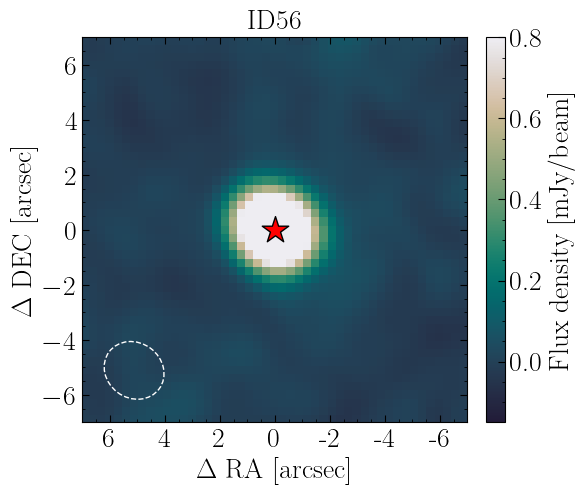}
    \caption{Continuum maps for the quasars without CO detection but continuum detection. Red stars mark the position of the quasar. }
    \label{fig:cont_maps}
\end{figure*}

\clearpage
\section{Companion galaxies} \label{sec:companion_galaxies}
\begin{figure*}[htbp]
    \centering

    \begin{minipage}{\linewidth}
        \centering
        \includegraphics[width=0.40\linewidth]{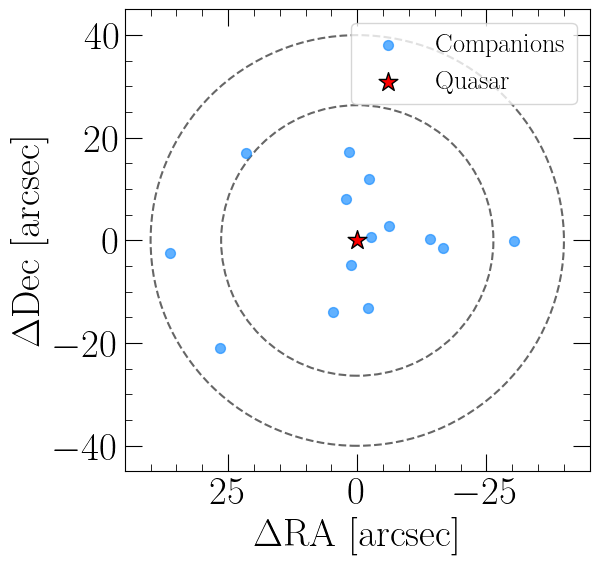}
        \includegraphics[width=0.50\linewidth]{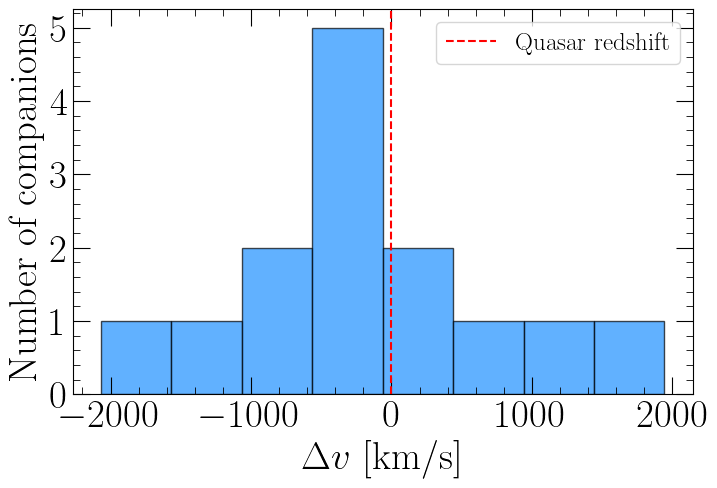}
    \end{minipage}

    \caption{\textit{Left:} Sky distribution of the potential CO(4-3) companion galaxies around the quasar for the 37 fields combined. The outer dashed circle corresponds to the radius where the sensitivity reaches 20\% of the maximum. The inner circle represents the theoretical ALMA primary beam, defined by the half-power beam width (HPBW), which is roughly $53''$ at $z\sim3.2$ (corresponding to a radius of 205\,kpc). \textit{Right:} The velocity offset distribution of the CO(4–3)-emitting companions relative to the central quasars. }
    \label{fig:companions_distr}
\end{figure*}

In this Appendix, we show the flux maps and spectra (Figure \ref{fig:flux_maps_spectra_companions}) of the CO-emitting companions detected in the 37 quasar fields, report their properties (Table \ref{tab:companion_properties}), and show their distribution on the sky and in line-of-sight velocity with respect to the quasar (Figure \ref{fig:companions_distr}).

\begin{figure*}[htbp]
    \centering

    \begin{minipage}{0.61\linewidth}
        \centering
        \includegraphics[width=0.46\linewidth]{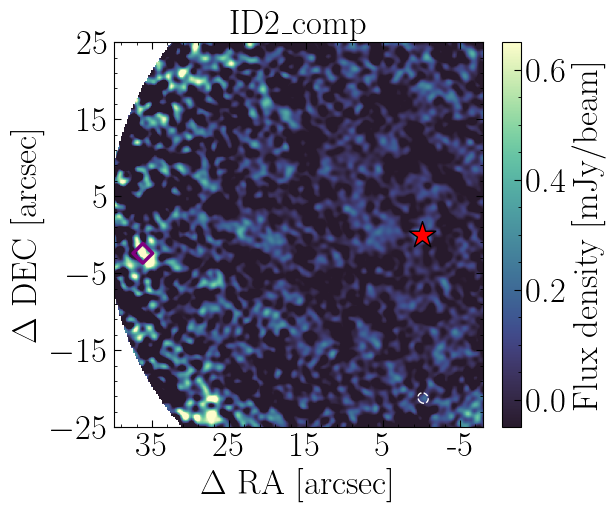}
        \includegraphics[width=0.50\linewidth]{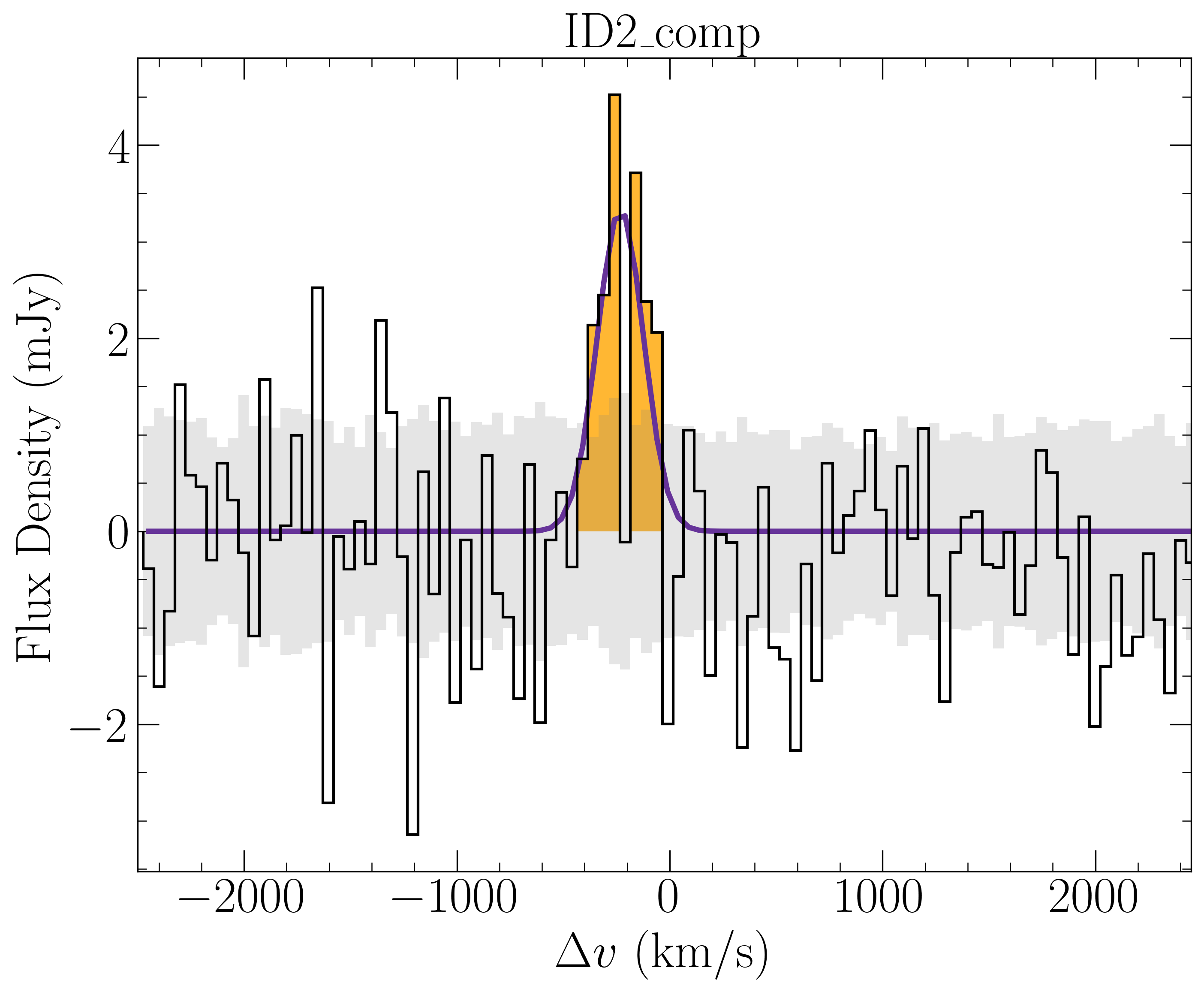}
    \end{minipage}

    \begin{minipage}{0.61\linewidth}
        \centering
        \includegraphics[width=0.46\linewidth]{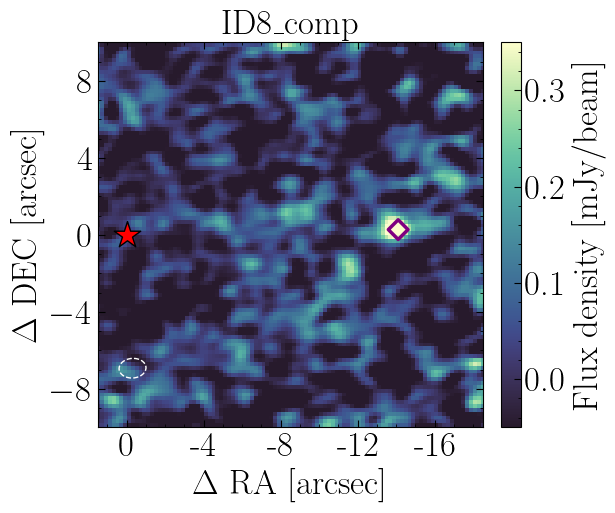}
        \includegraphics[width=0.50\linewidth]{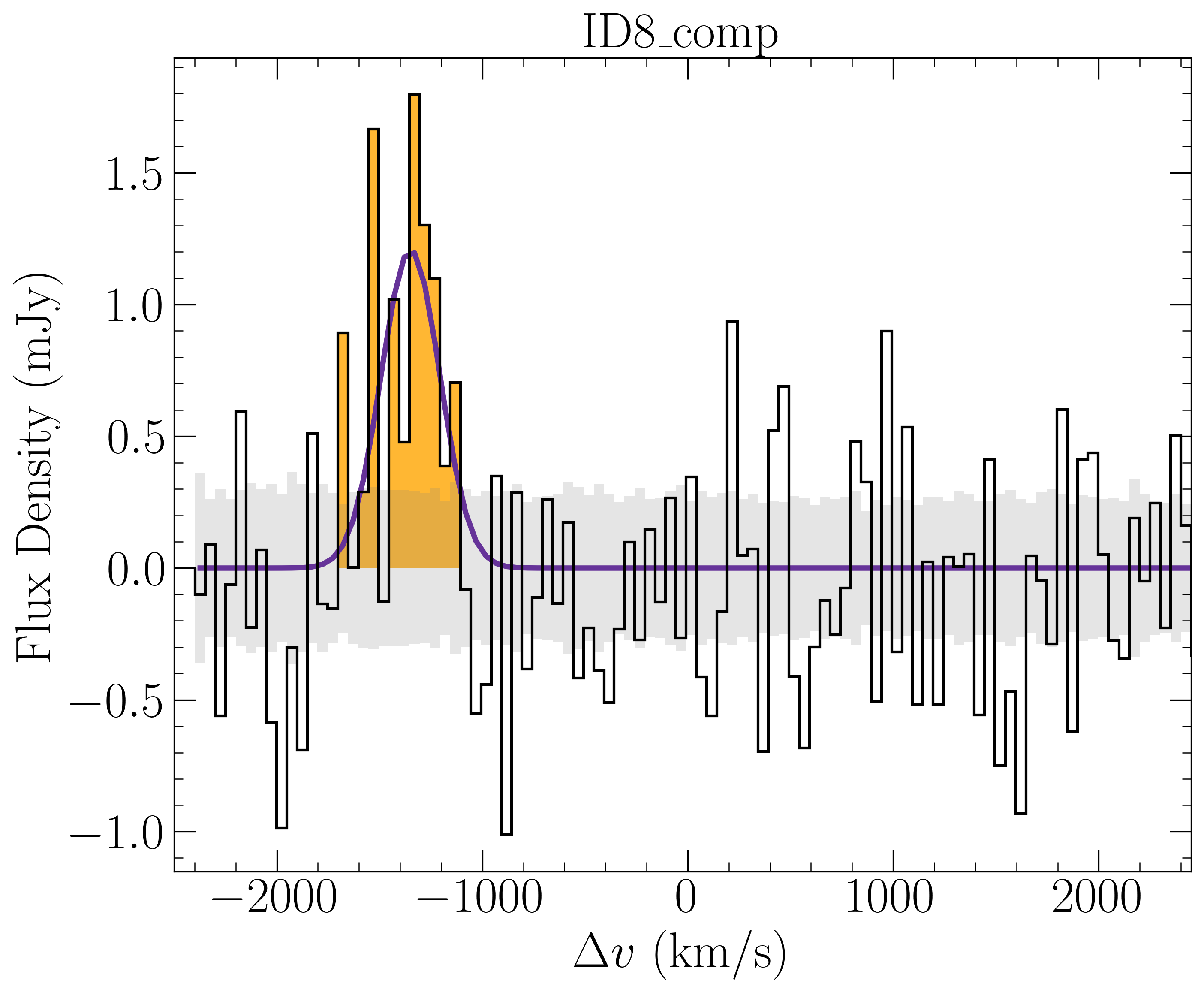}
    \end{minipage}

    \begin{minipage}{0.61\linewidth}
        \centering
        \includegraphics[width=0.46\linewidth]{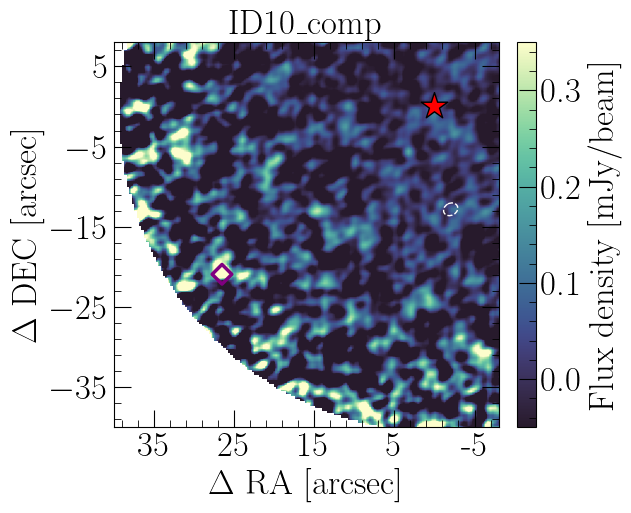}
        \includegraphics[width=0.50\linewidth]{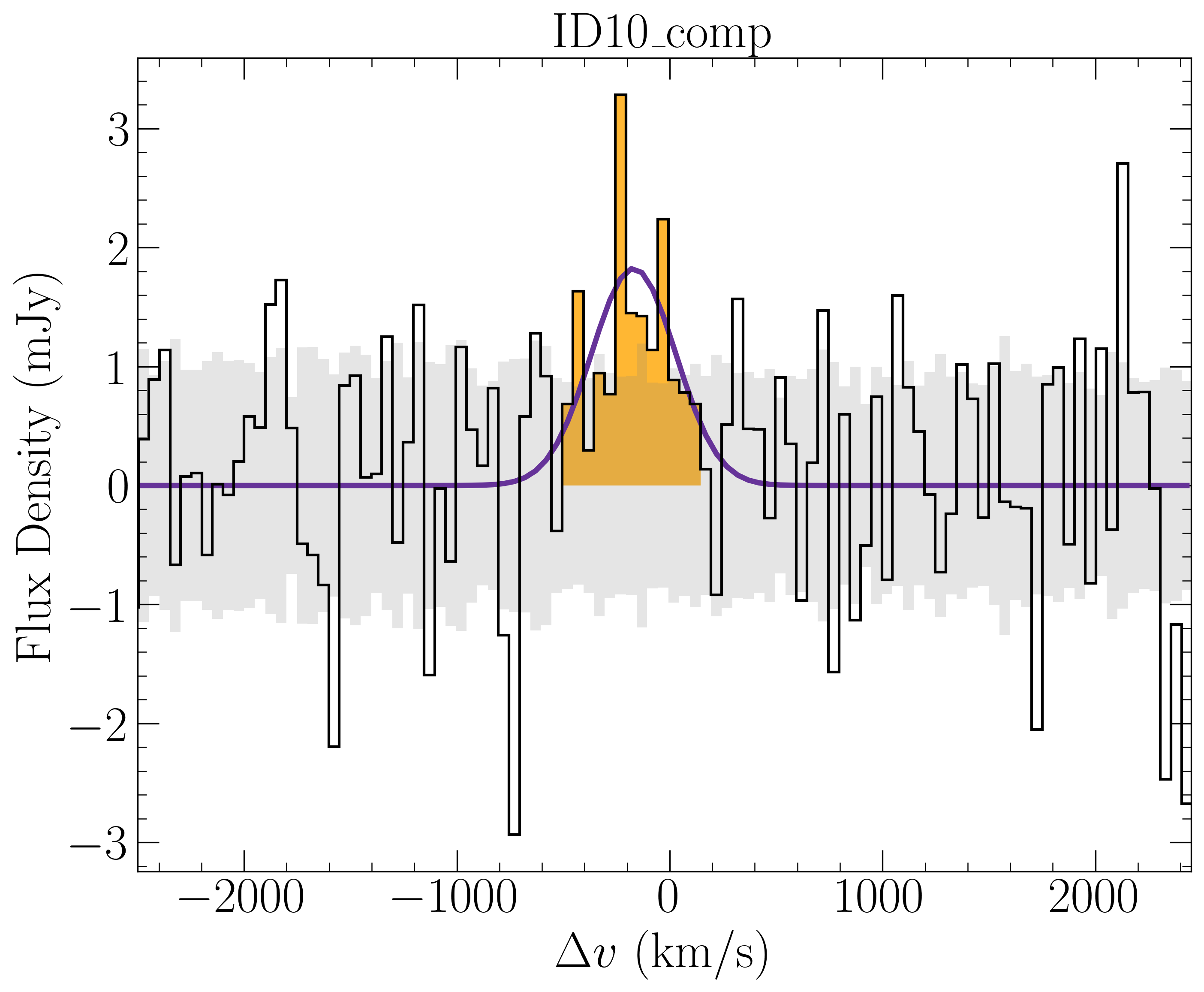}
    \end{minipage}

    \begin{minipage}{0.61\linewidth}
        \centering
        \includegraphics[width=0.46\linewidth]{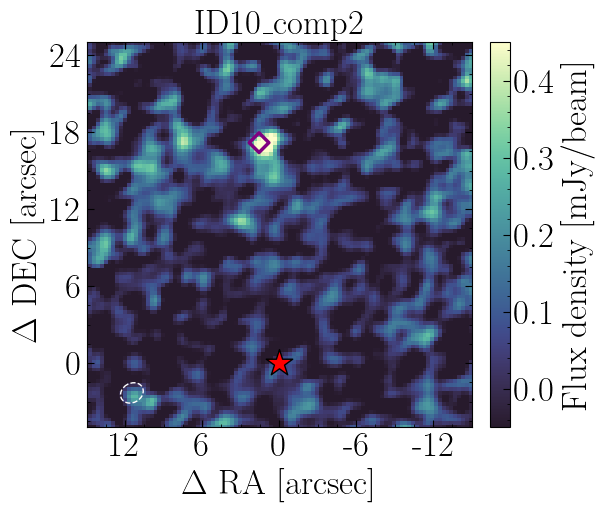}
        \includegraphics[width=0.50\linewidth]{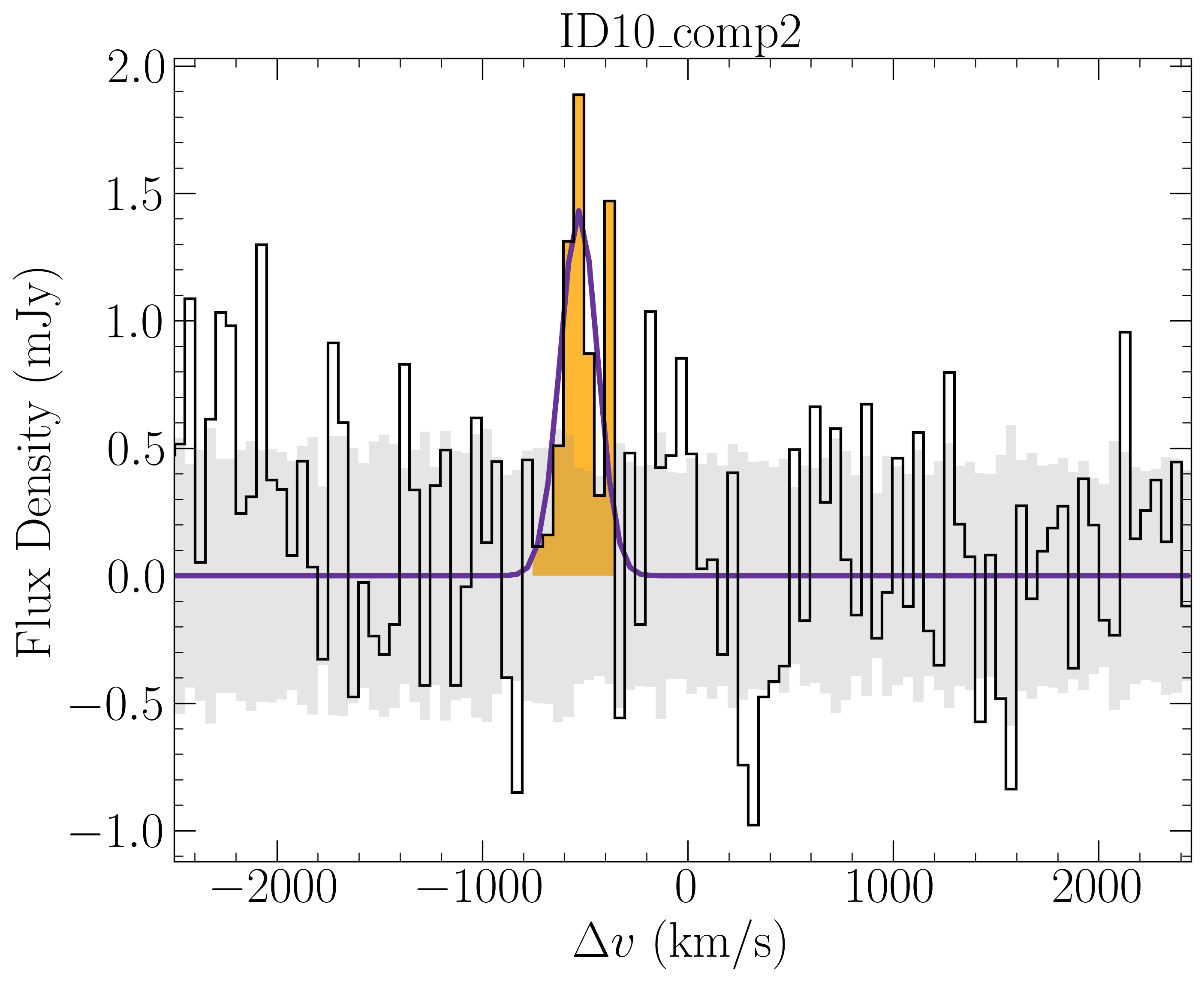}
    \end{minipage}

    \caption{\textit{Left panel}: Multifrequency synthesis (MFS) image of the CO(4-3) emission of the companion galaxies. Red stars mark the position of the quasars and purple diamonds the positions of the companion galaxies. Contours indicate the continuum emission (at 3, 5 and 10$\,\sigma$), plotted for those companions with continuum detection. The synthesized beam is shown as an ellipse. \textit{Right panel}: Spectrum of the CO(4-3) emission. $0\,\mathrm{km}\mathrm{s}^{-1}$ corresponds to the centroid of the CO(4-3) emission (determined from a first-moment analysis) of the central quasar, or if undetected in CO to the peak Ly$\alpha$ wavelength of the nebula \citep{GonzalezLobos2025}. The purple line indicates the Gaussian fit. The shaded orange regions highlight the line emissions.}
    \label{fig:flux_maps_spectra_companions}
\end{figure*}

\begin{figure*}[htbp]
    \addtocounter{figure}{-1}
    \centering
    \begin{minipage}{0.61\linewidth}
        \centering
        \includegraphics[width=0.46\linewidth]{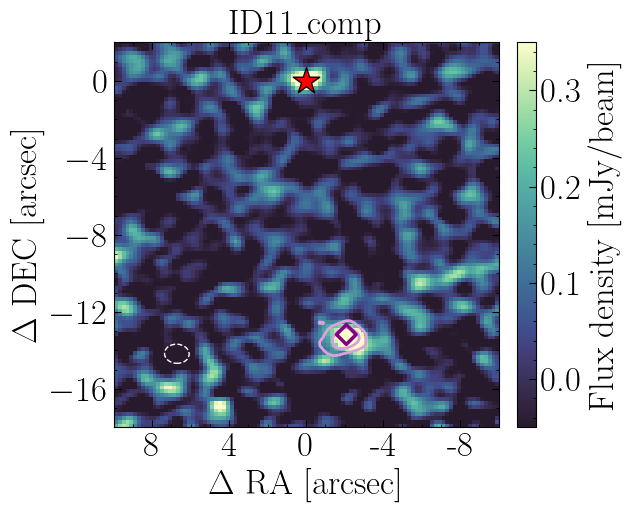}
        \includegraphics[width=0.50\linewidth]{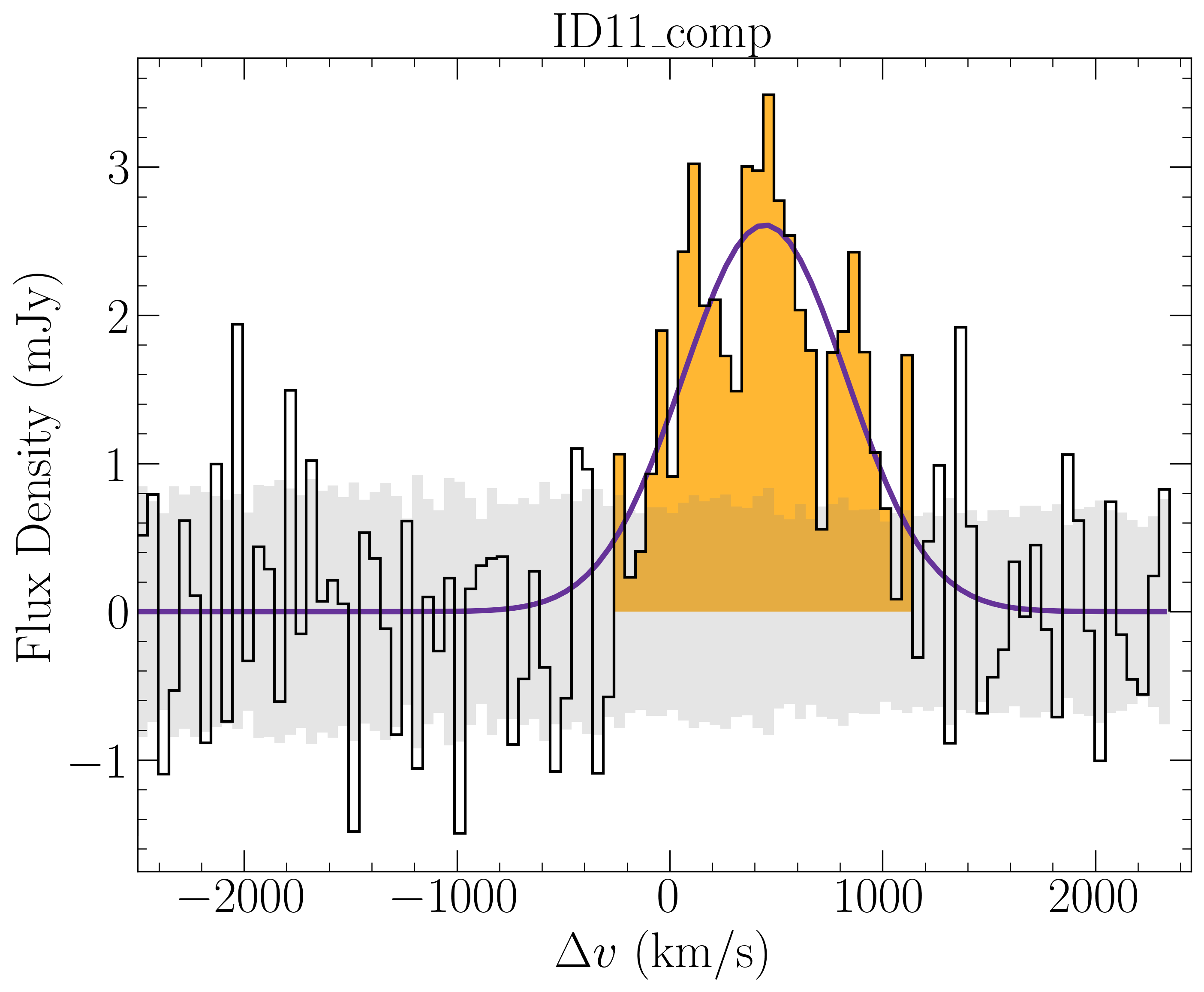}
    \end{minipage}

    \begin{minipage}{0.61\linewidth}
        \centering
        \includegraphics[width=0.46\linewidth]{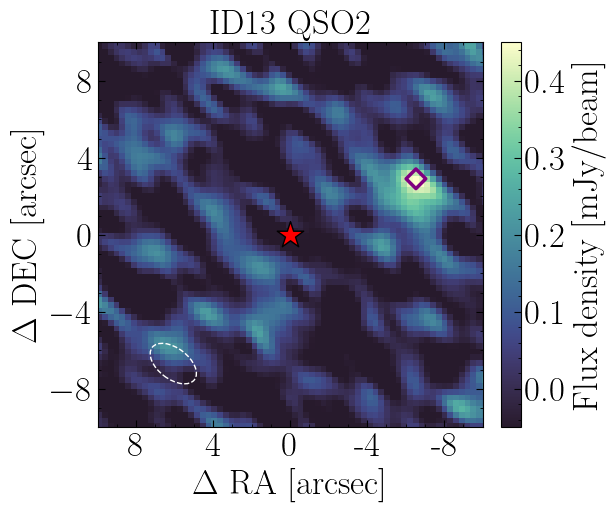}
        \includegraphics[width=0.50\linewidth]{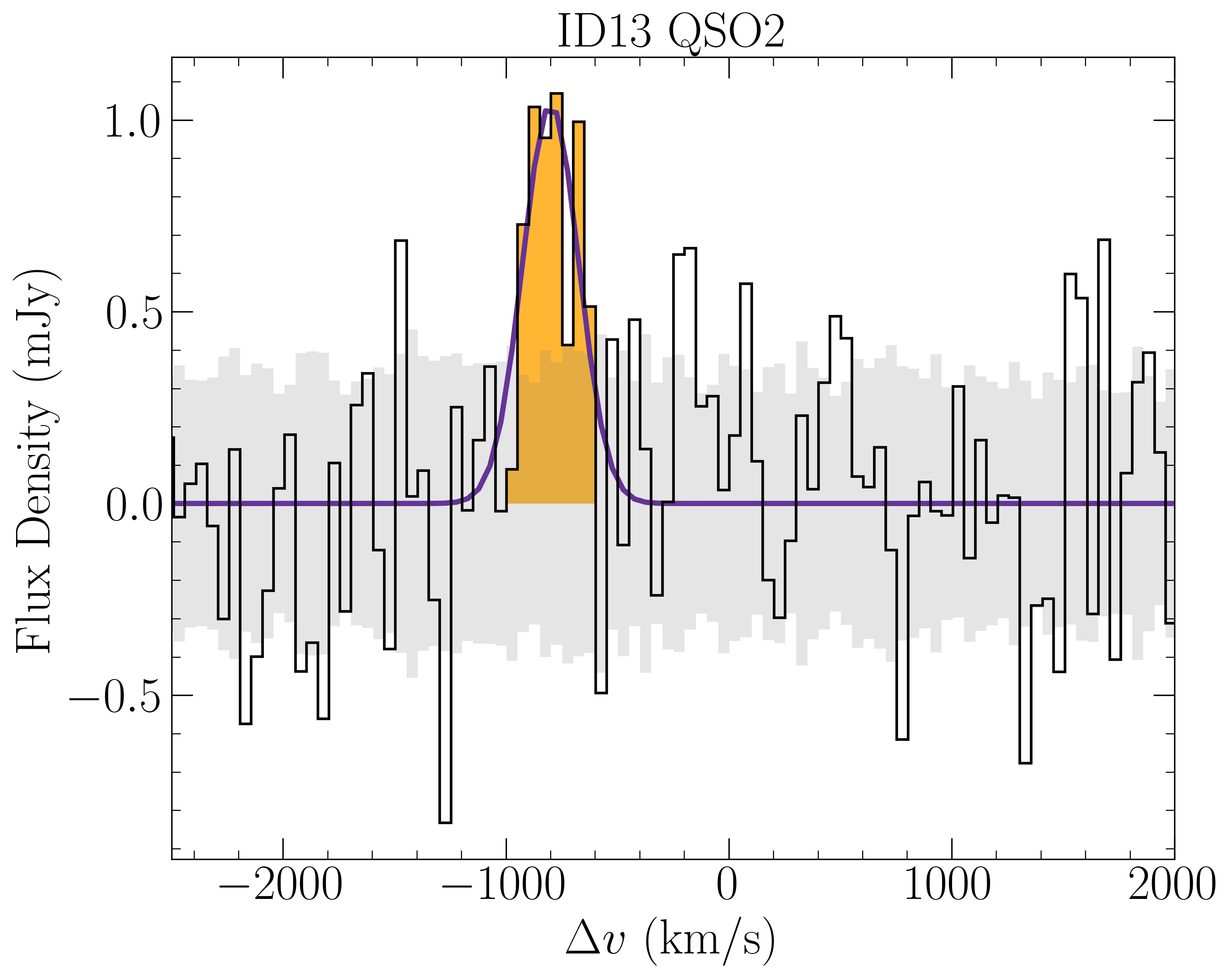}
    \end{minipage}

    \begin{minipage}{0.62\linewidth}
        \centering
        \includegraphics[width=0.46\linewidth]{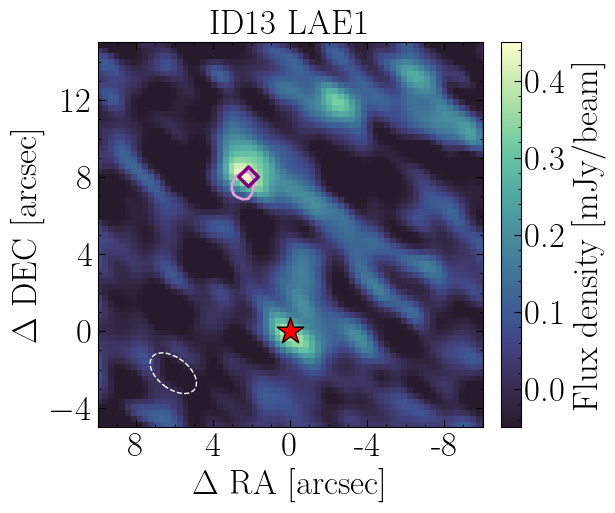}
        \includegraphics[width=0.50\linewidth]{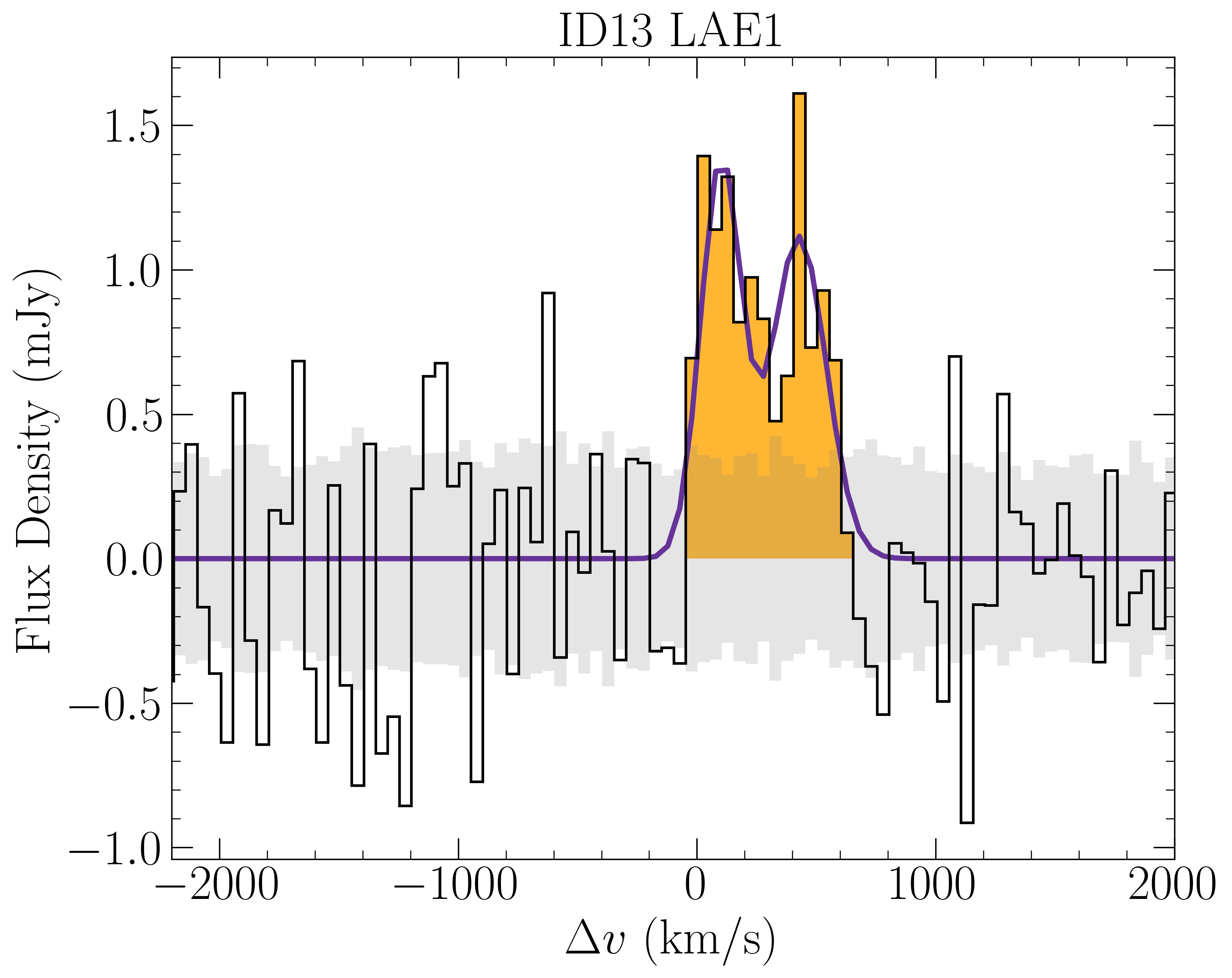}
    \end{minipage}

    \begin{minipage}{0.62\linewidth}
        \centering
        \includegraphics[width=0.46\linewidth]{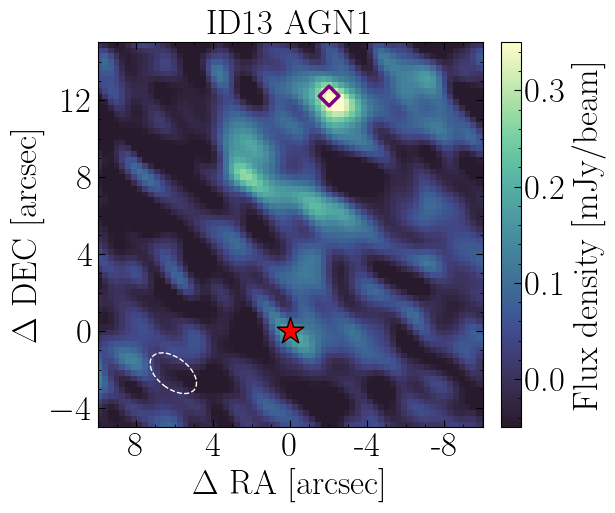}
        \includegraphics[width=0.50\linewidth]{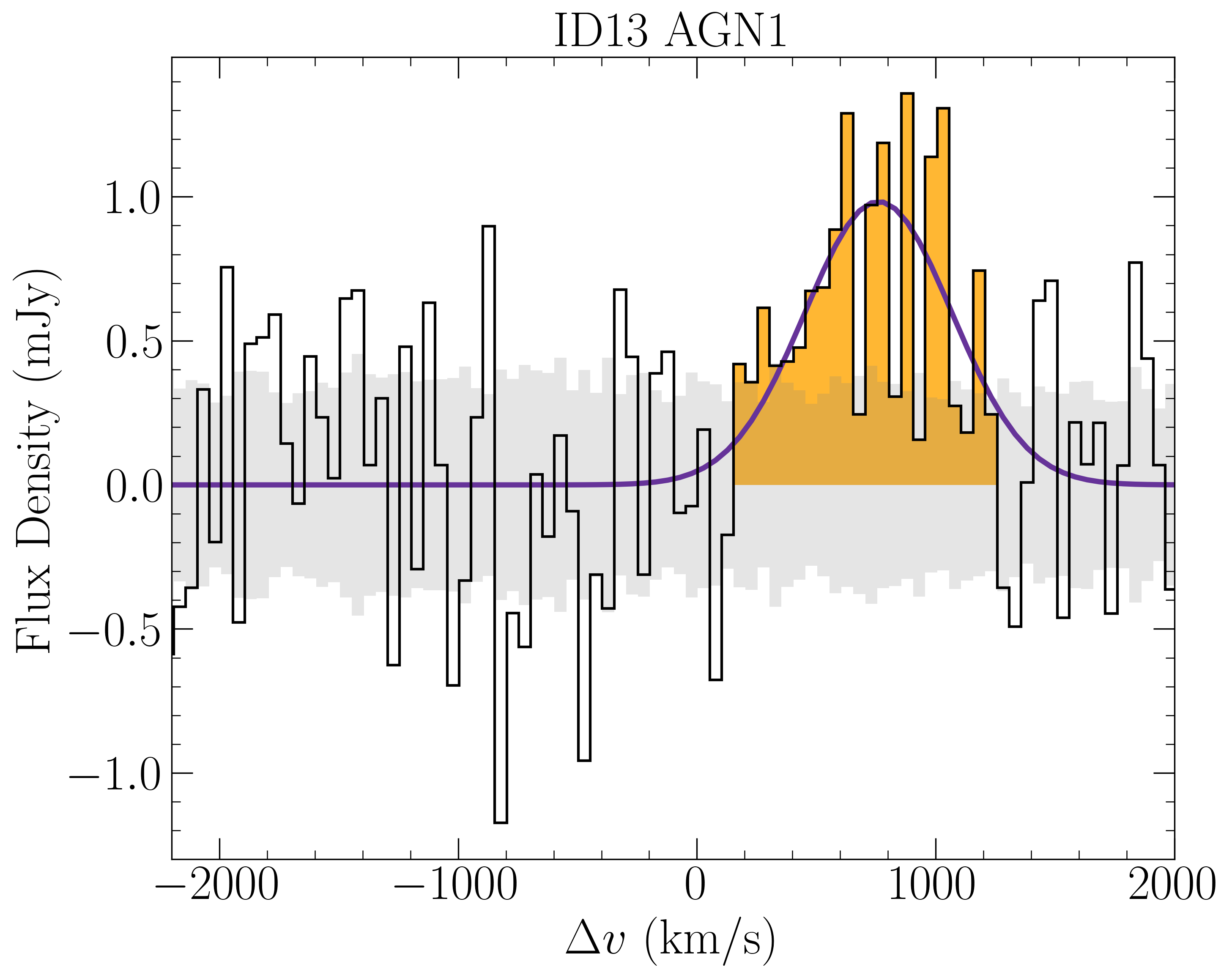}
    \end{minipage}

    \begin{minipage}{0.62\linewidth}
        \centering
        \includegraphics[width=0.46\linewidth]{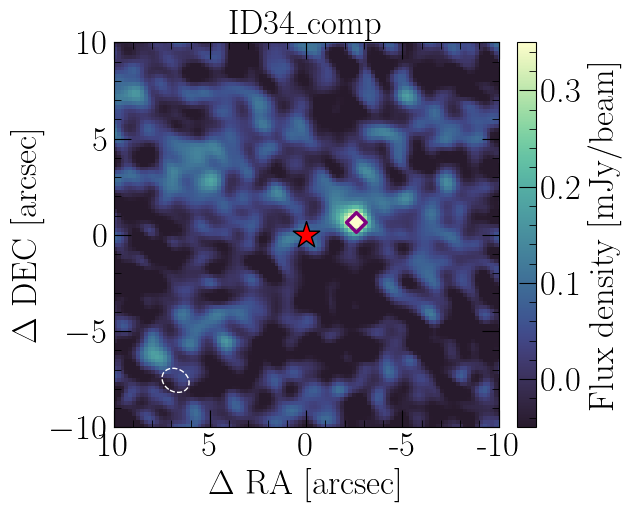}
        \includegraphics[width=0.50\linewidth]{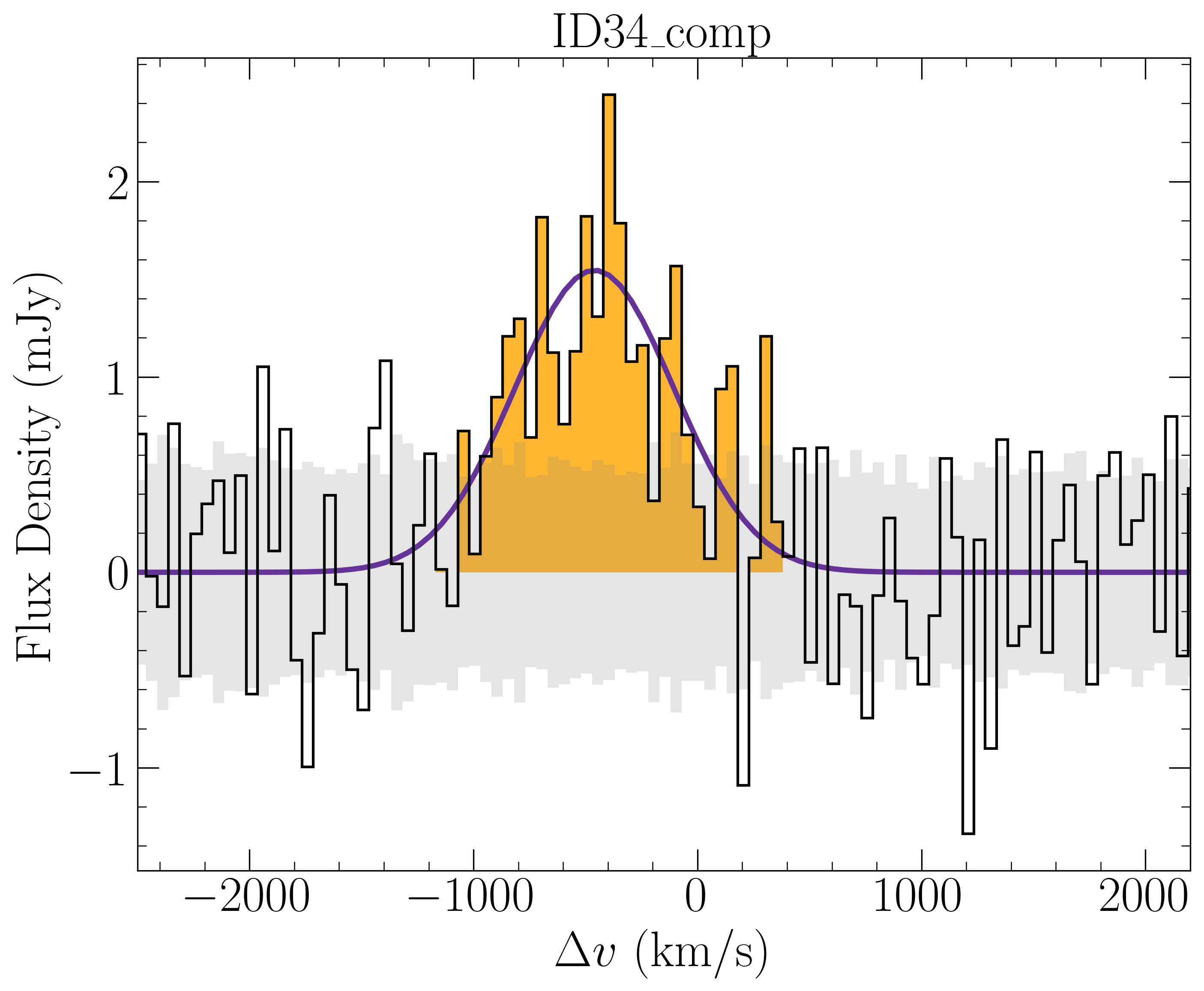}
    \end{minipage}

    \caption{continued.}
    \label{fig:flux_maps_spectra_companions_2}
\end{figure*}

\begin{figure*}[htbp]
    \addtocounter{figure}{-1}
    \centering

     \begin{minipage}{0.62\linewidth}
        \centering
        \includegraphics[width=0.46\linewidth]{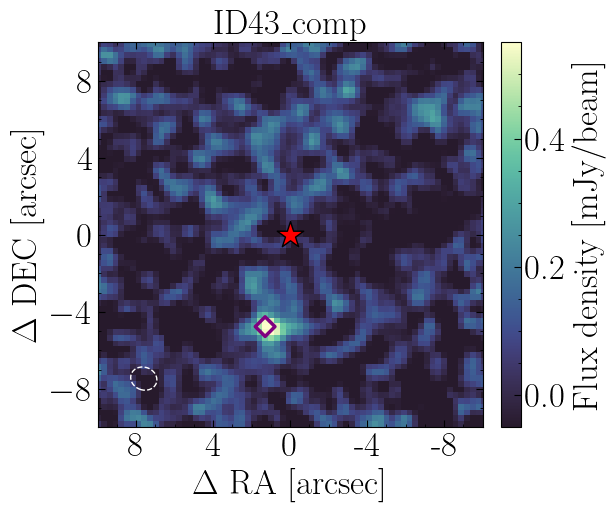}
        \includegraphics[width=0.50\linewidth]{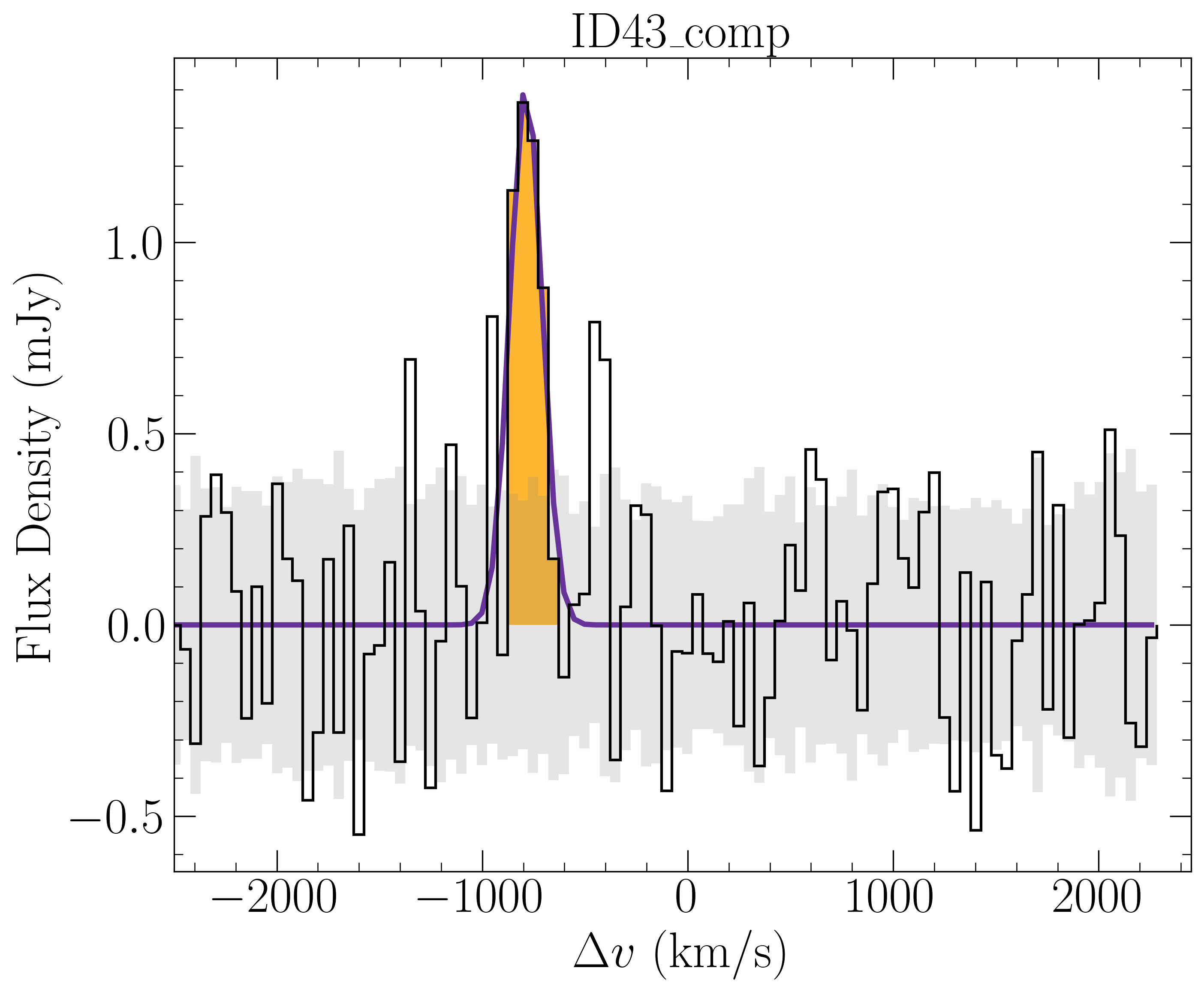}
    \end{minipage}
    
    \begin{minipage}{0.62\linewidth}
        \centering
        \includegraphics[width=0.46\linewidth]{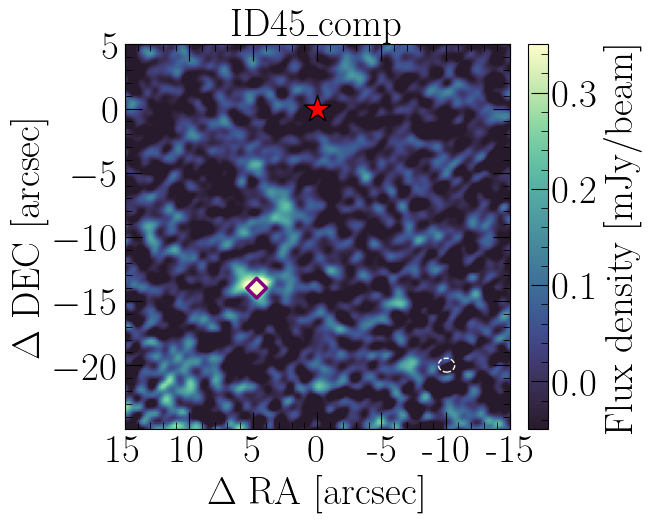}
        \includegraphics[width=0.50\linewidth]{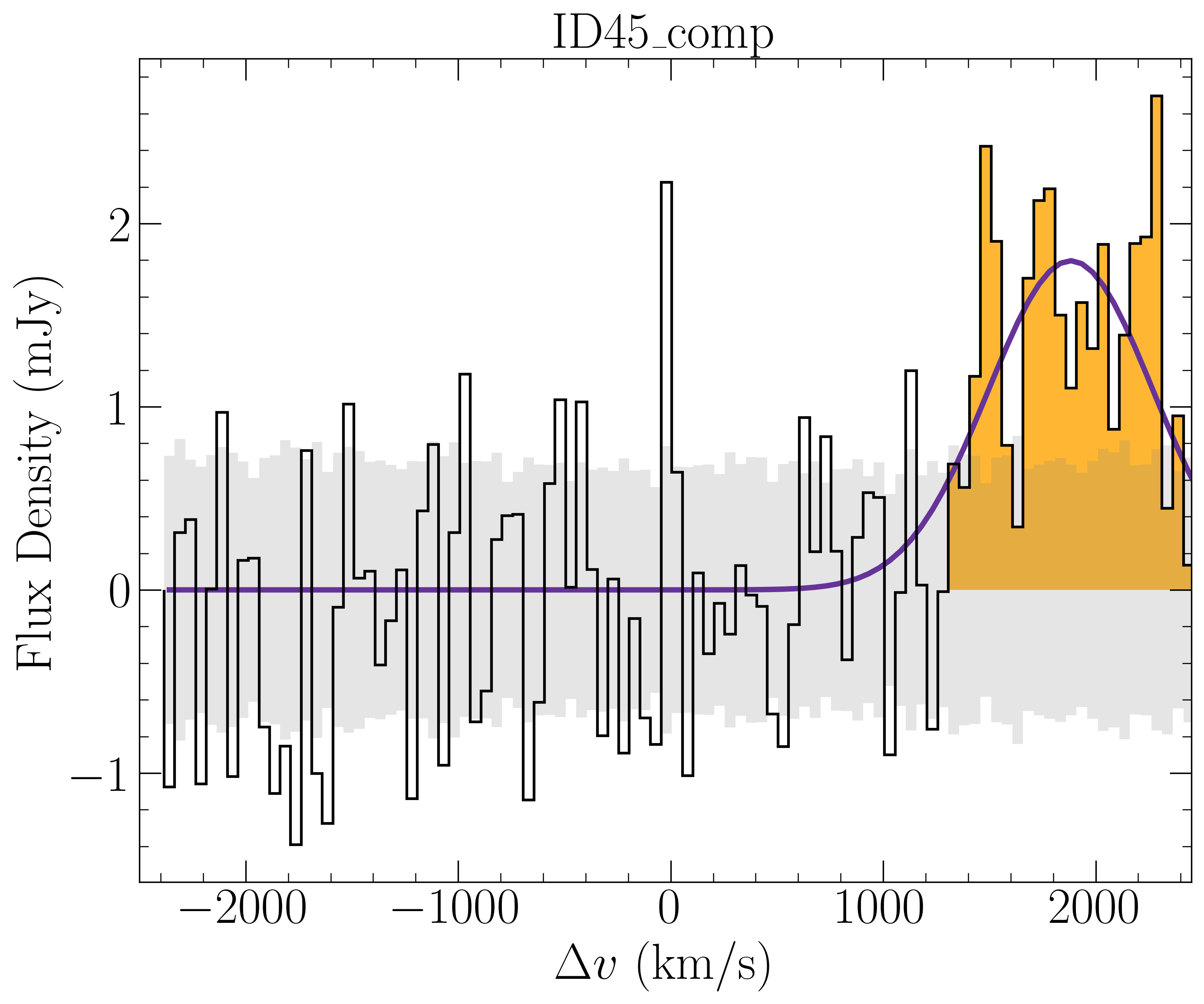}
    \end{minipage}

    \begin{minipage}{0.62\linewidth}
        \centering
        \includegraphics[width=0.46\linewidth]{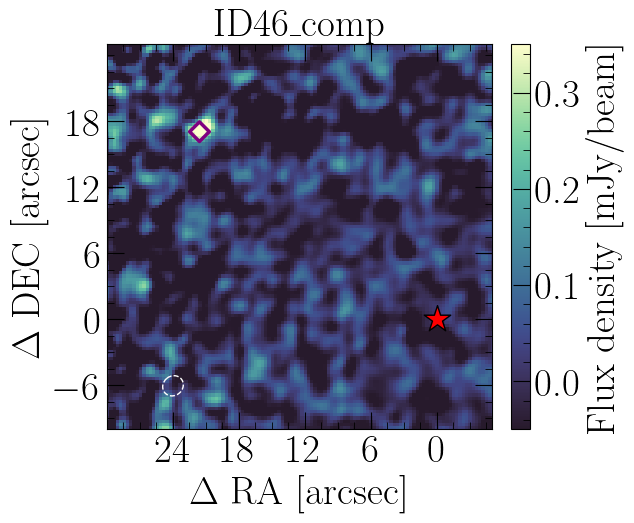}
        \includegraphics[width=0.50\linewidth]{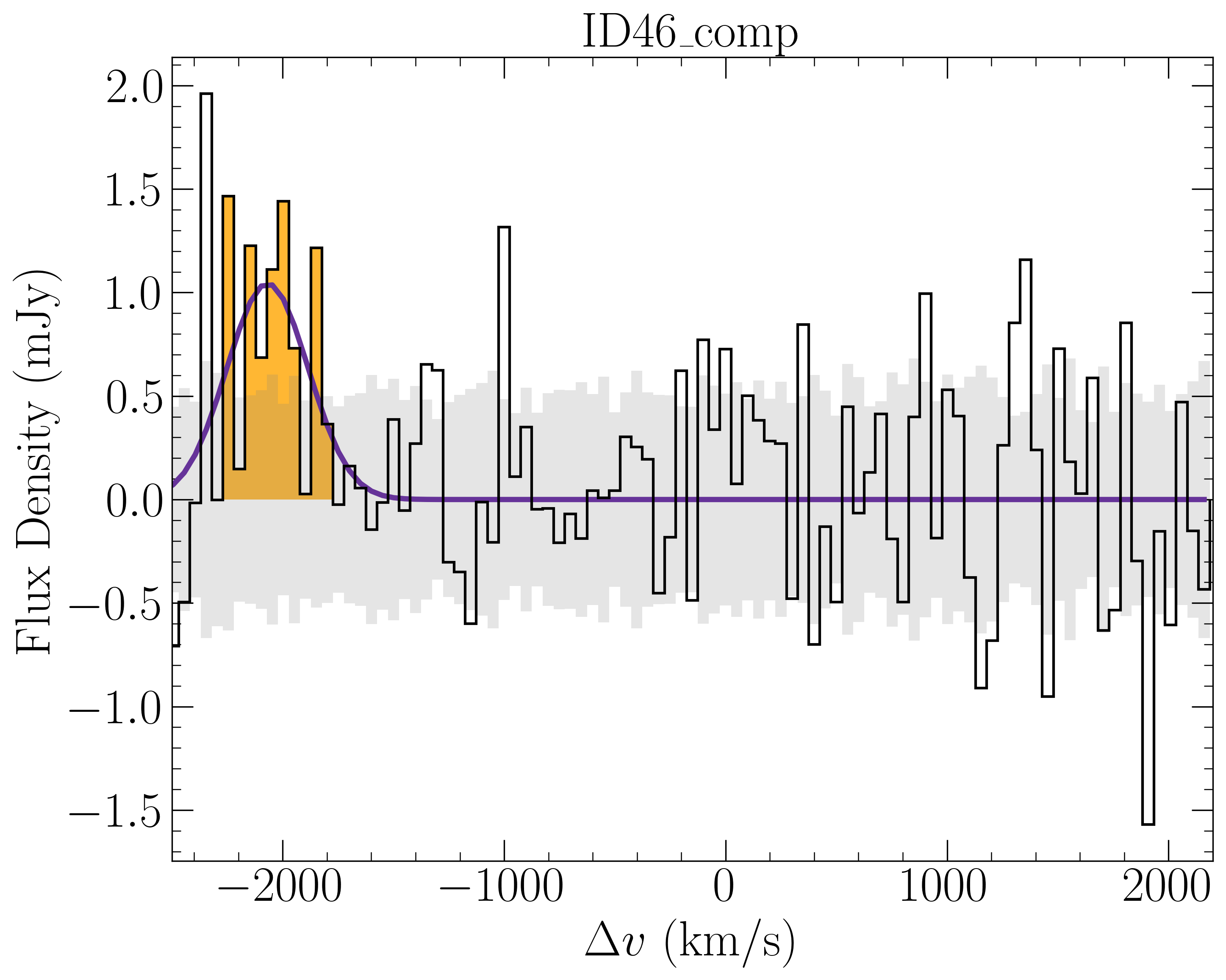}
    \end{minipage}

    \begin{minipage}{0.62\linewidth}
        \centering
        \includegraphics[width=0.46\linewidth]{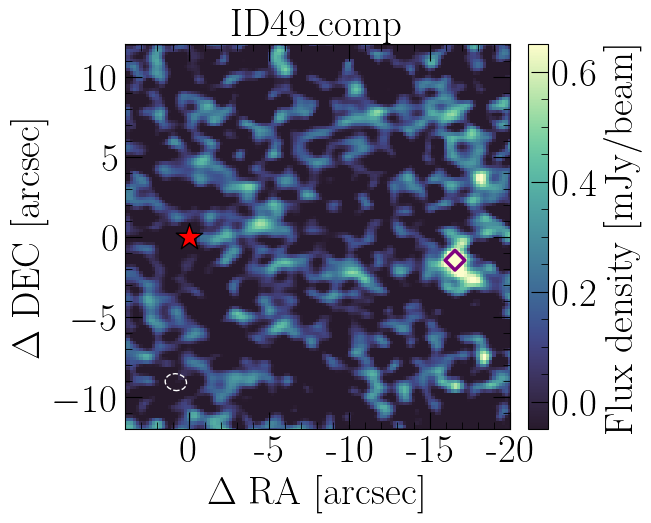}
        \includegraphics[width=0.50\linewidth]{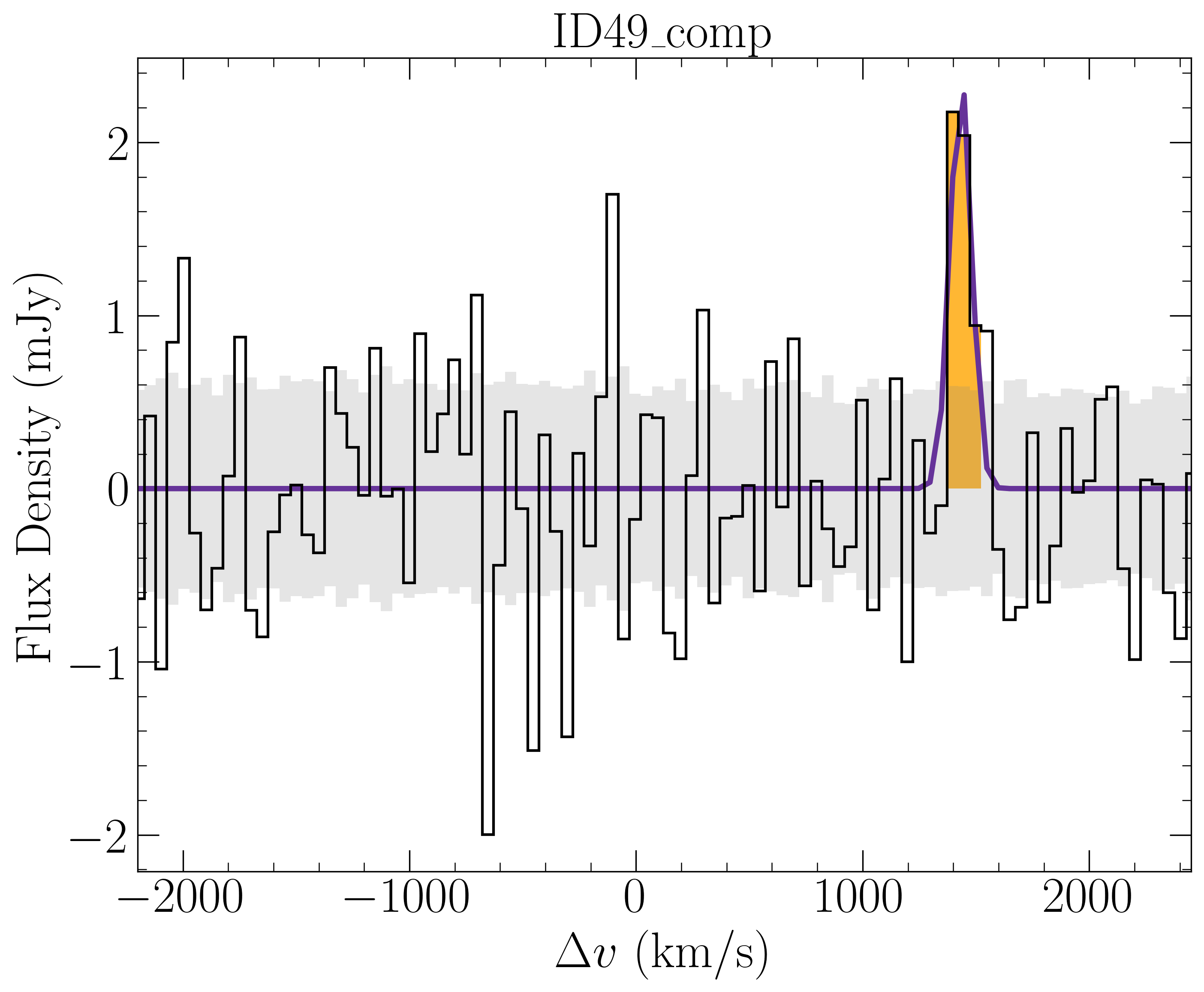}
    \end{minipage}

    \begin{minipage}{0.62\linewidth}
        \centering
        \includegraphics[width=0.46\linewidth]{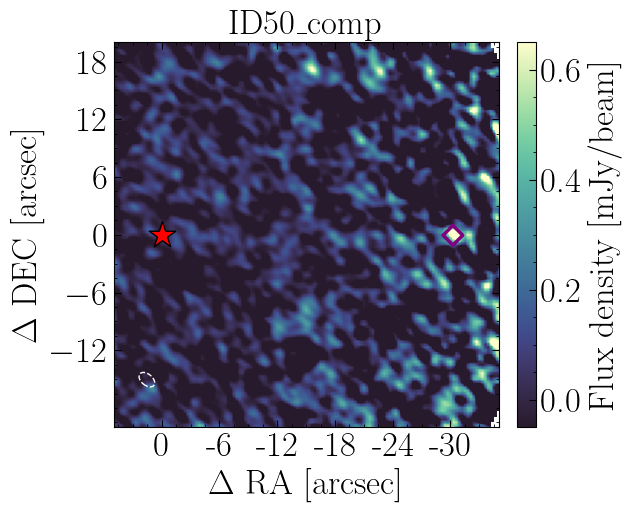}
        \includegraphics[width=0.50\linewidth]{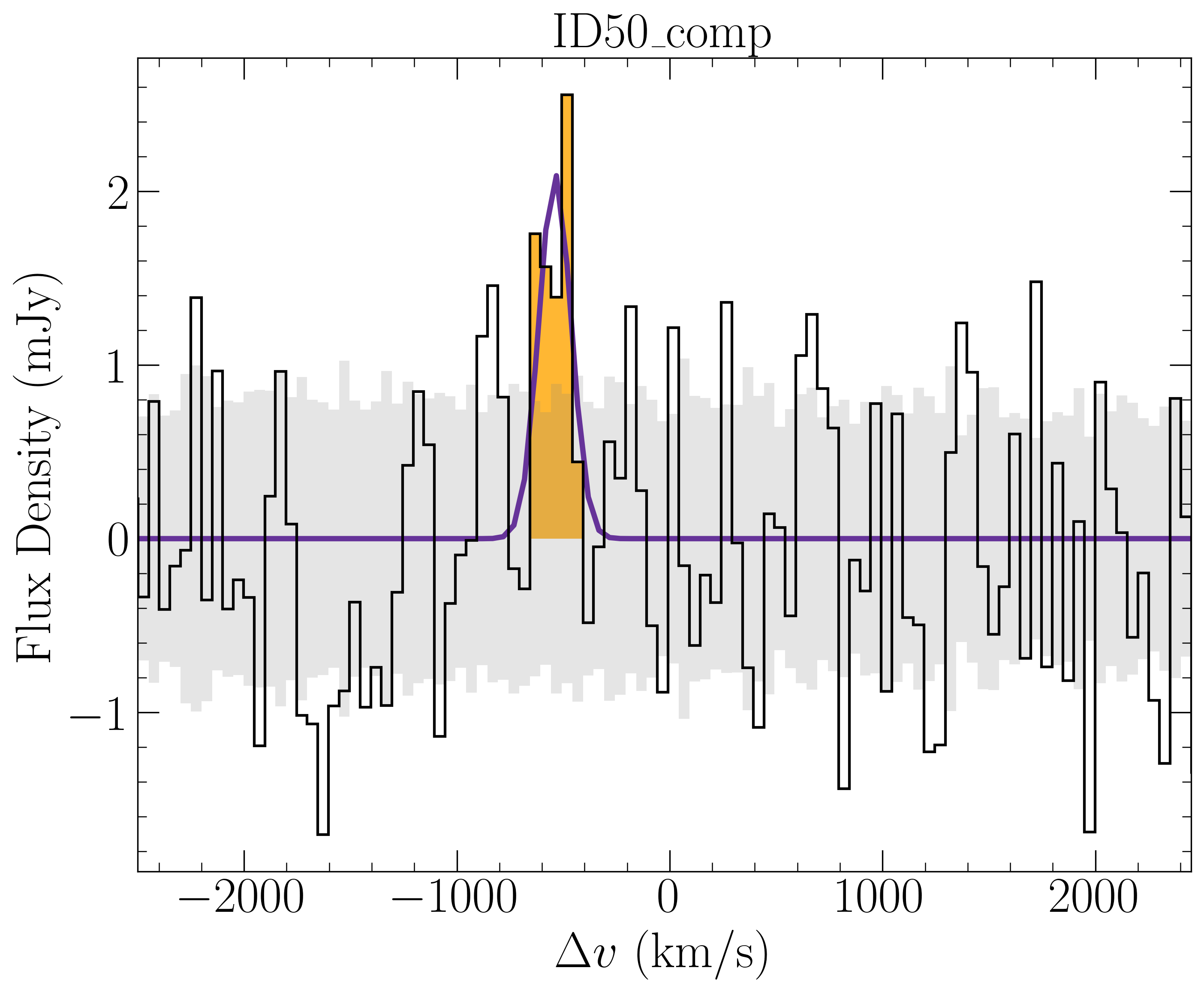}
    \end{minipage}

    \caption{continued.}
    \label{fig:flux_maps_spectra_companions_3}
\end{figure*}

\clearpage

\renewcommand{\arraystretch}{1.0}
\begin{table*}[htbp]
\centering
\begin{threeparttable}
\caption{Coordinates and properties of the CO(4-3) emission for the companion galaxies.}
\label{tab:companion_properties}

\setlength{\tabcolsep}{3.5pt} 
\small 

\begin{tabular}{|c|c|c|c|c|c|c|c|c|c|}
\hline
ID  & RA\tnote{a} & DEC\tnote{a} & $\Delta  d_{\mathrm{QSO}}$ \tnote{b} & $z_\mathrm{CO}$ \tnote{c} & $\Delta v_\mathrm{QSO}$ \tnote{d}& FWHM \tnote{e}  & $S\Delta v$ \tnote{f} & $L'_\mathrm{CO(4-3)}$  & Fid. \tnote{g} \\
 & [J2000] &  [J2000] & [kpc] &  & [$\mathrm{km\,s^{-1}}$] &[$\mathrm{km\,s^{-1}}$] &[Jy\,$\mathrm{km\,s^{-1}}$] &  [$10^9\,\mathrm{K\,km\,s^{-1}\,pc^2}$] & \\
\hline
ID2\_comp & 00:15:29.709 & +06:40:08.8 &  $\sim 282$ & $3.1591\pm 0.0006$ & $-250\pm14$ &$285\pm80$ & $0.88\pm0.16$ &  $24.9\pm4.4$ & 87\\
 ID8\_comp & 03:36:25.990 & -20:19:39.8 &  $\sim 110$ & $3.1117\pm 0.0003$ & $-1398\pm72$ & $338\pm85$ & $0.48\pm0.05$ &  $13.2\pm1.3$ & 100\\
 ID10\_comp1 & 10:25:11.407 & +04:52:25.9 &  $\sim 260$ & $3.242\pm0.001$& $-229\pm71$ & $467\pm135$ & $0.95\pm0.20$ &  $28.0\pm6.0$& 100 \\
 ID10\_comp2 & 10:25:09.741 & +04:53:04.0 &  $\sim 133$ & $3.2375\pm0.0004$& $-527\pm71$ & $214\pm60$ & $0.33\pm0.07$ &  $9.7\pm2.0$& 92\\
 ID11\_comp \tnote{\textdagger} & 02:46:33.984 & -30:05:08.0 &  $\sim 104$ & $3.1069\pm0.0006$ & $438\pm22$& $905\pm94$ & $2.62\pm0.20$ &  $71.8\pm5.5$ & 100\\
 ID13 QSO2 \tnote{*} & 10:20:09.573 & +10:40:05.6 &  $\sim 54$ & $3.1575\pm0.0005$& $-799\pm22$ & $295\pm69$ & $0.33\pm0.06$ &  $9.2\pm1.7$ & 99\\
 ID13 LAE1\tnote{*\textdagger} & 10:20:10.142 & +10:40:10.7 &  $\sim 64$ & $3.1724\pm0.0004$& $268\pm22$ & $413\pm67$ & $0.62\pm0.07$ &  $17.5\pm1.8$& 100 \\
 ID13 AGN1 \tnote{*}  & 10:20:09.837 & +10:40:14.6 &  $\sim 94$ & $3.1798\pm0.0007$& $799\pm22$ & $726\pm138$ & $0.79\pm0.09$ &  $22.5\pm2.4$& 100 \\
 ID34\_comp & 00:59:52.993 & -39:31:56.8 &  $\sim 21$ & $3.2489\pm0.0009$& $-467\pm70$ & $839\pm113$ & $1.43\pm0.16$ &  $42.1\pm4.6$& 100 \\
 ID43\_comp & 01:48:18.264 & -53:27:06.8 &  $\sim 39$ & $3.1193\pm0.0002$ & $-800\pm36$& $183\pm35$ & $0.28\pm0.04$ &  $7.8\pm1.2$& 100 \\
 ID45\_comp & 23:34:46.724 & -09:08:26.2 &  $\sim 112$ & $3.386\pm0.001$& $1944\pm28$ & $912\pm145$ & $1.86\pm0.20$ &  $58.8\pm6.3$& 100\\
 ID46\_comp & 23:58:09.980 & 01:25:24.3 &  $\sim 209$ & $3.3720\pm0.0008$& $-2069\pm27$ & $434\pm135$ & $0.53\pm0.10$ &  $16.5\pm3.0$& 100 \\
 ID49\_comp & 05:39:53.021 & -28:39:57.4 &  $\sim 130$ & $3.1583\pm0.0002$ & $1441\pm3$& $111\pm33$ & $0.26\pm0.05$ &  $7.3\pm1.4$& 88\\
 ID50\_comp & 08:19:38.538 & +08:23:57.9 &  $\sim 234$ & $3.1983\pm0.0005$ & $-548\pm71$& $177\pm65$ & $0.38\pm0.09$ &  $11.1\pm2.7$& 94 \\
\hline
\end{tabular}

\begin{tablenotes}
\footnotesize
\item[a] Right ascension and declination determined with \textsc{Lineseeker} \citep{Gonzalez-Lopez2017}.
\item[b] Projected distance between companion and quasar.
\item[c] CO redshift determined from the line centroid, measured via a first-moment analysis.
\item[d] Line-of-sight velocity offset relative to the quasar systemic redshift. In case the quasar is not detected in CO, the velocity offset is computed with respect to the peak Ly$\alpha$ wavelength of the nebula \citep{GonzalezLobos2025}.
\item[e] Line width.
\item[f] Line intensity.
\item[g] \textsc{Lineseeker} fidelity  (probability the candidate is real) from comparison with the negative data.
\item[*] Companion galaxies of ID13 discovered in \citet{ArrigoniBattaia2018}. We follow their naming conventions: QSO2 denotes a companion quasar of the main quasar, LAE1 a strong Ly$\alpha$ emitter and AGN1 a type-2 AGN.
\item[\textdagger] ID11\_comp and ID13 LAE1 are detected in dust continuum with a flux of $S_{\mathrm{cont}}=0.14\pm0.03\,\mathrm{mJy}$ and $S_{\mathrm{cont}}=0.08\pm0.04\,\mathrm{mJy}$, respectively.
\end{tablenotes}

\end{threeparttable}
\end{table*}

\clearpage

\end{appendix}
\end{document}